\documentclass{article}
\usepackage{spconf,amsmath,graphicx}
\usepackage{times}
\usepackage{epsfig}
\usepackage{graphicx}
\usepackage{amsmath}
\usepackage{amssymb}
\usepackage{cite}
\usepackage{multicol}

\def\swfour{0.23\linewidth}
\def\swfiveh{0.18\linewidth}

\def\sweight{0.11\linewidth}

\newcommand{\reffig}[1]{Fig.\ref{#1}}

\newcommand{\reftable}[1]{Table.\ref{#1}}

\title{Low-dose CT Denoising Using a Structure-Preserving\\ Kernel Prediction Network}
\name{Lu Xu$^{1*}$ \space Yuwei Zhang $^{2*}$ \thanks{*Equal contribution. This work was done when Lu Xu was an intern at SenseTime. $\dagger$Corresponding to jimmy.sj.ren@gmail.com, Weijingwei2014@ia.ac.cn and yezhaoxiang@163.com. This work is supported by National Natural Science Foundation of China under Grant No.82001917, 81930053.} Ying Liu$^{2}$ Daoye Wang$^{3}$ \space Mu Zhou$^{4}$ Jimmy Ren$^{5,6\dagger}$ \space Jingwei Wei $^{7\dagger}$ \space Zhaoxiang Ye$^{2\dagger}$ }

\address{$^{1}$The Chinese University of Hong Kong \space
	$^{2}$ Tianjin Medical University Cancer Institute and Hospital\\
	$^{3}$ ETH Zurich \space
	$^{4}$ SenseBrain Technology Limited LLC\space
	$^{5}$ SenseTime Research\\
	$^{6}$ Qing Yuan Research Institute, Shanghai Jiao Tong University\\
	$^{7}$ Key Laboratory of Molecular Imaging, Institute of Automation, Chinese Academy of Sciences
}
\begin{document}
	%
	\maketitle
	\begin{abstract}
		Low-dose CT has been a key diagnostic imaging modality to reduce the potential risk of radiation overdose to patient health. Despite recent advances, CNN-based approaches typically apply filters in a spatially invariant way and adopt similar pixel-level losses, which treat all regions of the CT image equally and can be inefficient when fine-grained structures coexist with non-uniformly distributed noises. To address this issue, we propose a Structure-preserving Kernel Prediction Network (StructKPN) that combines the kernel prediction network with a structure-aware loss function that utilizes the pixel gradient statistics and guides the model towards spatially-variant filters that enhance noise removal, prevent over-smoothing and preserve detailed structures for different regions in CT imaging. Extensive experiments demonstrated that our approach achieved superior performance on both synthetic and non-synthetic datasets, and better preserves structures that are highly desired in clinical screening and low-dose protocol optimization.
	\end{abstract}
	\section{Introduction}
	Low-dose computed tomography (LDCT) is of great clinical importance to reduce radiation risk to patients in clinical screening and diagnosis 
	.
	Extensive efforts have been made to develop computational approaches for low-dose image reconstruction and denoising. Examples include sinogram filtering \cite{balda2012ray,wang2006penalized}, iterative reconstruction \cite{sidky2008image,xu2012low,cai2014cine}, and image post-processing \cite{li2014adaptive,kang2013image,chen2013improving} algorithms. However, these studies tend to show a slow computational convergence with over-smoothing appearances.
	Furthermore, preserving high-quality image characteristics is still challenging from noisy inputs and corresponding CT-based interpretation is yet to be elucidated.

	Recent deep learning approaches have demonstrated their superiority for LDCT  denoising and post-processing.
	For example,
	Chen \emph{et al.} proposed RED-CNN \cite{chen2017low} with convolution and deconvolution layers and skip connections for image denoising.  
	Kang \emph{et al.} \cite{kang2017deep} proposed applying U-net \cite{ronneberger2015u} for denoising on the wavelet coefficients.
	Despite advances, 
	we recognize that over-smoothed outputs and potential artifacts are commonly seen in these studies.
	%
	%
	Existing CNN-based methods~\cite{chen2017low}~\cite{kang2017deep} approximate the input-dependent and spatially variant mapping between noisy and ground truth images with an overcomplicated non-linear filter, constructed by a sequence of spatial invariant kernels and non-linear activation functions in convolution layers.
	When the noise distribution is close to uniform across the image, a sequence of convolution layers can give a close approximation of the desired mapping.
	It is known that CT images contain a relatively narrow span of contents representing specific regions of the human organs. However, the noise patterns are highly variant in different regions within an image and across different images.
	In addition, fine-grained structures are widely seen in the targeted normal dose CT (NDCT) images.
	It thus becomes increasingly challenging to perform accurate discrimination between these structures and the spatially variant noises.

	Dispite CNN models can form more complicated mappings by increasing the number of filter channels and convolution layers, the efficiency of such an approximation approach remains questionable.
	%
	%
	Furthermore, during model training, these CNN networks usually apply the same loss to every pixel in the image. Such uniform losses guide the network towards the same objective for inputs with different characteristics, which can result in sub-optimal convergence and can cause over-smoothing, artifacts, and less detailed contents. 
	%
	%
	
	To address the above challenges, we propose a Structure-preserving Kernel Prediction Network, termed as StructKPN, to achieve strong performance on low-dose medical image denoising. Our approach draws inspiration by kernel prediction network (KPN)\cite{mildenhall2018burst}, a kernel-based network promises to generate a dedicated filter for every pixel of the input, offering superior ability for approximating input dependent filters. In our task, our analysis combines two key insights that (1) using kernel weights on pixels greatly improves local pattern consistency; (2) unlike applying non-linear transformations to the input image, our approach specializes in predicting kernel weights and makes the network input-dependent by nature. Furthermore, we introduce a novel structure-aware objective function by utilizing gradient statistics of local neighborhoods and assigning weights to different loss functions. 
	We demonstrated that StructKPN can generate spatially-variant filters to significantly enhance noise removal in flat areas, prevent over-smoothing at sharp edges and preserve details in fine-structured regions of CT images.

	
	
	
	\begin{figure}[!t]
		\includegraphics[width=\linewidth]{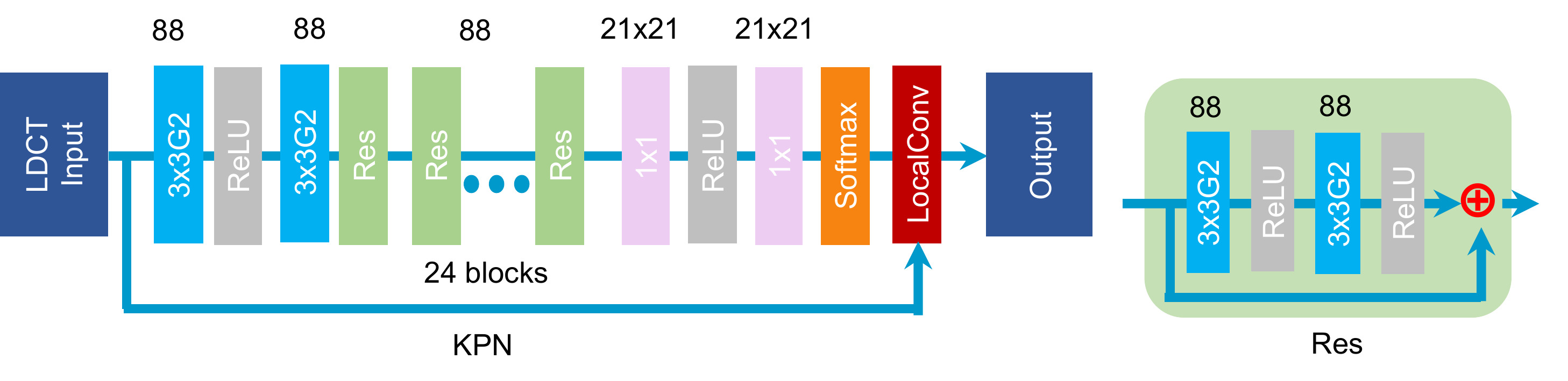}
		
		\caption{Overall structure of the kernel prediction network. Numbers on the top of the layers denote numbers of output channels. $3 \times 3$ and $1 \times 1$ denote the convolution layers with kernel size $3 \times 3$ and $1 \times 1$. G2 denotes convolution with 2 groups. Res denotes the residual blocks. LocalConv denotes the Local Convolution Layer that filters input pixels with dedicated kernels according to Eq.~\ref{eq:kpn}. } \label{networkfig}
	\end{figure}

	\section{Methods}
	
	\setlength{\tabcolsep}{1.4pt}
	\begin{figure}[!t]
		\begin{center}
			\begin{tabular}{cccc}
				\includegraphics[width=\swfour]{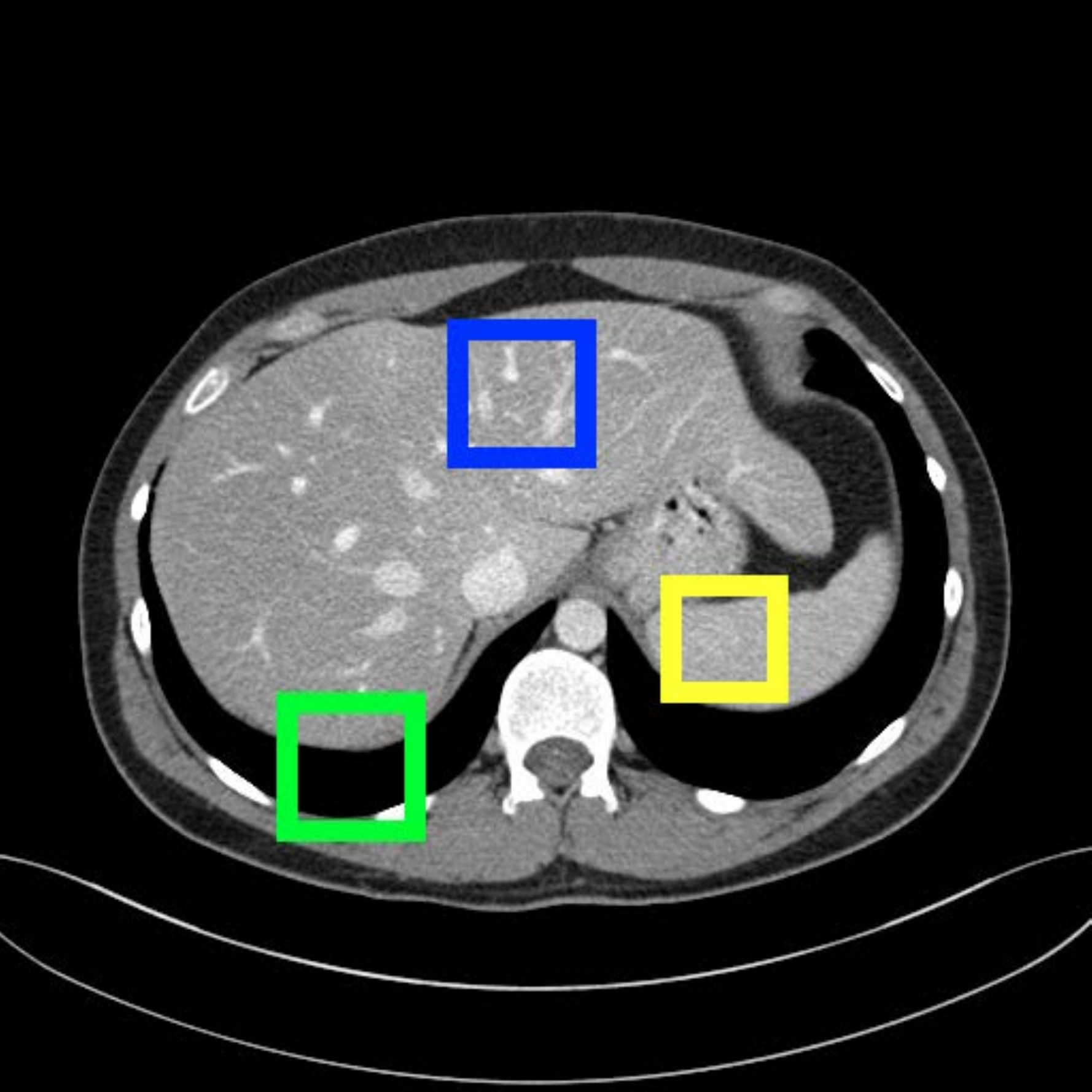} &
				\includegraphics[width=\swfour]{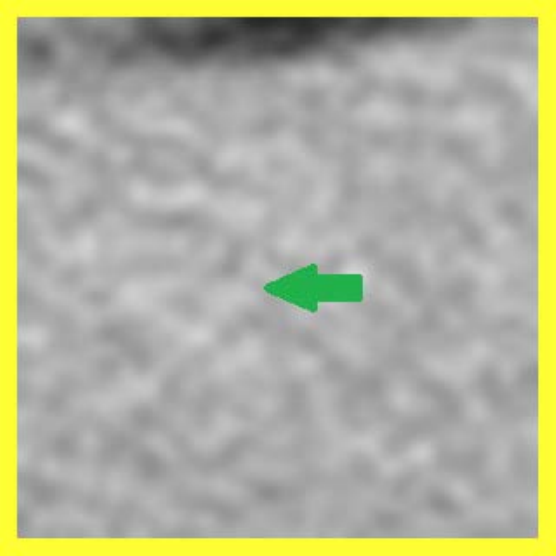} &
				\includegraphics[width=\swfour]{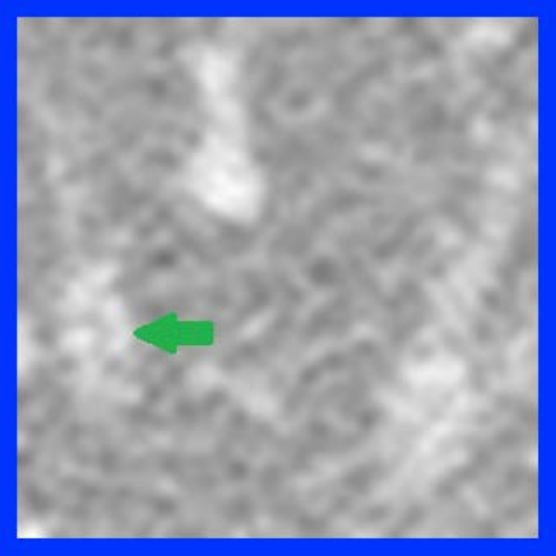} &
				\includegraphics[width=\swfour]{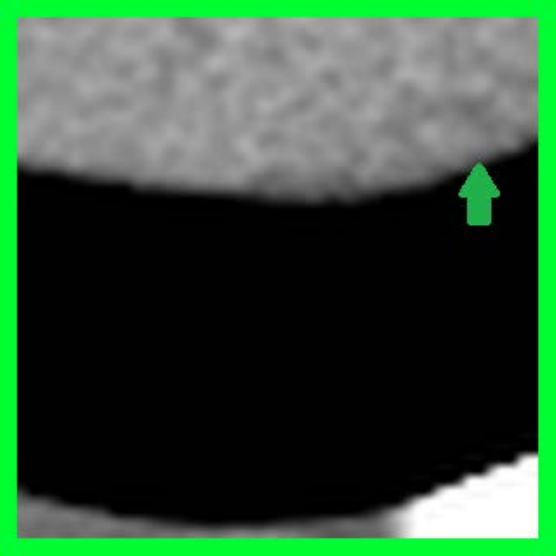}
				\\
				(a) &	(b)  & (c)  & (d)\\
			\end{tabular}
			\\
			\begin{tabular}{cccc}
				\includegraphics[width=0.18\linewidth]{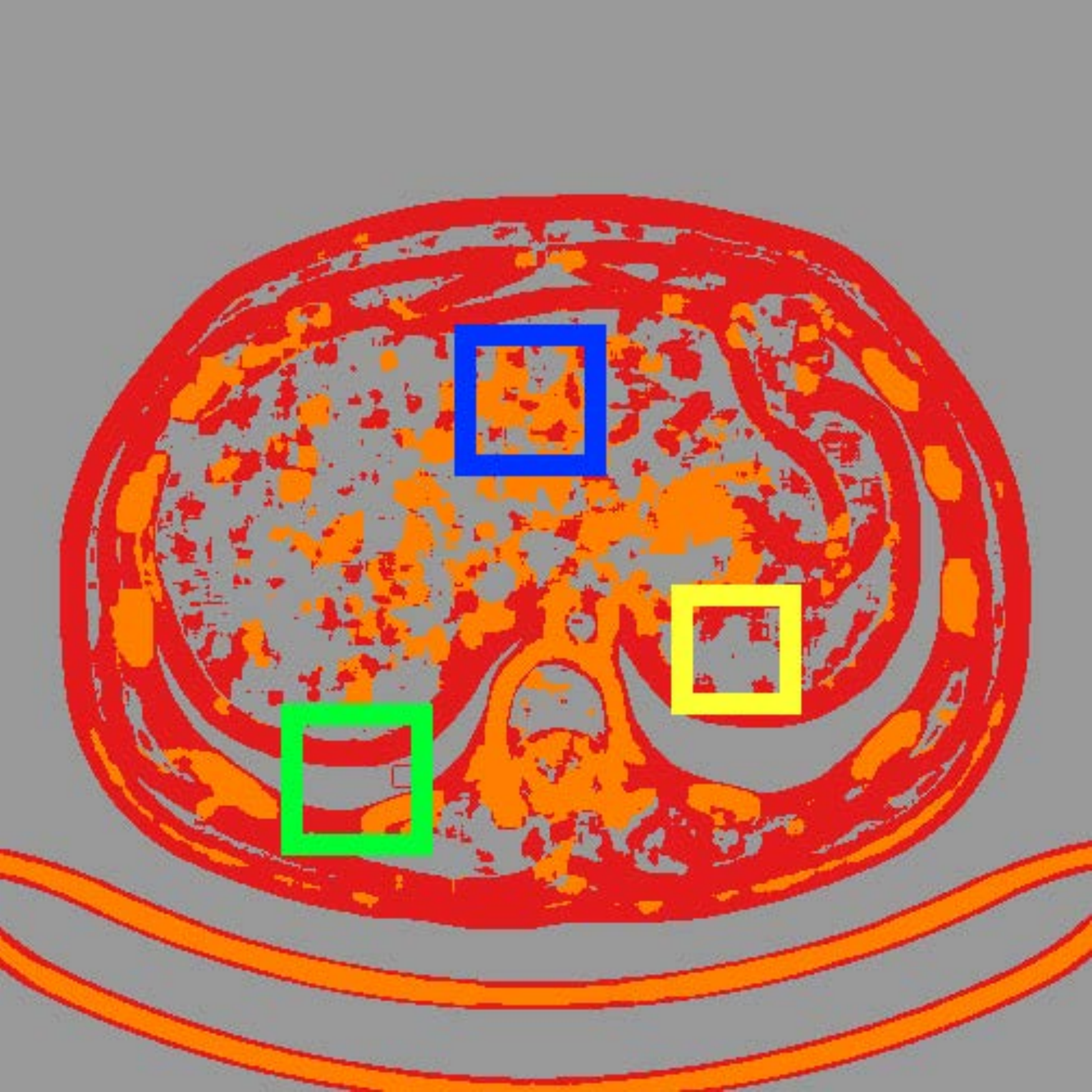} &
				\includegraphics[width=0.26\linewidth]{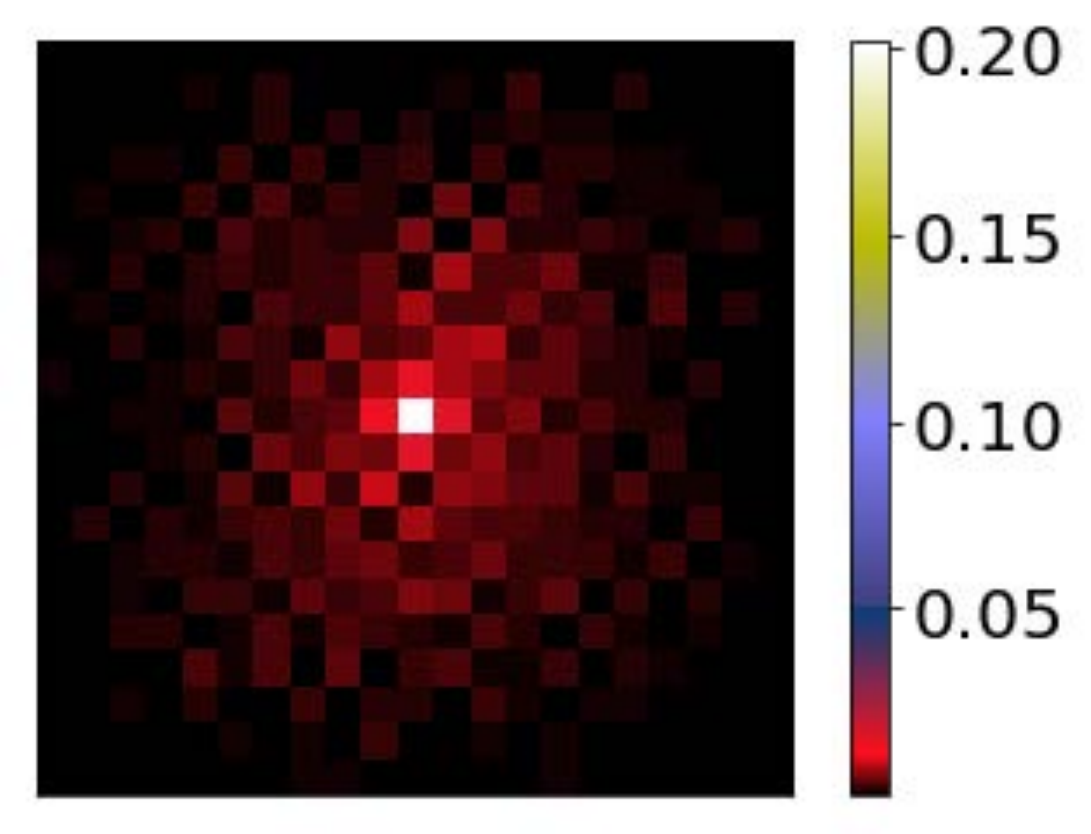} &
				\includegraphics[width=0.24\linewidth]{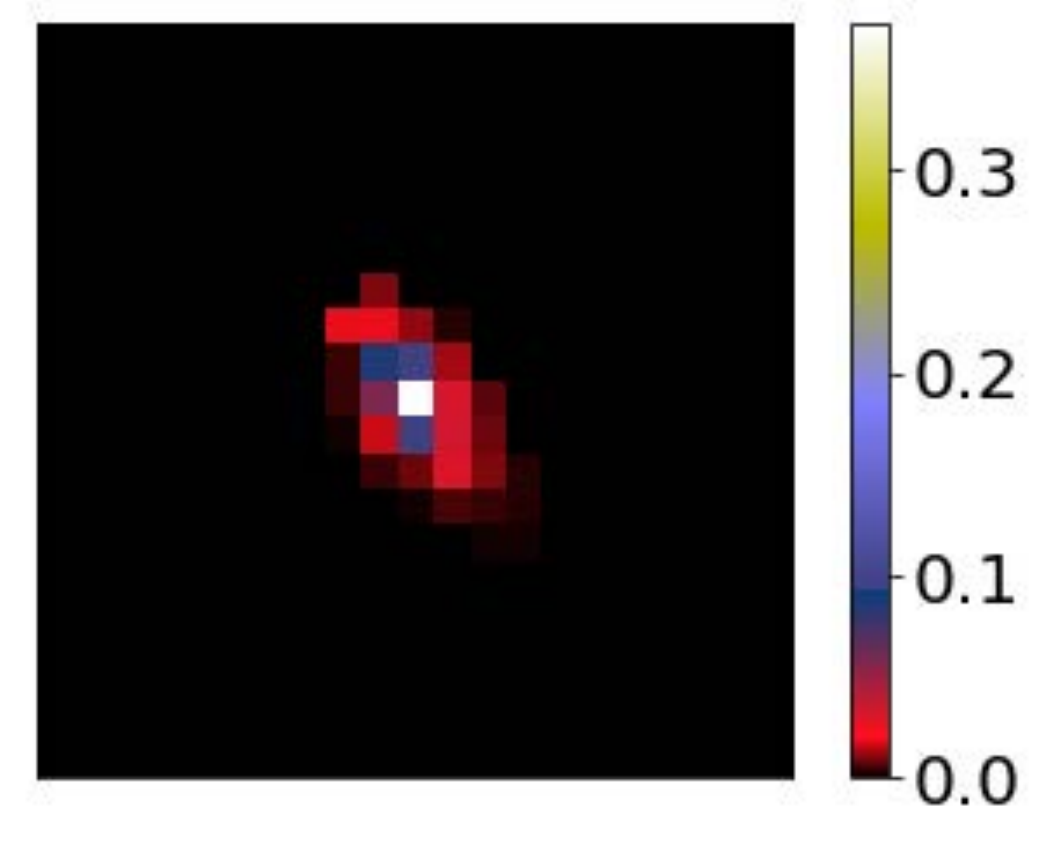} &
				\includegraphics[width=0.24\linewidth]{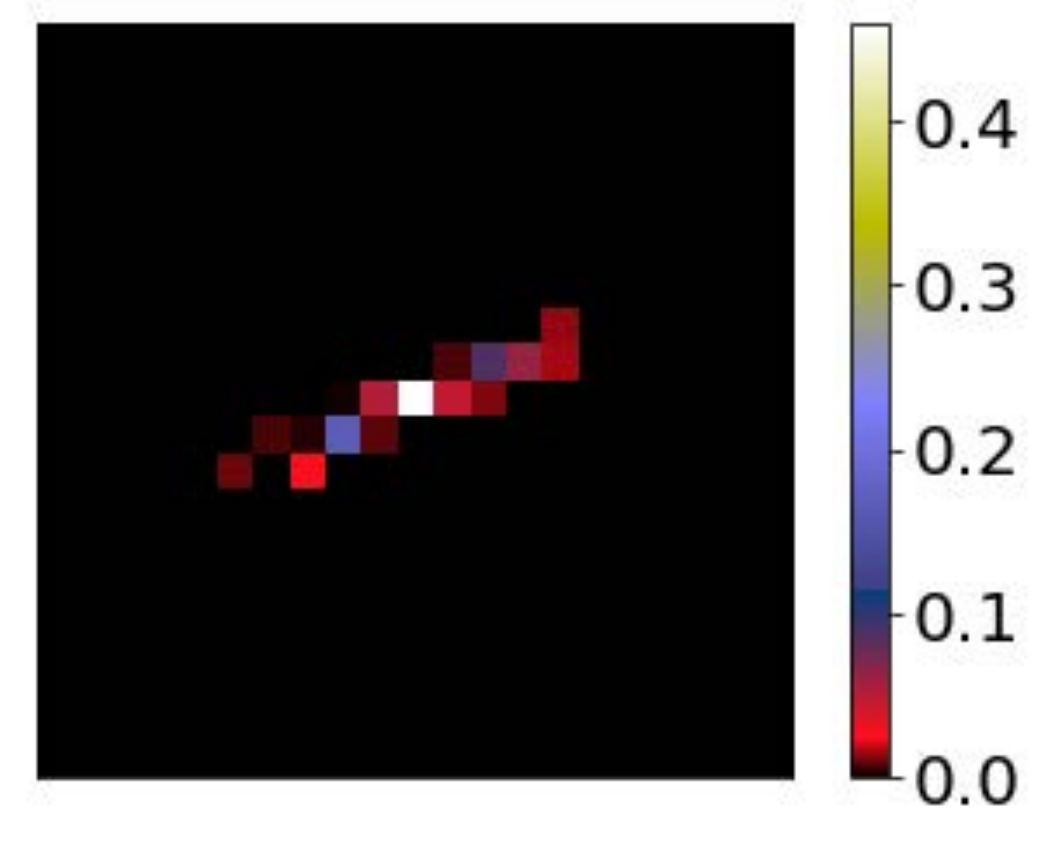}
				\\
				(e) &  (f) & (g) & (h)\\
			\end{tabular}
		\end{center}
		\caption{
			(a) An NDCT image from the NIH dataset. (e): The corresponding components of Structure-aware Loss that have the largest weight, where red, orange and gray represent L2, SSIM and L1 loss respectively. (b)(c)(d) show the enlarged regions marked in (a). The green arrows indicate where the kernels are taken for visualization. (f)(g)(h) show visualization of representative kernels produced by the proposed StructKPN in regions (b)(c)(d) respectively. At flat regions, StructKPN tends to smooth over a large neighborhood of the center pixel. At fine-grained structures, StructKPN enhances structure preservation by learning asymmetric filters that fit specific local structures.  
		}
		\label{fig:structloss}
	\end{figure}
	
	\begin{figure}[!t]
		\begin{center}
			\begin{tabular}{cccc}
				\includegraphics[width=\swfiveh]{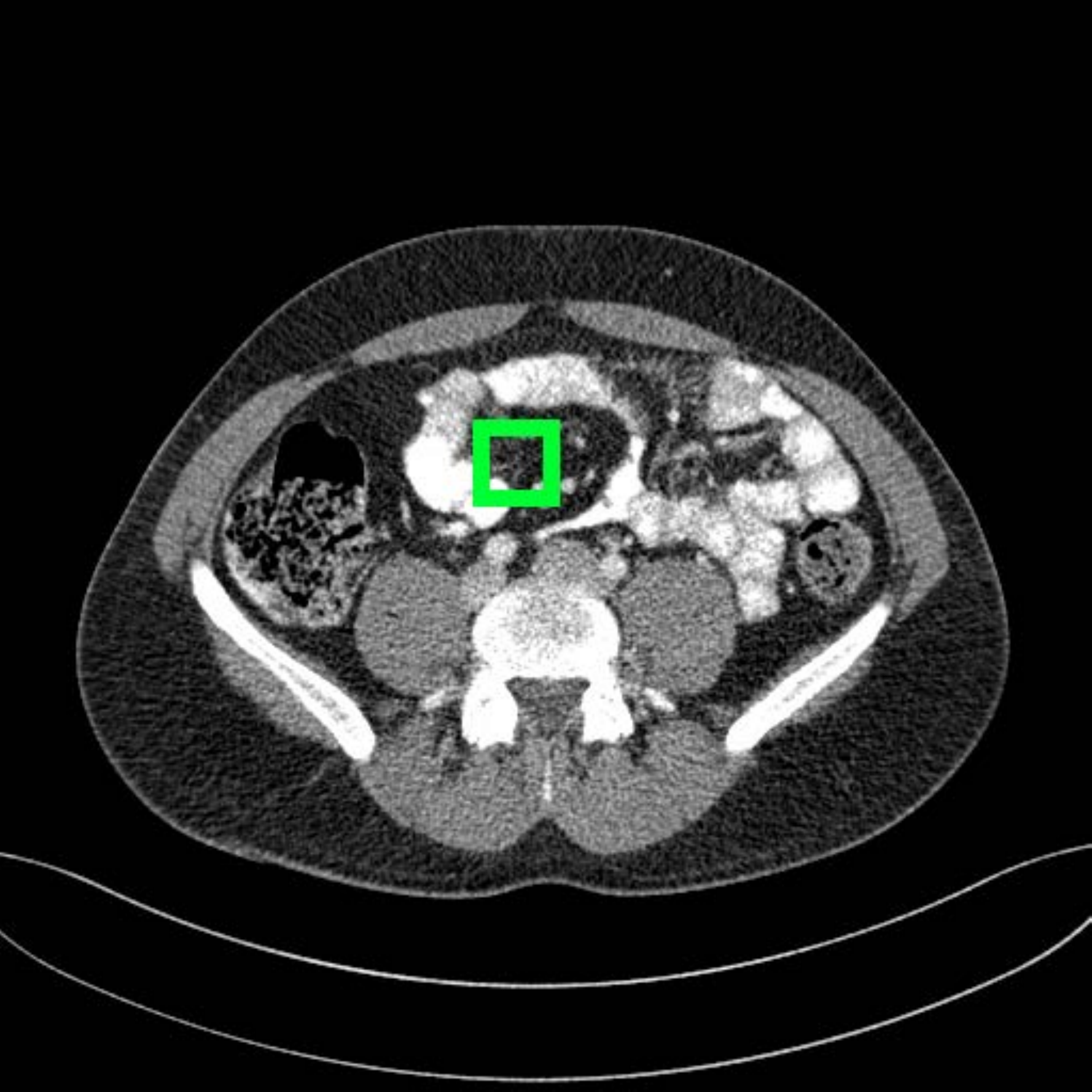} &
				\includegraphics[width=\swfiveh]{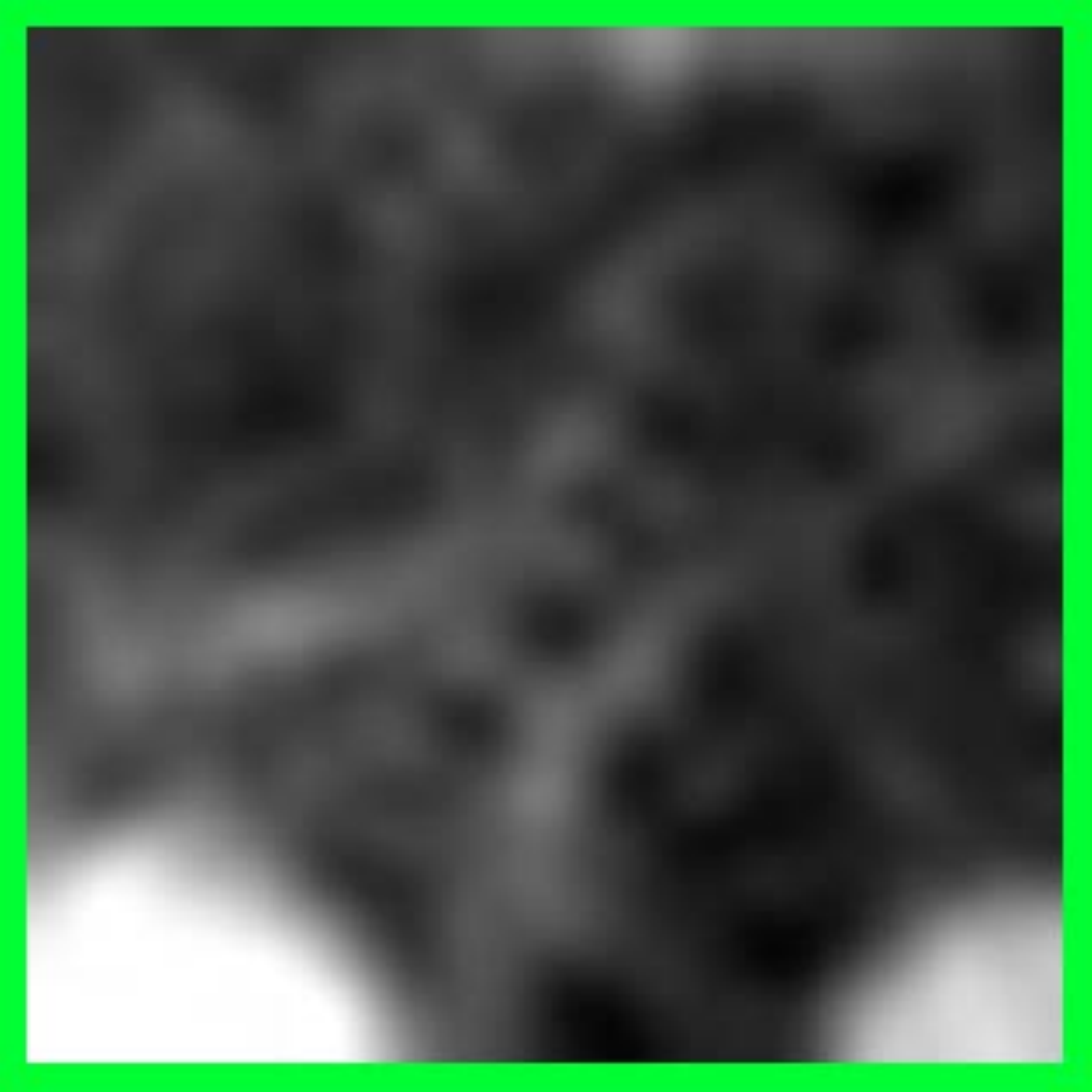} &
				\includegraphics[width=\swfiveh]{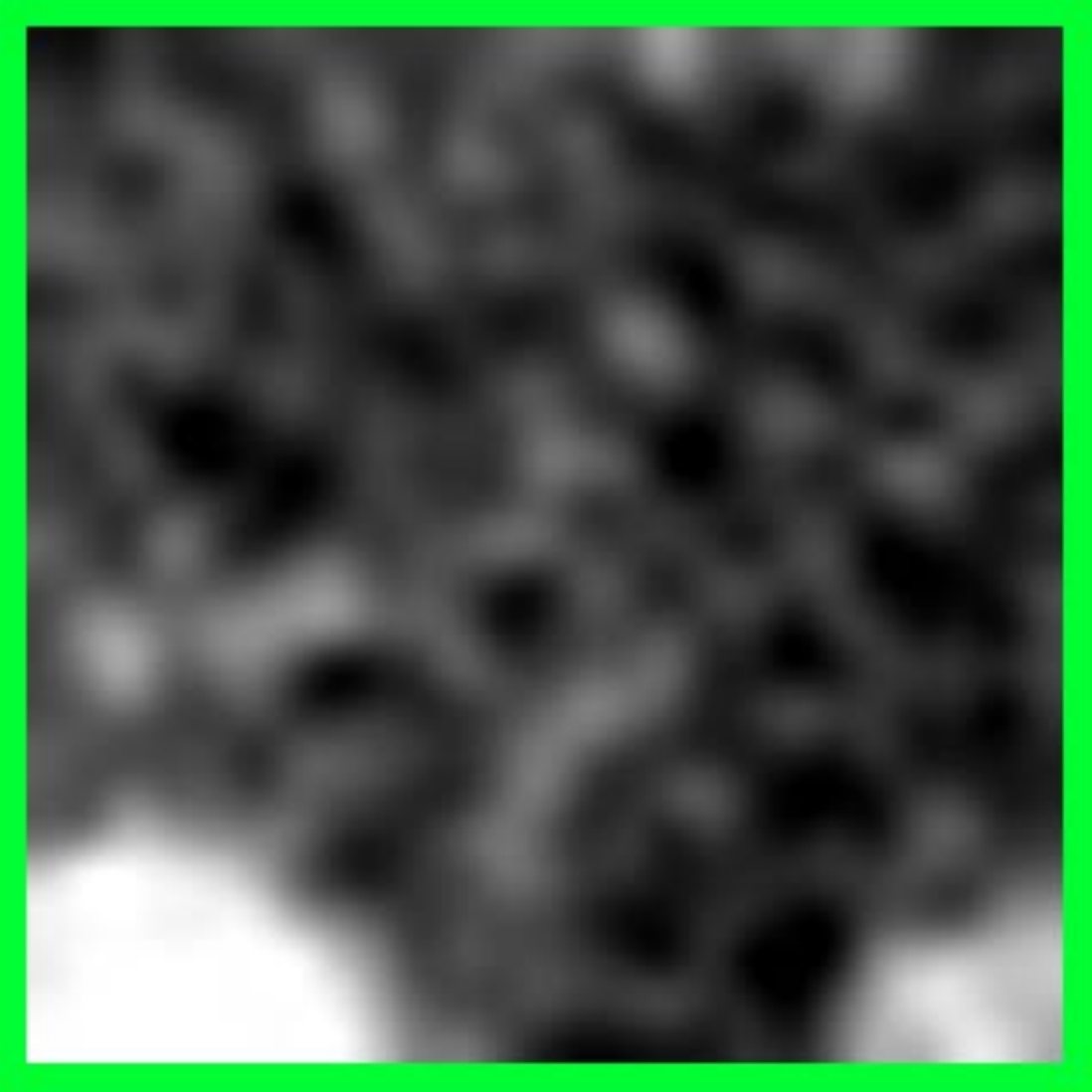} &
				\includegraphics[width=\swfiveh]{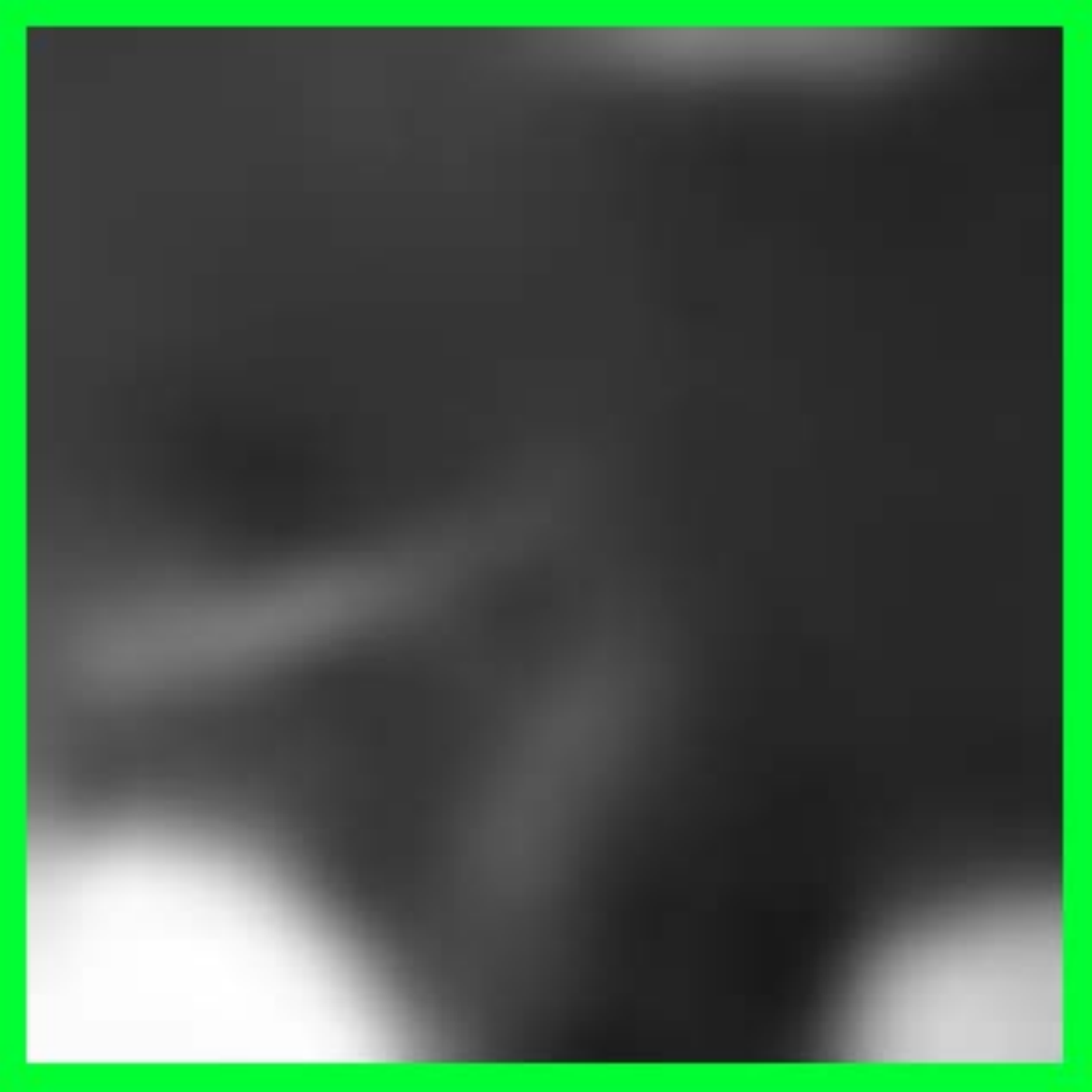} 
				\\
				full image & NDCT & LDCT & K-SVD
				\\
				\includegraphics[width=\swfiveh]{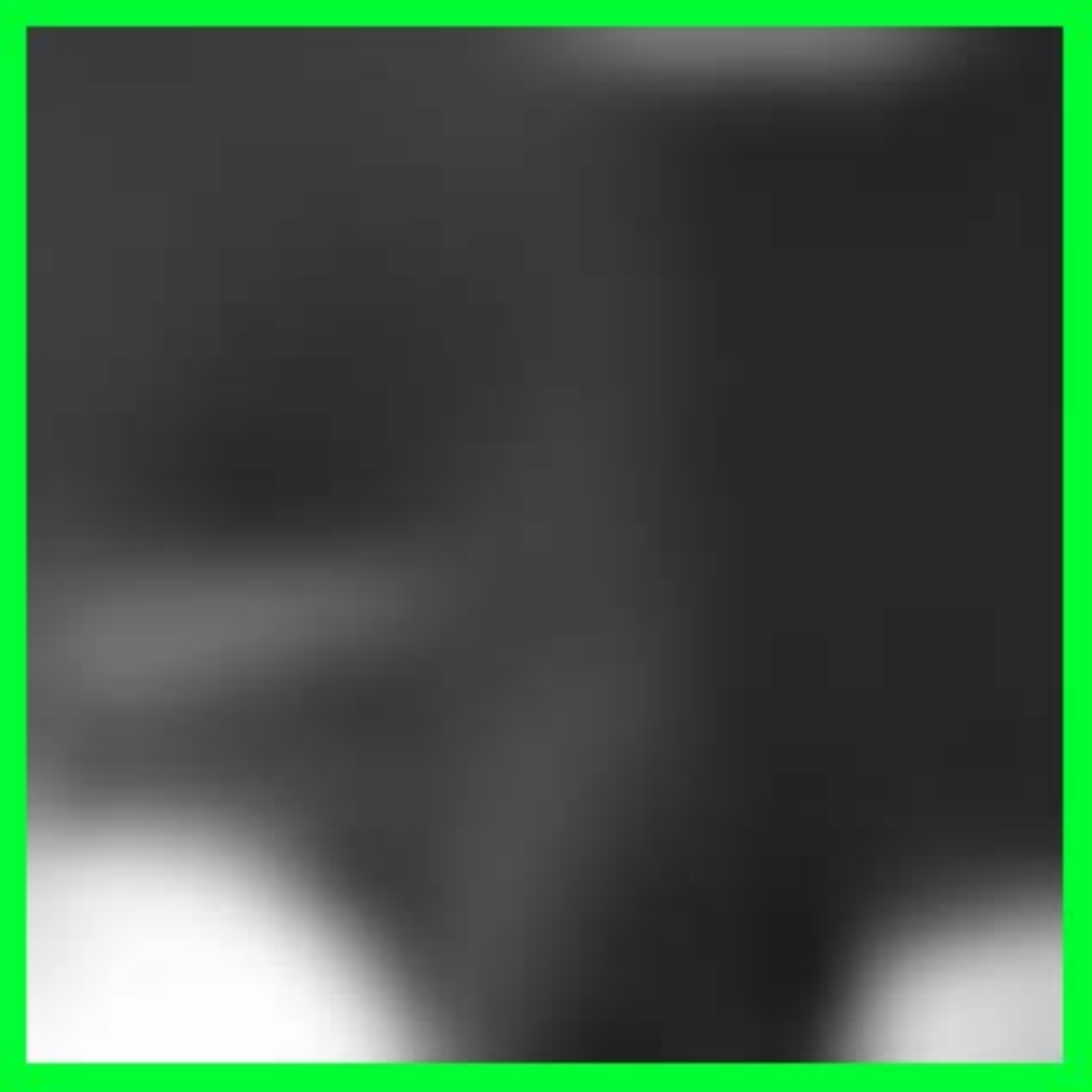} &
				\includegraphics[width=\swfiveh]{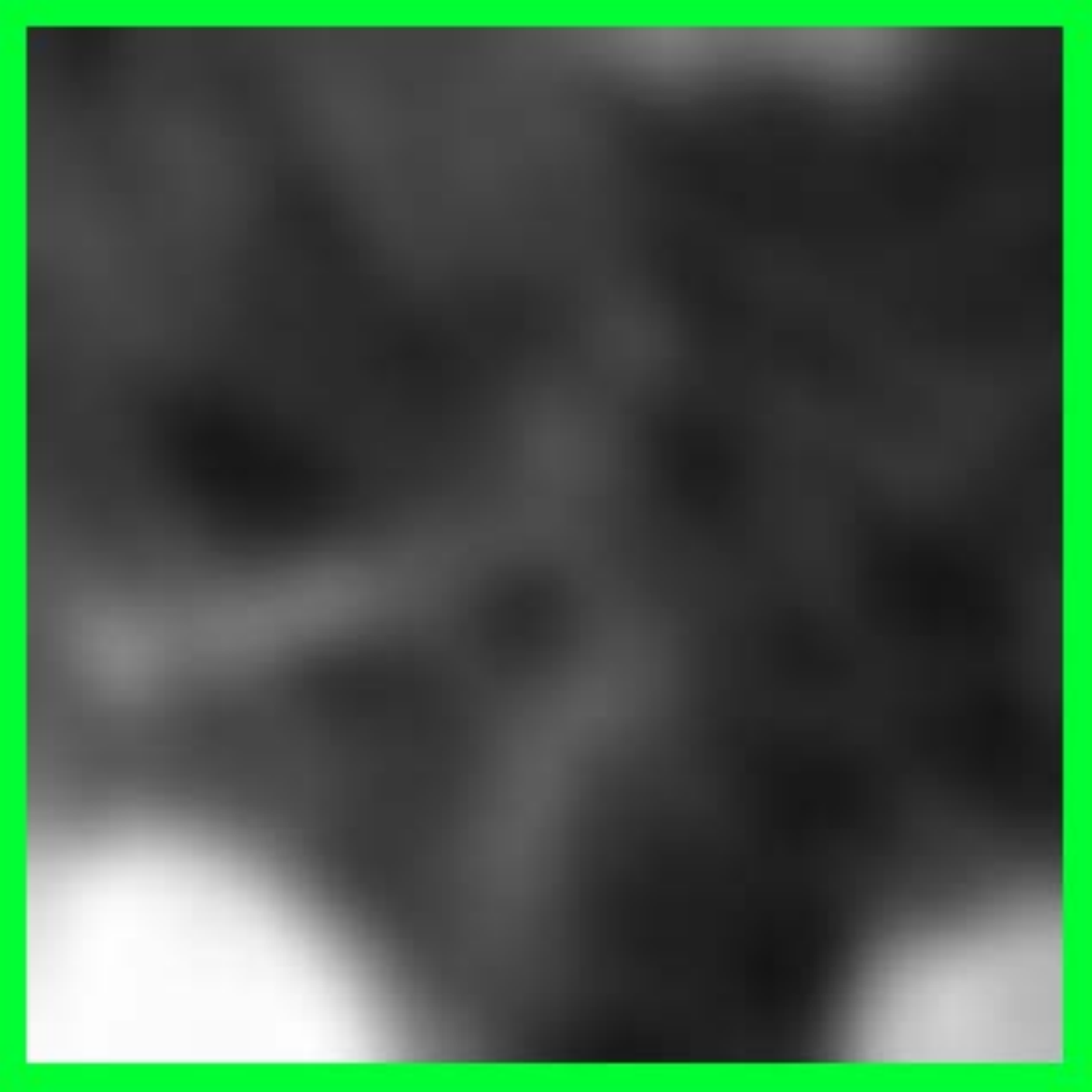} &
				\includegraphics[width=\swfiveh]{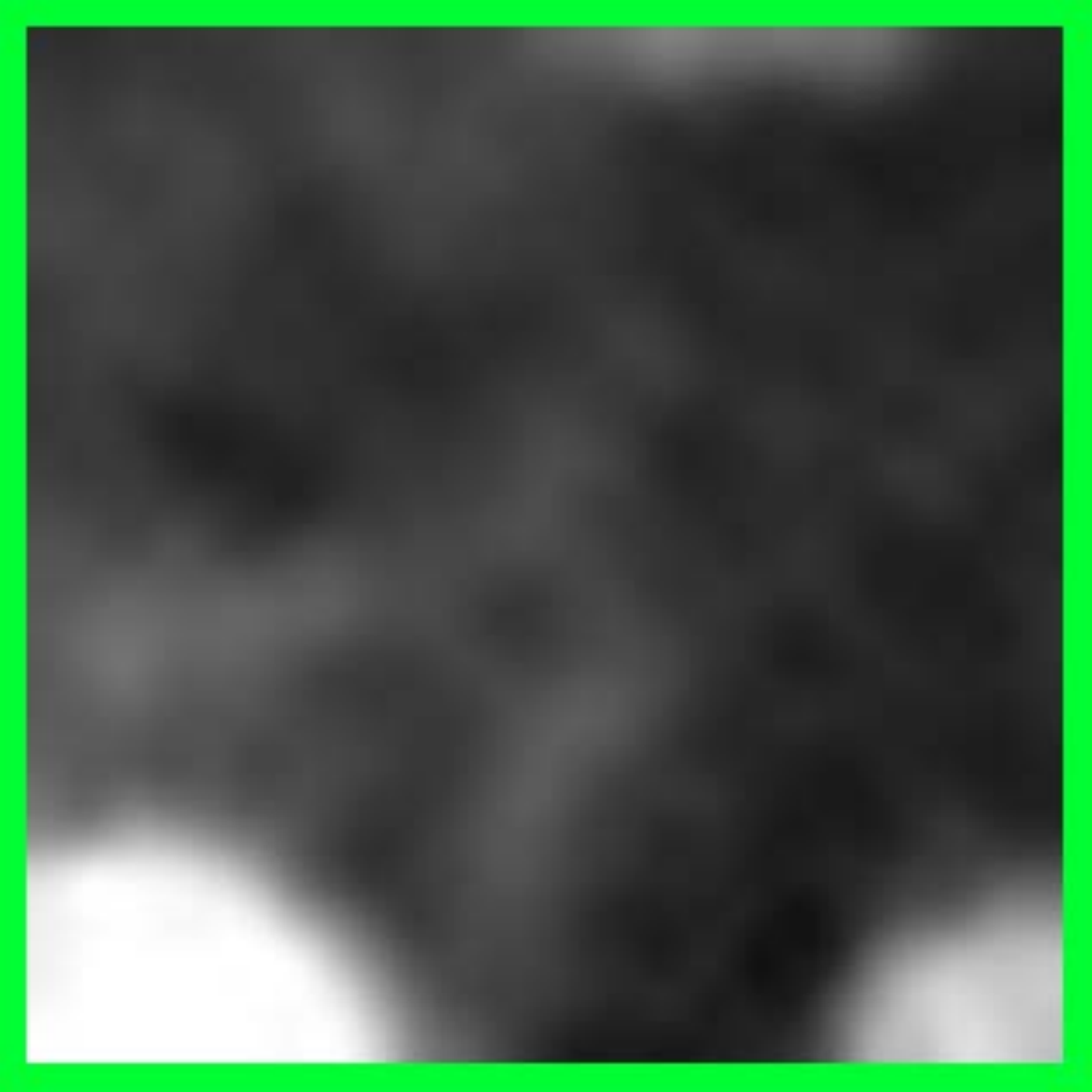} &
				\includegraphics[width=\swfiveh]{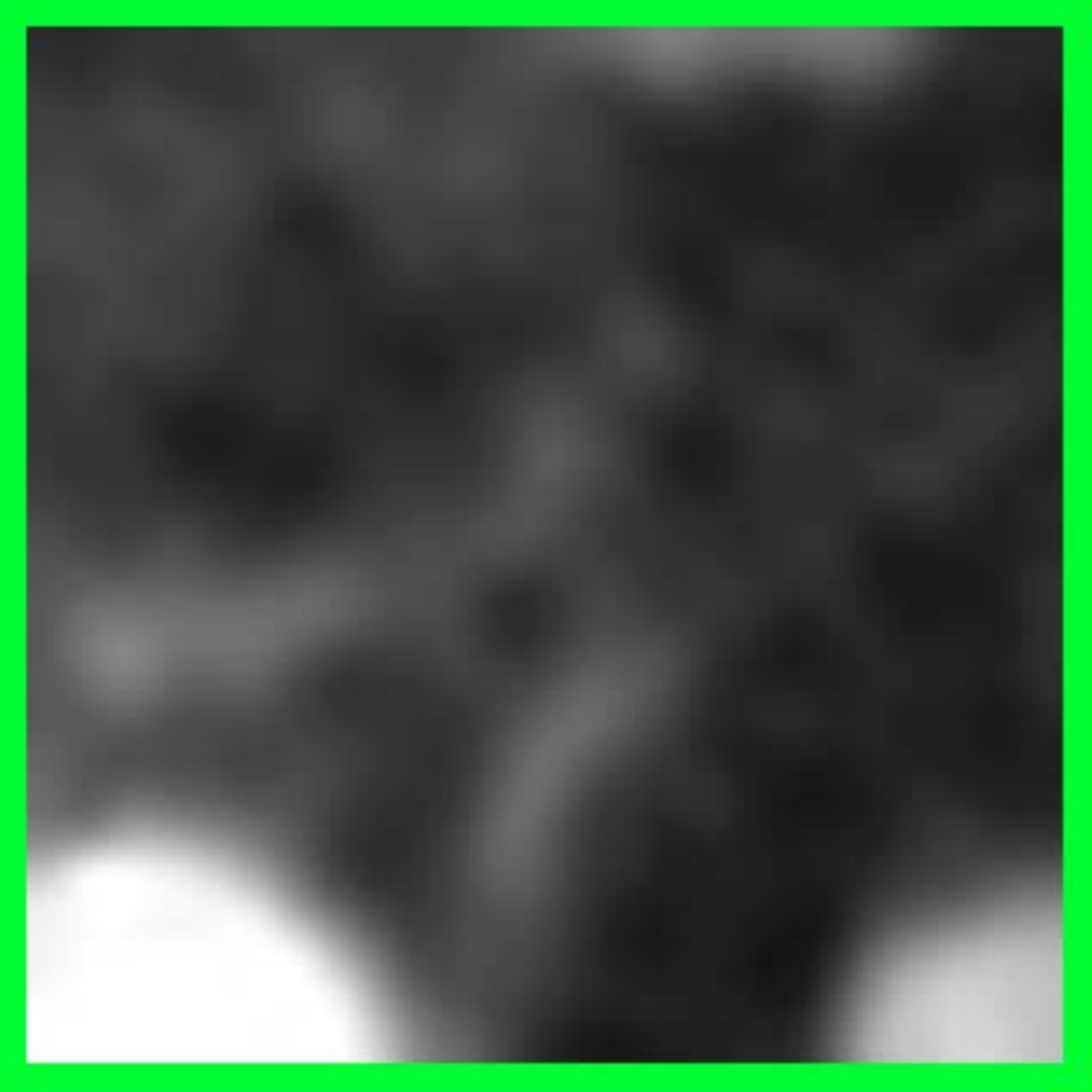}\\
				BM3D & WavResNet & RED-CNN & StructKPN
			\end{tabular}
		\end{center}
		\caption{
			Visual comparison on the NIH dataset. StructKPN recovers clearer details of the mesenteric vein, which is important for discriminating acute mesenteric ischemia from acute abdomen.
		}
		\label{fig:aapmres}
	\end{figure}

	\subsection{Kernel Prediction Network} \label{section:kpn}
	In this work, we focus on the single frame LDCT denoising problem.
	To better estimate the spatially variant filters for the LDCT denoising problem, we adopt an efficient model based on Kernel Prediction Network (KPN). For an input image of height $H$, width $W$, instead of regressing the output pixels with one same network, KPN uses a deep network $\mathbf{F}$ to generate a set of $k\times k$ kernels for each pixel position where
	$k=2r+1$ is the output kernel size,
	$r$ is the half kernel size
	. 
	%
	%
	By applying the kernels to neighborhood patches of size $k\times k$, the output pixels are computed in the Local Convolution Layer as
	\vspace{-3mm}
	\begin{eqnarray}
	\label{eq:kpn}
	&&\hat{Y}_{m,n} = \hat{Q}(m, n, c)\nonumber\\&&=\sum_{s=-r}^{r} \sum_{t=-r}^{r}[ X_{m-s,n-t} \times V_{m,n,((s+r)\cdot k + t+r+1)}]
	\end{eqnarray}

	As can be seen in Eq.~\ref{eq:kpn}, KPN is explicitly input-dependent. Thus, the supervision for the kernels is directly depending on input, output and ground truth pixels, which makes KPN a suitable candidate to be guided by specially designed loss functions based on image gradient statistics, such as the one we will introduce in \ref{section:lossfunc}. Furthermore, since one can examine the denoising process by directly visualizing the output kernels, KPN is more interpretable than CNN networks, which is highly desired for clinical usage. 
	
	In this work, we use a KPN architecture with residual block structure 
	to generate the kernel weights. The detailed network structure is shown in Fig.~\ref{networkfig}.

	\subsection{Structure-aware Loss Function} \label{section:lossfunc}
	To guide KPN to generate kernels according to different types of neighborhood, we design a structure-aware loss that combines three commonly-used losses based on neighborhood gradient statistics of NDCT images. 
	Inspired by RAISR \cite{romano2016raisr} which divides the image super-resolution problem into sub-problems for different local structures, we compute the gradient statistics:  strength $\lambda$ and coherence $\mu$ for the patch $P_{m,n}$ of size $k_{r}$ surrounding every pixel position $(m,n)$ in the ground truth image:
	\begin{equation}
	\lambda_{m,n} = \lambda_{1,m,n}
	\end{equation}
	\begin{equation}
	\mu_{m,n}=\frac{\sqrt{\lambda_{1,m,n}}-\sqrt{\lambda_{2,m,n}}}{\sqrt{\lambda_{1,m,n}}+\sqrt{\lambda_{2,m,n}}} \\
	\end{equation}
	
	where for the size $ k_{r} \cdot k_{r} \times 2$ flattened 2D patch gradient matrix $G$, $\lambda_{1}, \lambda_{2}$ are its eigenvalues in decreasing order. %
	Eigenvalues are computed using Singular Value Decomposition (SVD) in our implementation.
	Typically, for a pixel neighborhood patch, a large strength means abrupt changes of pixel values, which indicates clear structures in the patch. A large coherence suggests coherent directions in patches that contain edges or stripes. Accordingly, we categorize the patches into three types based on $\lambda$ and $\mu$: a) edges, having high strength and coherence, b) fine-grained structures, having high strength but lower coherence than edges, and c) flat regions, low in both strength and coherence. 
	
	For edges, good denoising should preserve the sharpness and direction of the edge. L2 loss, which increases the penalty for larger differences and reduces over-smoothing, is suitable for such type of patches. For fine-grained structures, we choose SSIM loss \cite{wang2004image} to enhance the structural similarity between denoised images and ground truth. Following the findings by Zhao et.al \cite{zhao2016loss} that using L1 loss in denoising tasks leads to less splotchy artifact than L2 and SSIM loss, we apply L1 loss to flat regions. For a pixel position (x,y), the structure-aware loss is expressed as

	\begin{table}[!t]
		\begin{center}
			\caption{Denoising performance (PSNR/SSIM) on the NIH dataset by different methods. Best results are displayed in bold.}
			\label{table:comparison}
			\resizebox{\linewidth}{!}{
				\begin{tabular}{l|llllll}\hline\hline
					Dataset     & LDCT       & K-SVD   & BM3D & WavResNet & RED-CNN & StructKPN   \\ \hline
					NIH      & 27.93/0.829 & 31.11/0.860 & 31.48/0.871 & 32.32/0.876 & 32.35/0.887 & \textbf{32.56/0.893}  \\ \hline\hline
			\end{tabular}}
			
		\end{center}
	\end{table}
	\setlength{\tabcolsep}{1.4pt}
	\begin{figure}[!t]
		\begin{center}
			\begin{tabular}{cccc}
				\includegraphics[width=0.178\linewidth]{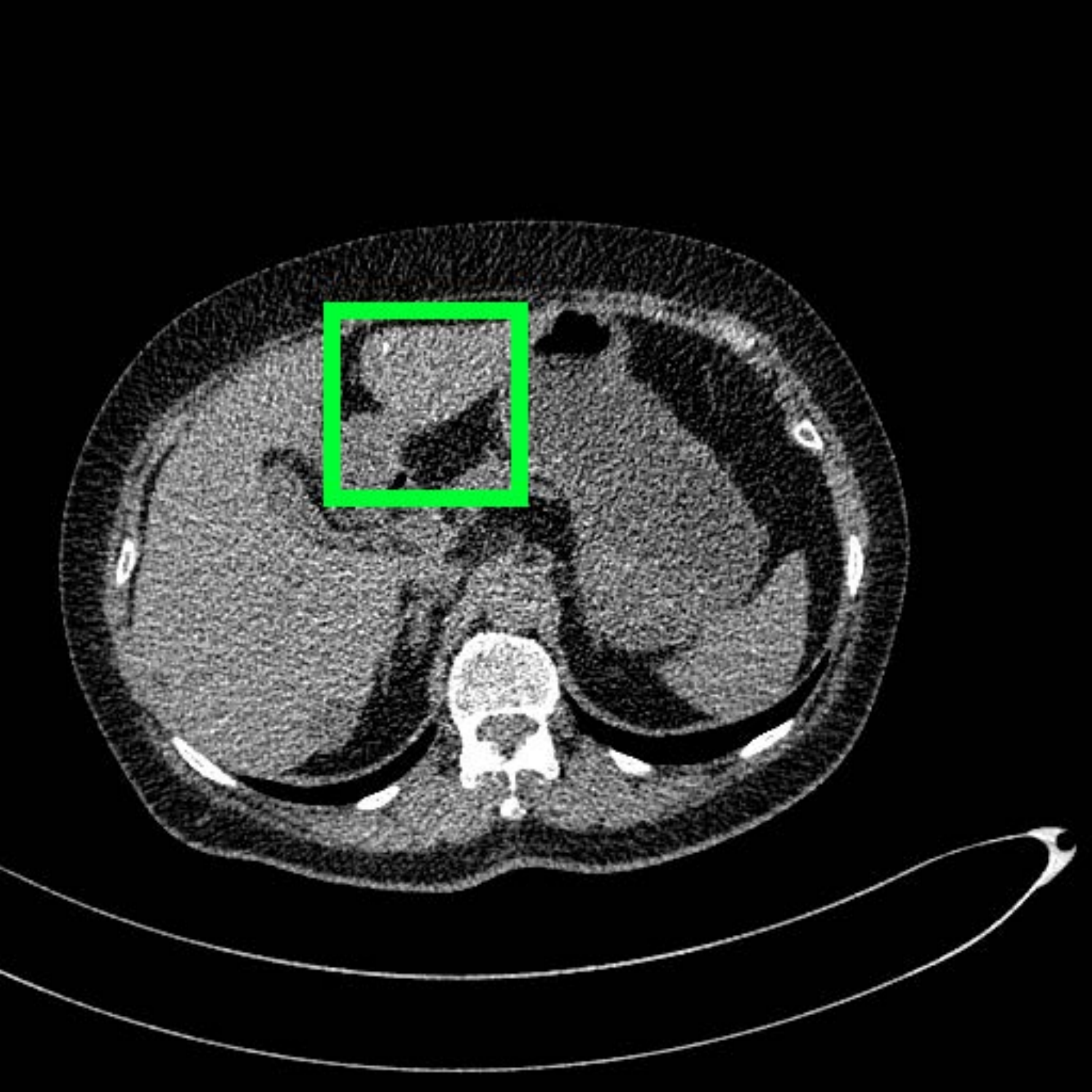} &
				\includegraphics[width=0.185\linewidth]{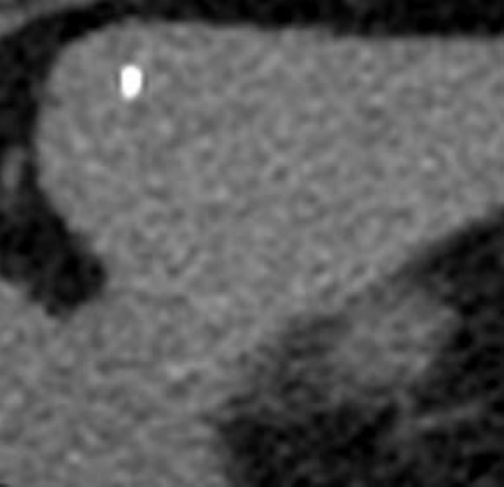} &
				\includegraphics[width=\swfiveh]{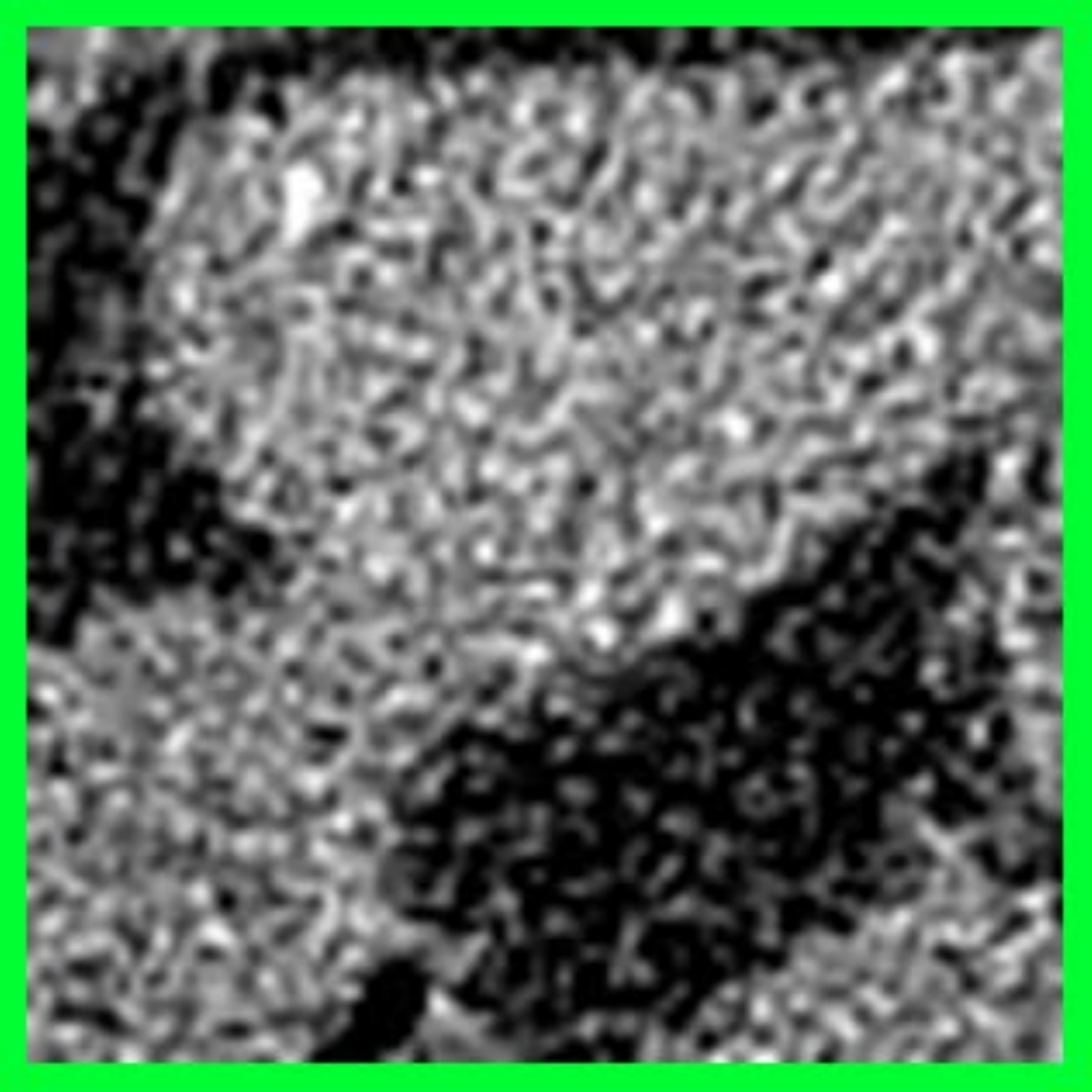} &
				\includegraphics[width=\swfiveh]{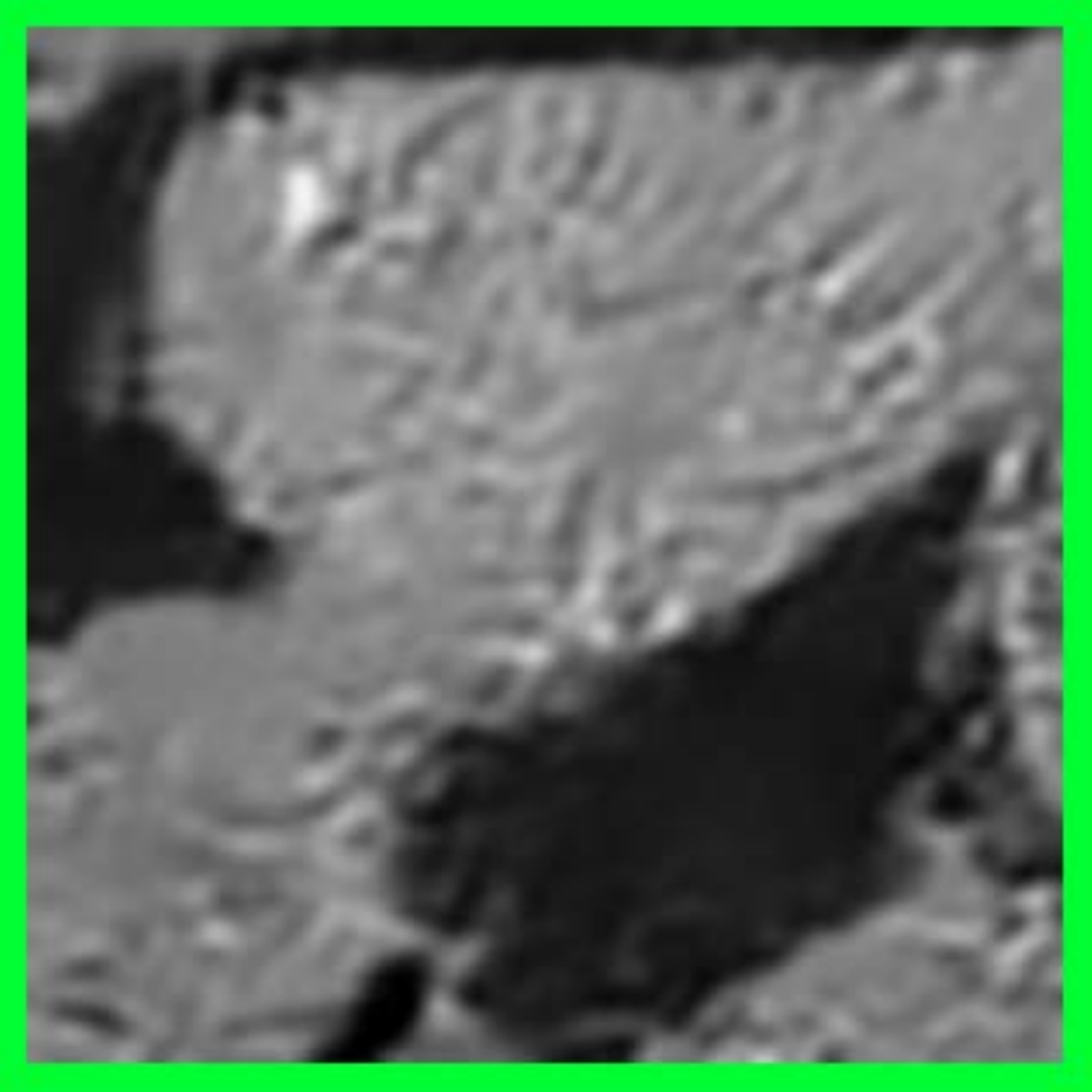}
				\\
				full image & NDCT & LDCT & K-SVD 
				\\
				\includegraphics[width=\swfiveh]{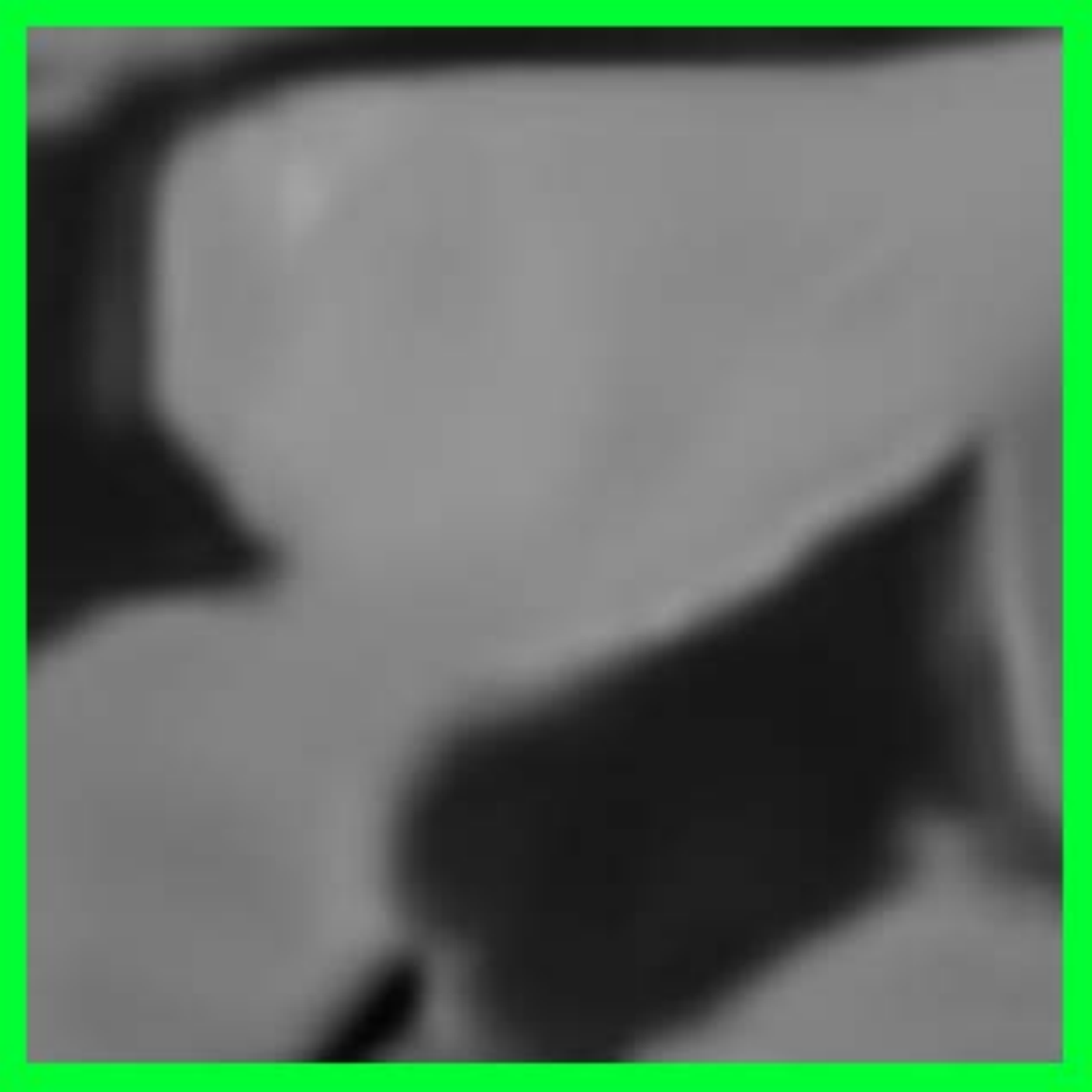} &
				\includegraphics[width=\swfiveh]{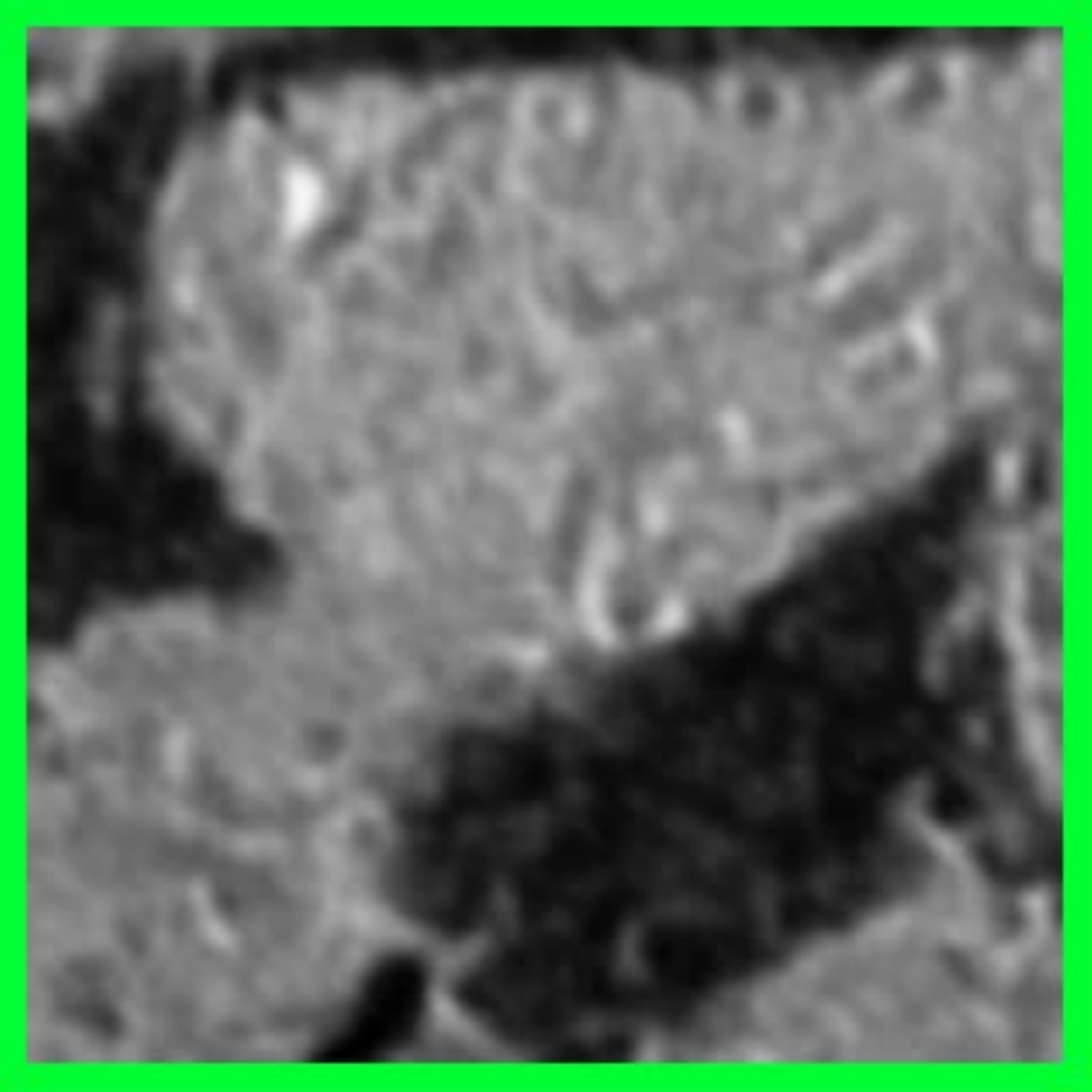} &
				\includegraphics[width=\swfiveh]{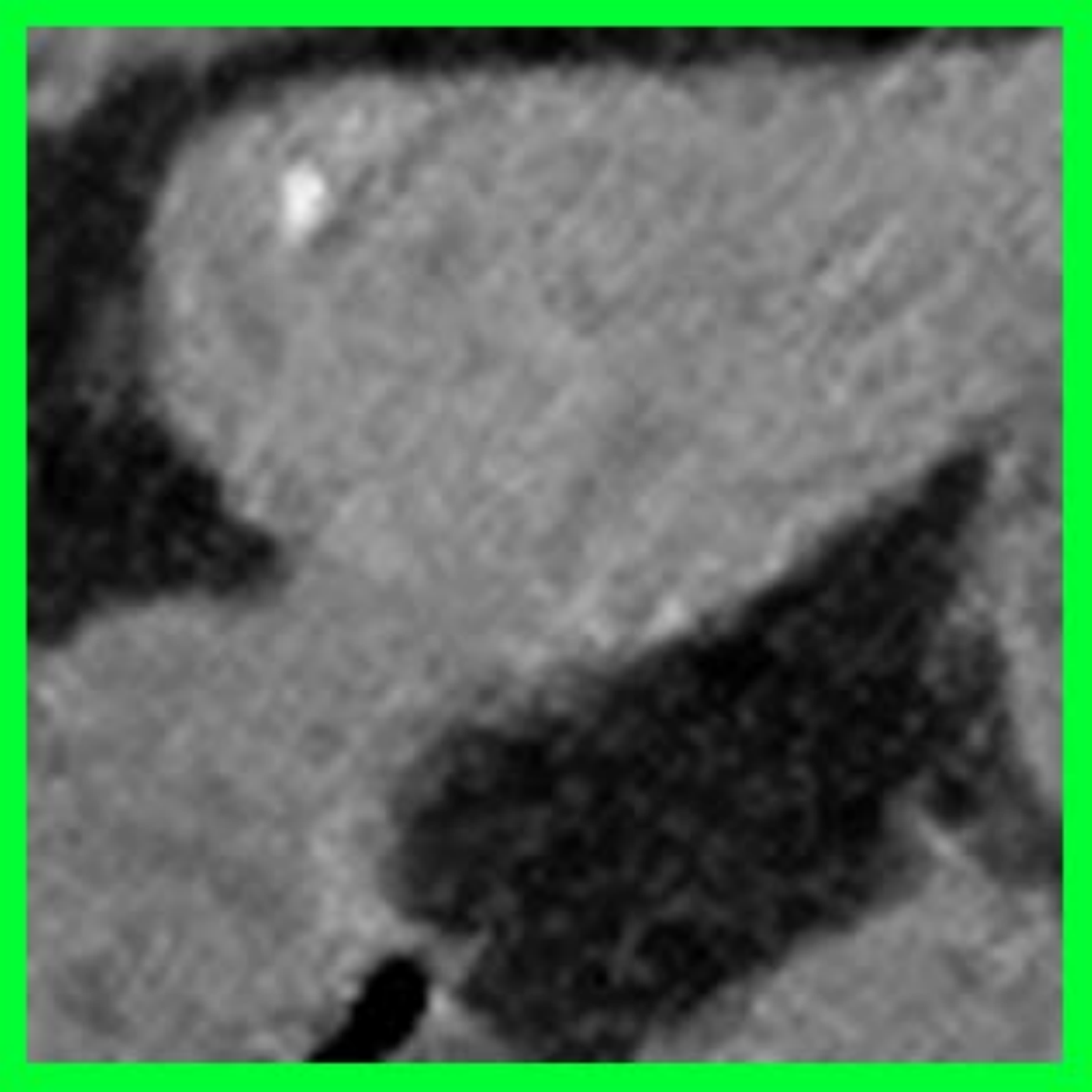} &
				\includegraphics[width=\swfiveh]{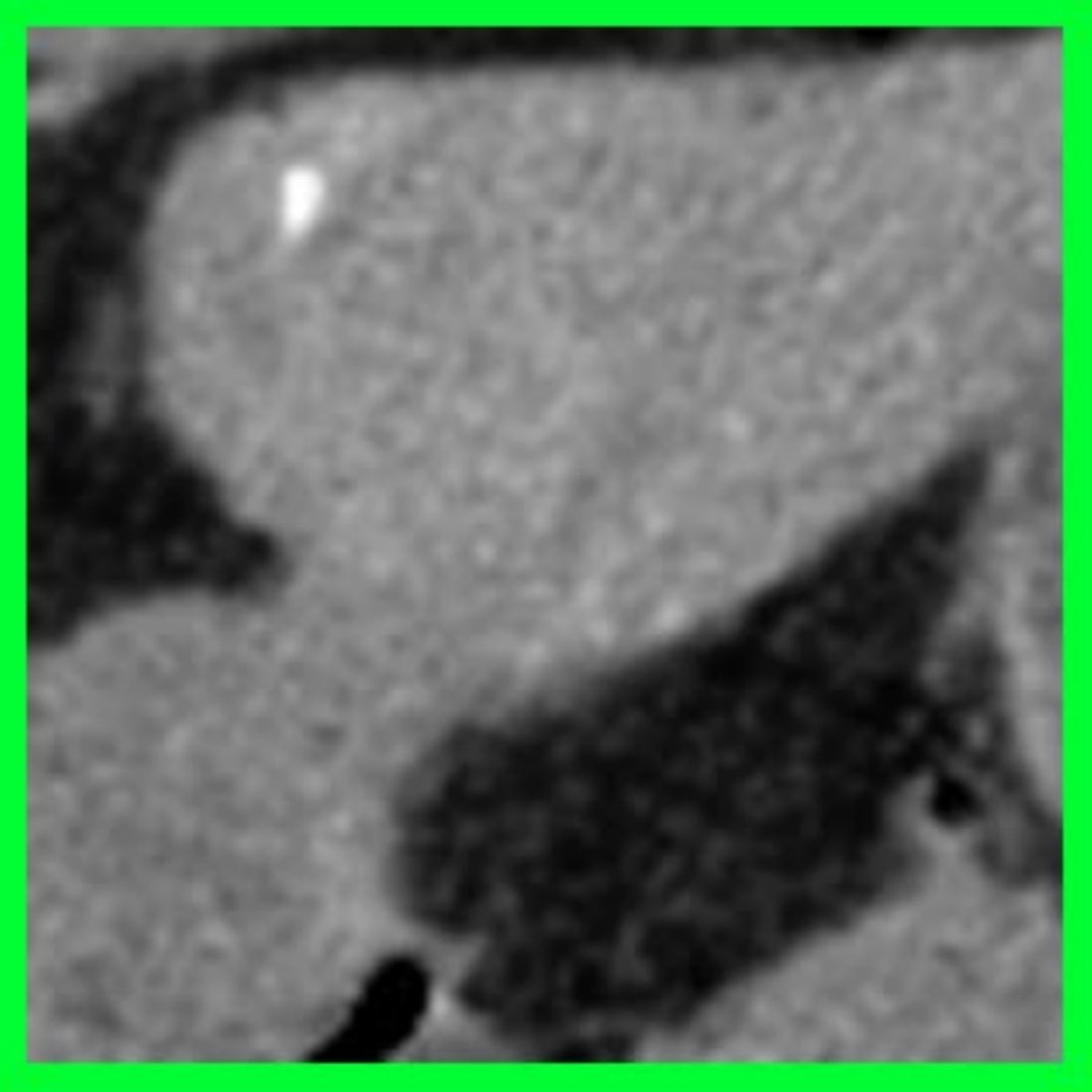} \\
				BM3D & WavResNet  & RED-CNN  & StructKPN
			\end{tabular}
		\end{center}
		\caption{
			Visual comparison on the DOSE dataset. The enlarged region contains a calcification in the left lobe of liver. The NDCT image is picked by the radiologist and only serves as a reference of desired beam-hardening effect recovery and texture of liver parenchyma as the NDCT is not perfectly aligned with LDCT. Compared to RED-CNN  \cite{chen2017low}, StructKPN shows better beam-hardening effect recovery. K-SVD \cite{chen2013improving} and WavResNet \cite{cho2017wavelet} suffer from artifacts, and BM3D \cite{sheng2014denoised} has severe over-smoothing.
		}
		\label{fig:doseres}
	\end{figure}
	\vspace{-3mm}
	\begin{eqnarray}
	L_{S}(\hat{Y},Y) = \sum_{m=1}^{H}\sum_{n=1}^{W} \frac{L_{w}(m,n)+L1(\hat{Y}_{m,n},Y_{m,n})}{2H\cdot W}
	\end{eqnarray}
	\vspace{-3mm}
	\begin{align}
	L_{w}(m,n)
	& = \gamma_{1} \cdot L2(\hat{Y}_{m,n},Y_{m,n})+ \gamma_{2} \cdot L1(\hat{Y}_{m,n},Y_{m,n})\nonumber
	\\& -\gamma_{3} \cdot SSIM(P_{m,n},Q_{m,n})
	\end{align}
	\vspace{-3mm}
	\begin{equation}
	[\gamma_{1}, \gamma_{2}, \gamma_{3}] = Softmax([\mu_{m,n}*\sigma_{L2},\sigma_{L1},\lambda_{m,n}])
	\end{equation}

	where $Q_{m,n}$ is the NDCT patch corresponding to $P_{m,n}$, $\gamma_{1}, \gamma_{2}, \gamma_{3}$ are the sum-to-one weights for three loss components, and $\sigma_{L2},\sigma_{L1} $ are constant hyper-parameters to control the relative importance of three losses. In this work, we choose $k_r = 11, \sigma_{L2}=1.8, \sigma_{L1}=0.35$. An illustration of the combination of three losses and the corresponding local structures is shown in \reffig{fig:structloss} (a)-(e).

	\section{Experiments and Results}
	\subsection{Datasets}
	In this work, we utilize three datasets for evaluation. The 2016 NIH-AAPM-Mayo LDCT dataset\cite{mccollough2016tu} (denoted by NIH in the rest of the paper) contains 2378 aligned pairs of 3mm LDCT-NDCT images from ten patients with simulated noise. We use 1156 pairs from the first five patients as the training set, and the rest are reserved for evaluation. 
	To assess the quality of denoised images, we compare the average peak signal to noise ratio (PSNR) and structural similarity index measure (SSIM) against NDCT images.

	\begin{figure}[!t]
		\includegraphics[width=\linewidth]{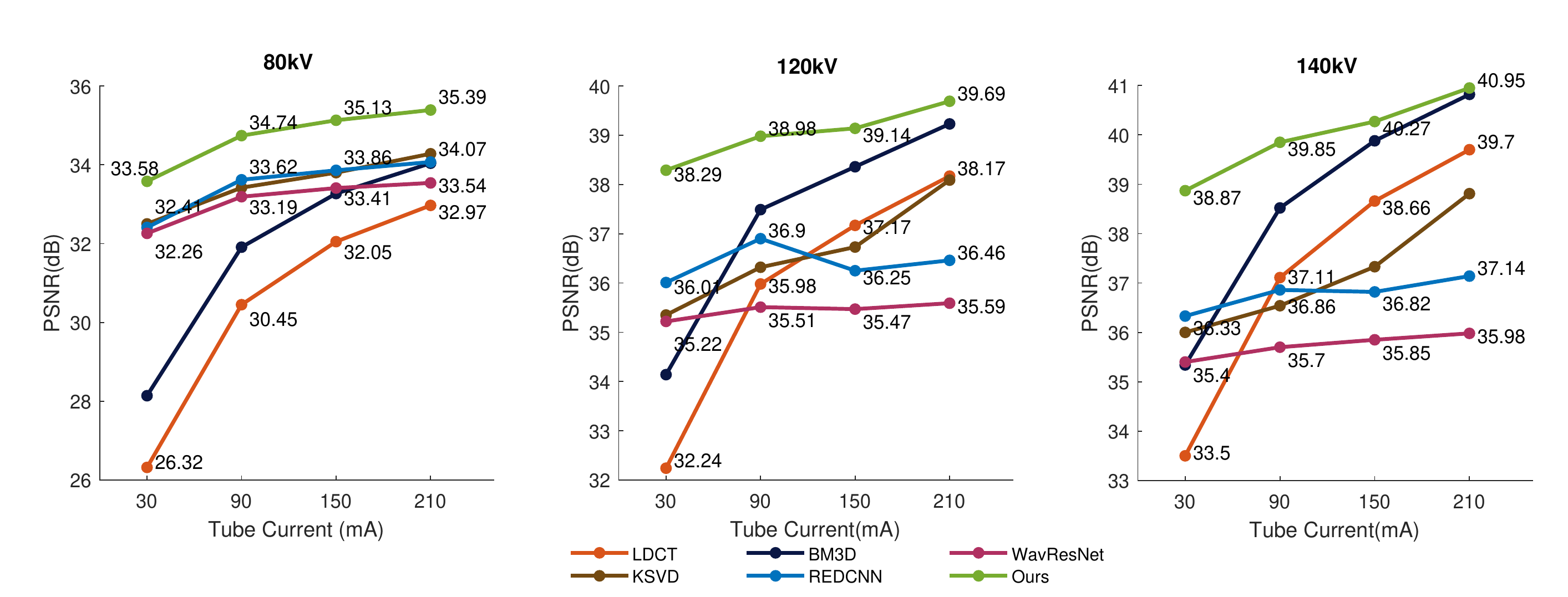}
		
		\caption{Denoising performance of different methods on the PHANOM dataset under different LDCT tube voltages and currents. In all experiments of the three tube voltages, the CT images under tube current 270mA and the corresponding tube voltages are used as NDCT images for computing the PSNR.} 
		\label{phantomcurve}
	\end{figure}

	To further validate the generalization ability and clinical value of the proposed method, we used two additional non-synthetic image datasets for evaluation. 
	Firstly, the PHANTOM dataset contains 15 sets of scans under varied radiation dose levels by different combinations of three tube voltages (80, 120, 140kV) and five tube currents (30, 90, 150, 210, 270mA). 
	The scans are taken on a commercially available CT phantom that consists of the basic chest and epigastric structures. 
	The images with different radiation doses are well aligned.
	We choose radiation dose $\mathbf{d}_{ND} = $[270mA] as NDCT current.
	Secondly, the DOSE dataset \cite{du2019methods} contains 60 scans from 30 patients who enrolled in a population-based lung cancer screening. The patients received preoperative NDCT scans within a month after the LDCT scans, therefore accurately anatomical correspondence could not be provided in the dataset. Nevertheless, we invite radiologists to examine the dataset and pick image pairs that contain organ parts with close correspondence that are important for diagnosis.

	\begin{figure}[!t]
		\begin{center}
			\begin{tabular}{ccccccc}
				\includegraphics[width=0.22\linewidth]{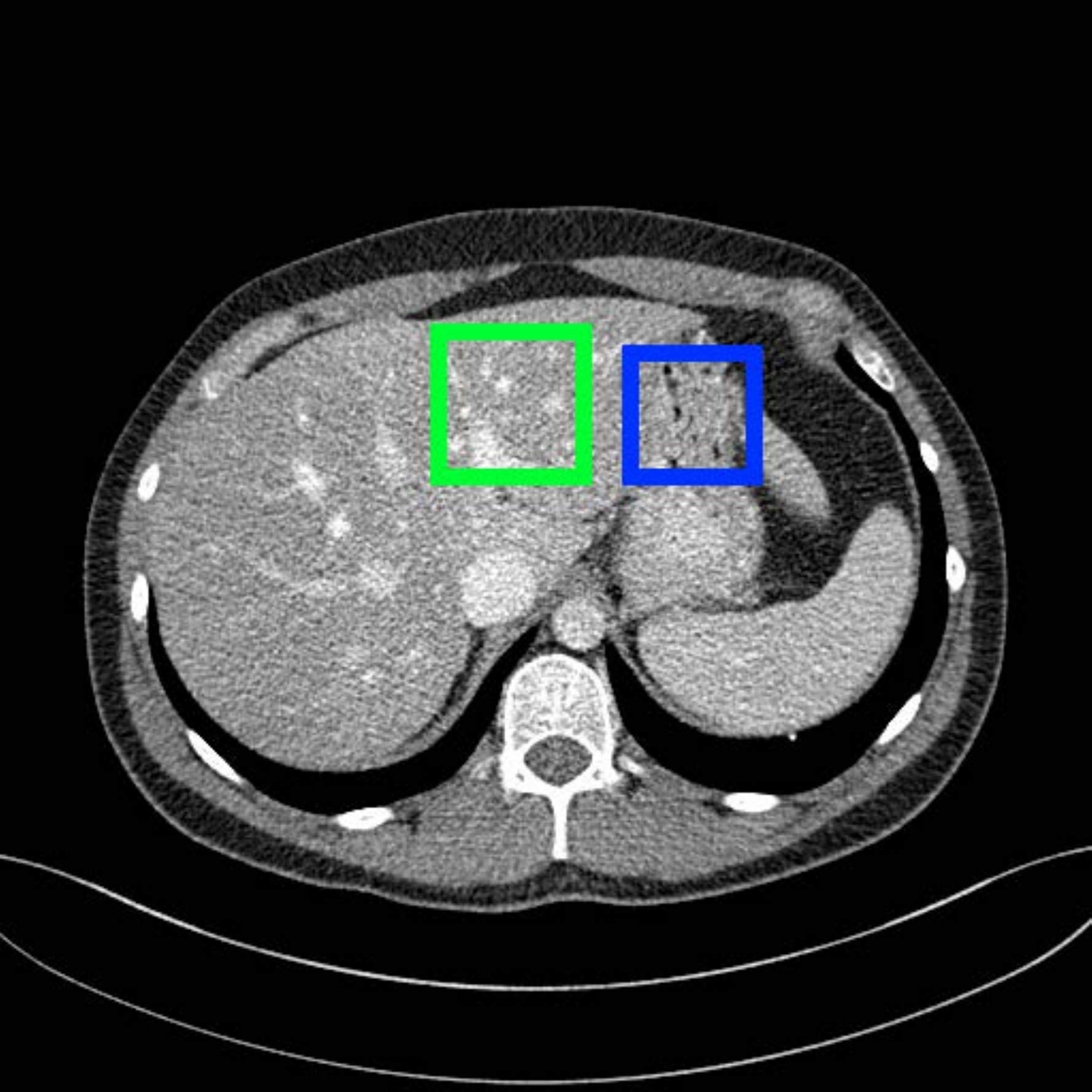} &
				\includegraphics[width=\sweight]{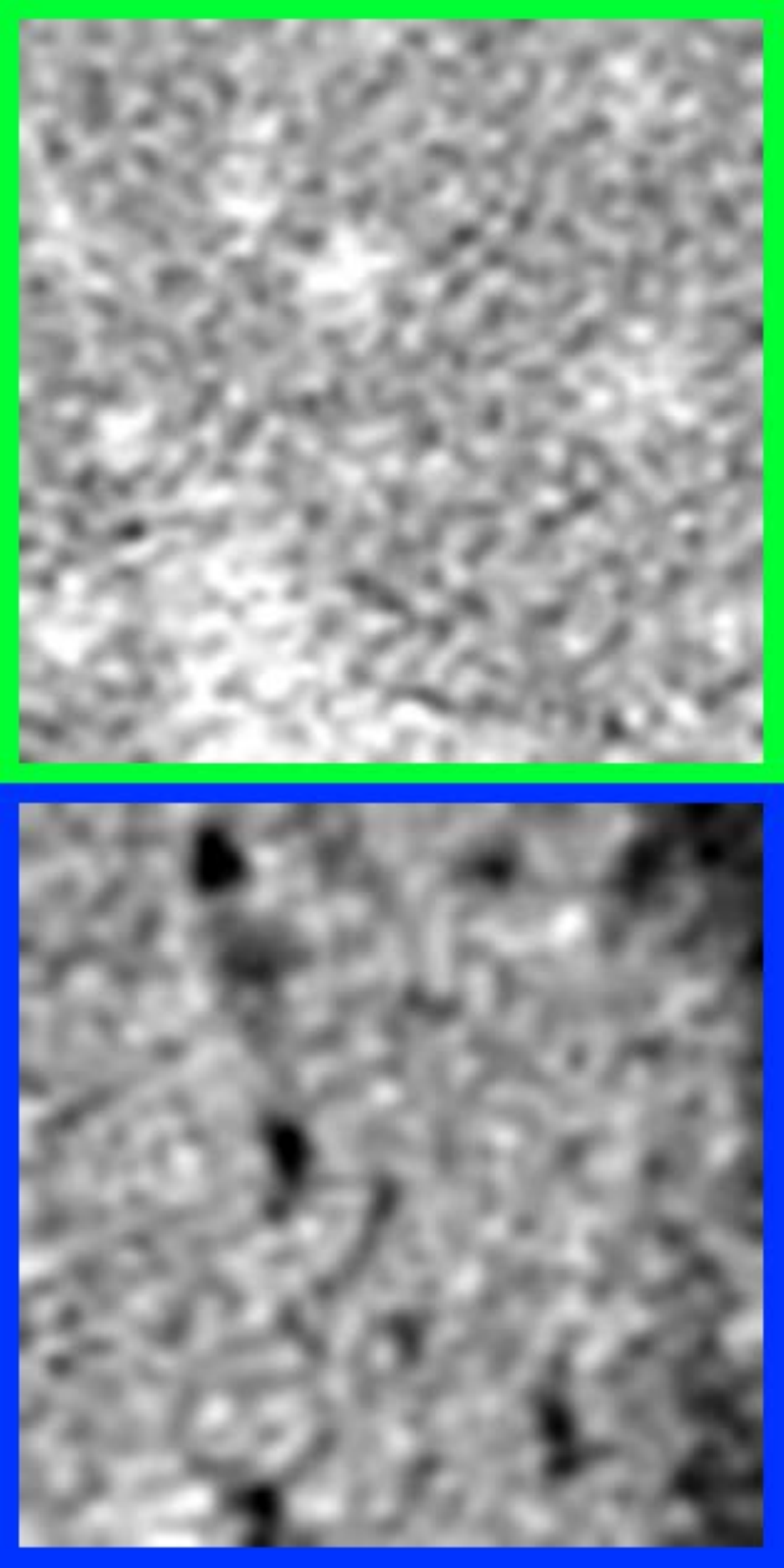} &
				\includegraphics[width=\sweight]{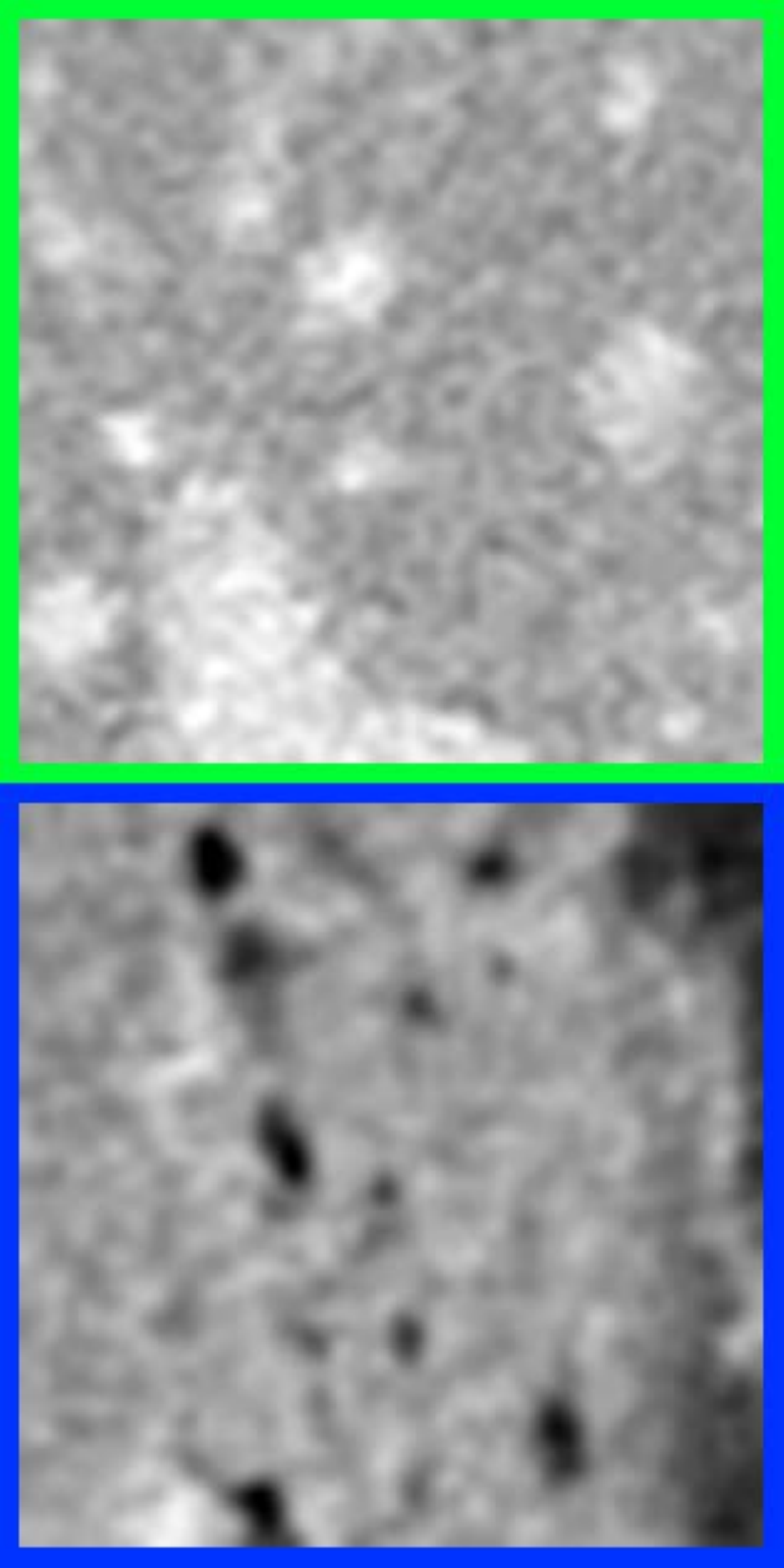} &
				\includegraphics[width=\sweight]{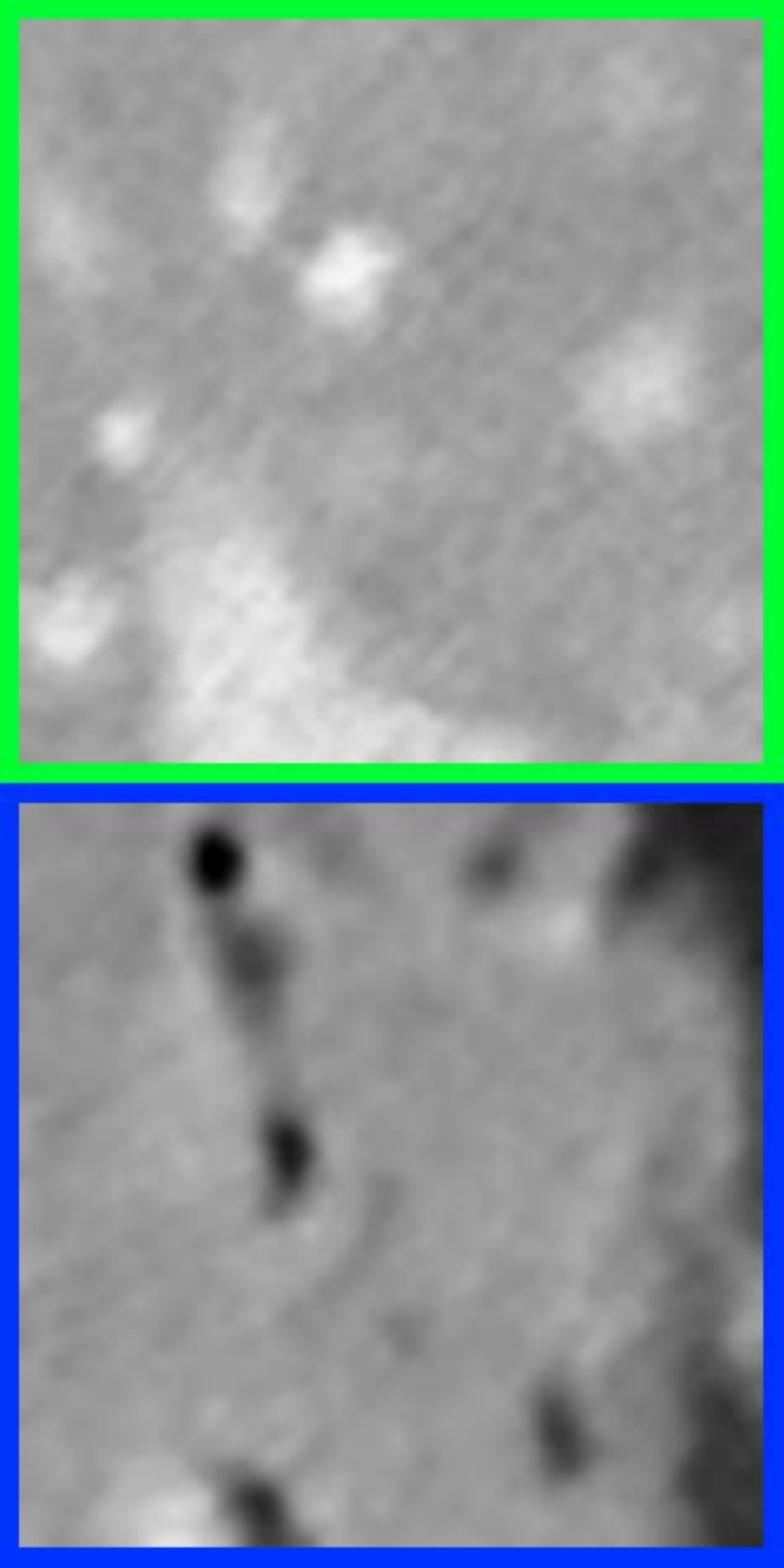} &
				\includegraphics[width=\sweight]{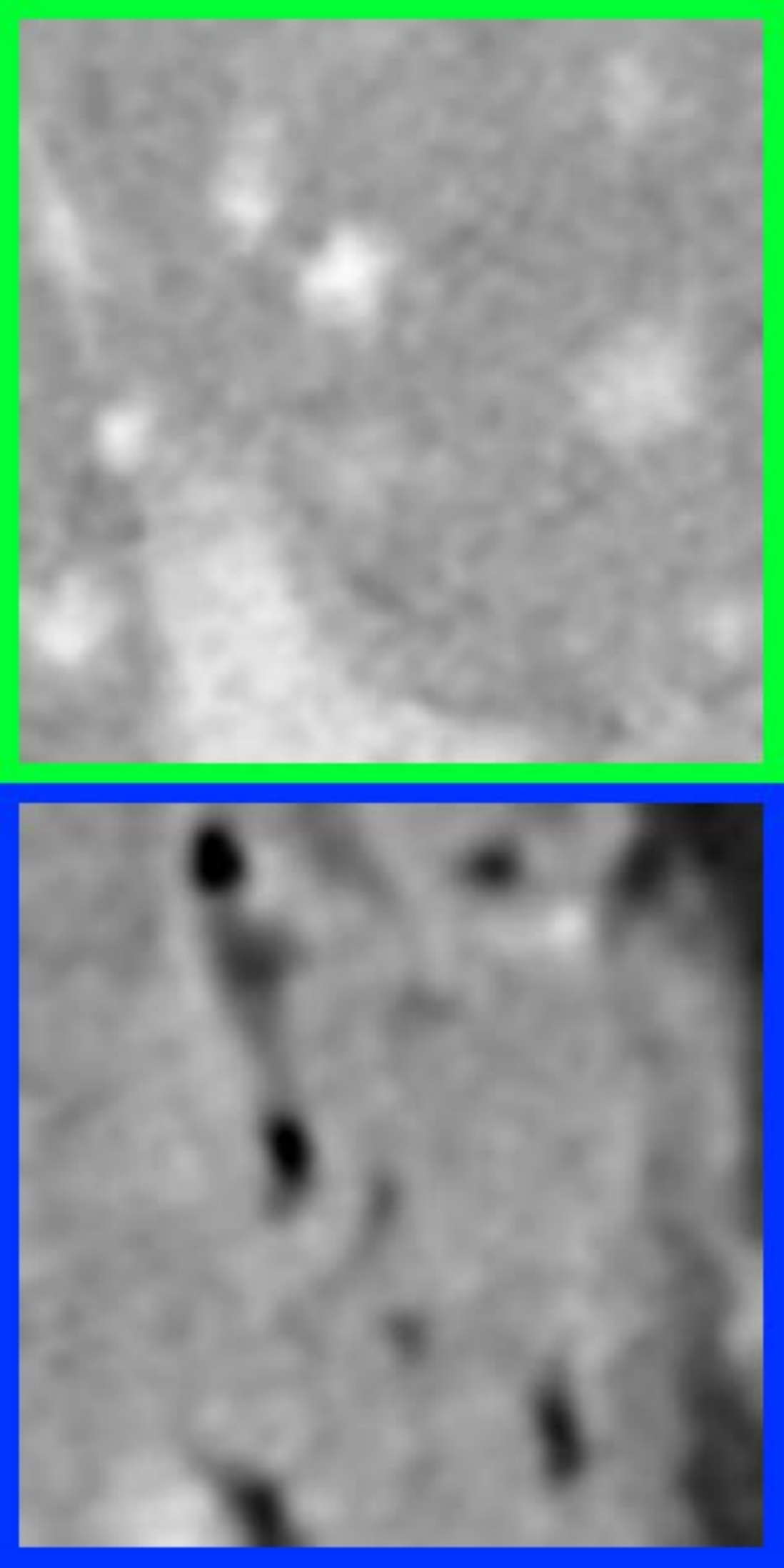} &
				\includegraphics[width=\sweight]{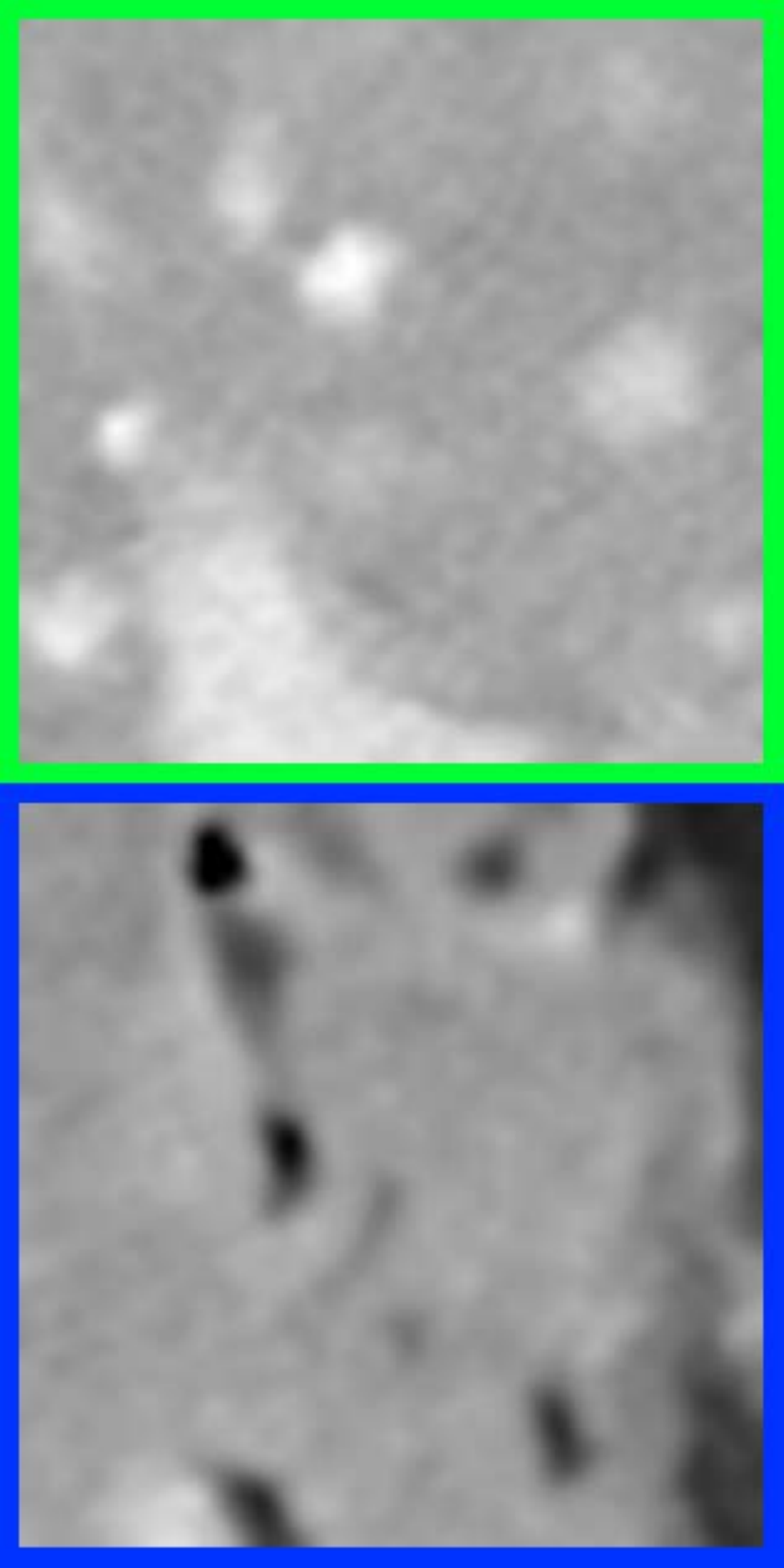} &
				\includegraphics[width=\sweight]{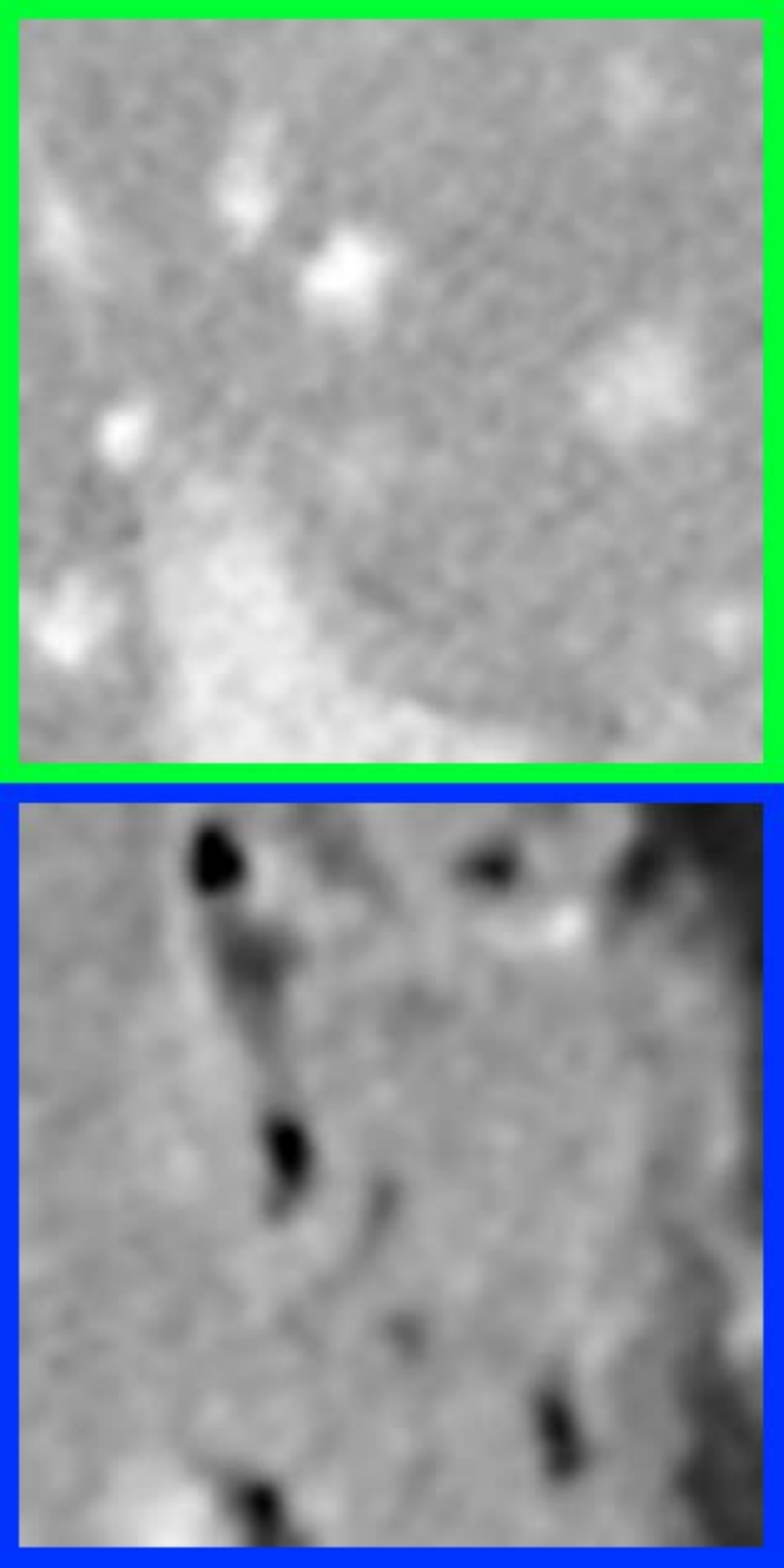}\\
				(a) & (b)  & (c)  & (d)  & (e) & (f) & (g)\\
			\end{tabular}
		\end{center}
		\caption{
			(a): A ground truth image in NIH dataset, (b): enlarged LDCT image, (c): enlarged NDCT image, and the denoising results by (d) RED-CNN \cite{chen2017low} with L1 loss, (e) RED-CNN with $L_{S}$, (f) KPN with L1 loss (g) KPN with $L_{S}$.
		}
		\label{fig:ablation}
	\end{figure}
	
	\subsection{Baseline Methods}
	We choose four methods to compare with our proposed model.
	Two of them are classical methods that do not require training data.
	K-SVD\cite{aharon2006k} generates denoised image patches by linear combinations of an overcompleted low-rank dictionary obtained from the noisy image.
	Block Matching 3D Filtering (BM3D)\cite{dabov2007image} is a nonlocal self-similarity method that groups similar 2D patches into 3D groups and performs linear transforms.
	For selecting the hyperparameters of BM3D and K-SVD, we follow \cite{sheng2014denoised} and \cite{chen2013improving} respectively.
	RED-CNN \cite{chen2017low} and WavResNet \cite{cho2017wavelet} are two neural network baselines that adopt residual encoder-decoder and wavelet domain CNN.

	\subsection{Results and Analysis}
	
	On the NIH dataset, our method performs favorably against the baseline methods (see Table.\ref{table:comparison}). Visually, BM3D and K-SVD have severe over-smoothing problems. As in Fig~\ref{fig:aapmres}, the proposed StructKPN recovers clearer veins, whereas RED-CNN and WavResNet produce more blurry results. 
	
	We also perform visual analysis for the DOSE dataset. Since the LDCT and NDCT images were from CT scans at different dates, no aligned LDCT-NDCT pair can be obtained. Nevertheless, the quality of reconstructed images can be qualitatively assessed with corresponding LDCT images as Fig~\ref{fig:doseres}. 

	On the PHANTOM dataset, StructKPN results in better performance on different radiation doses (see \reffig{phantomcurve}). Due to both the spatially invariant filter architectures and the uniform pixel-level loss functions, RED-CNN and WavResNet perform denoising less effectively when the LDCT voltage and current are high, resulting in over-smoothed images. In contrast, StructKPN has good denoising performance when testing on various radiation doses under low and high LDCT voltages and currents. The explicitly input-dependent design of KPN and the Structure-aware loss enable the proposed method to produce robust denoising filters without requiring massive training datasets that cover all the radiation doses in the test data. We provide representative visual comparisons in the supplementary materials. 
	\setlength{\tabcolsep}{4pt}
	\begin{table}[!t]
		\begin{center}
			\caption{Denoising performance (PSNR/SSIM) by RED-CNN and KPN models trained with different loss functions. L1 denotes the L1 loss. $L_{S}$ denotes the structure-aware loss described in \ref{section:lossfunc}.}
			\label{table:ablation}
			\resizebox{\linewidth}{!}{
				\begin{tabular}{l|llllll}\hline\hline
					Model      & RED-CNN+L1 & RED-CNN+$L_{S}$ & KPN+L1 & KPN+ $L_{S}$  \\\hline
					NIH    &  32.35/0.887 & 32.42/0.890 & 32.45/0.888 & 32.56/0.893  \\\hline\hline
				\end{tabular}
			}
			
		\end{center}
	\end{table}
	\setlength{\tabcolsep}{1.4pt}
	
	\subsection{The Effectiveness of Structure-aware Loss}
	We conduct an ablation experiment to study the effectiveness of using the structure-aware loss function for guiding the network towards better behavior at different types of regions.
	As seen in \reftable{table:ablation}, adopting the structure-aware loss improves the performance for both RED-CNN and KPN models, whereas KPN gains more increase in PSNR, which is probably due to the supervision signals that are directly propagated to the KPN's predicted kernels. Visual comparison is shown in \reffig{fig:ablation}, where after applying the structure-aware loss, the liver texture is closer to the NDCT image and less blurry.

	\section{Conclusion and Discussion}
	In this work, we have proposed StructKPN, an efficient approach that combines kernel prediction network and a structure-aware non-uniform loss function to exploit the spatially-variant characteristics for LDCT denoising. Both quantitative and qualitative evaluations on synthetic and non-synthetic LDCT datasets provide superior performance against prior methods, revealing strong structure preservation and noise removal capability of StructKPN. By evaluating on datasets under various radiation doses, we show that StructKPN trained with public synthetic data can be used for robust denoising of real LDCT images.

\bibliographystyle{IEEEbib}
\bibliography{refs}

\begin{thebibliography}{10}

\bibitem{balda2012ray}
Michael Balda, Joachim Hornegger, and Bjoern Heismann,
\newblock ``Ray contribution masks for structure adaptive sinogram filtering,''
\newblock {\em IEEE TMI}, vol. 31, no. 6, pp. 1228--1239, 2012.

\bibitem{wang2006penalized}
Jing Wang, Tianfang Li, Hongbing Lu, and Zhengrong Liang,
\newblock ``Penalized weighted least-squares approach to sinogram noise
  reduction and image reconstruction for low-dose x-ray computed tomography,''
\newblock {\em IEEE TMI}, vol. 25, no. 10, pp. 1272--1283, 2006.

\bibitem{sidky2008image}
Emil~Y Sidky and Xiaochuan Pan,
\newblock ``Image reconstruction in circular cone-beam computed tomography by
  constrained, total-variation minimization,''
\newblock {\em Physics in Medicine \& Biology}, vol. 53, no. 17, pp. 4777,
  2008.

\bibitem{xu2012low}
Qiong Xu, Hengyong Yu, Xuanqin Mou, Lei Zhang, Jiang Hsieh, and Ge~Wang,
\newblock ``Low-dose x-ray ct reconstruction via dictionary learning,''
\newblock {\em IEEE TMI}, vol. 31, no. 9, pp. 1682--1697, 2012.

\bibitem{cai2014cine}
Jian-Feng Cai, Xun Jia, Hao Gao, Steve~B Jiang, Zuowei Shen, and Hongkai Zhao,
\newblock ``Cine cone beam ct reconstruction using low-rank matrix
  factorization: algorithm and a proof-of-principle study,''
\newblock {\em IEEE TMI}, vol. 33, no. 8, pp. 1581--1591, 2014.

\bibitem{li2014adaptive}
Zhoubo Li, Lifeng Yu, Joshua~D Trzasko, David~S Lake, Daniel~J Blezek, Joel~G
  Fletcher, Cynthia~H McCollough, and Armando Manduca,
\newblock ``Adaptive nonlocal means filtering based on local noise level for ct
  denoising,''
\newblock {\em Medical Physics}, vol. 41, no. 1, pp. 011908, 2014.

\bibitem{kang2013image}
Dongwoo Kang, Piotr Slomka, Ryo Nakazato, Jonghye Woo, Daniel~S Berman, C-C~Jay
  Kuo, and Damini Dey,
\newblock ``Image denoising of low-radiation dose coronary ct angiography by an
  adaptive block-matching 3d algorithm,''
\newblock in {\em Medical Imaging 2013: Image Processing}. International
  Society for Optics and Photonics, 2013, vol. 8669, p. 86692G.

\bibitem{chen2013improving}
Yang Chen, Xindao Yin, Luyao Shi, Huazhong Shu, Limin Luo, Jean-Louis
  Coatrieux, and Christine Toumoulin,
\newblock ``Improving abdomen tumor low-dose ct images using a fast dictionary
  learning based processing,''
\newblock {\em Physics in Medicine \& Biology}, vol. 58, no. 16, pp. 5803,
  2013.

\bibitem{chen2017low}
Hu~Chen, Yi~Zhang, Mannudeep~K Kalra, Feng Lin, Yang Chen, Peixi Liao, Jiliu
  Zhou, and Ge~Wang,
\newblock ``Low-dose ct with a residual encoder-decoder convolutional neural
  network,''
\newblock {\em IEEE TMI}, vol. 36, no. 12, pp. 2524--2535, 2017.

\bibitem{kang2017deep}
Eunhee Kang, Junhong Min, and Jong~Chul Ye,
\newblock ``A deep convolutional neural network using directional wavelets for
  low-dose x-ray ct reconstruction,''
\newblock {\em Medical Physics}, vol. 44, no. 10, pp. e360--e375, 2017.

\bibitem{ronneberger2015u}
Olaf Ronneberger, Philipp Fischer, and Thomas Brox,
\newblock ``U-net: Convolutional networks for biomedical image segmentation,''
\newblock in {\em MICCAI}, 2015, pp. 234--241.

\bibitem{mildenhall2018burst}
Ben Mildenhall, Jonathan~T Barron, Jiawen Chen, Dillon Sharlet, Ren Ng, and
  Robert Carroll,
\newblock ``Burst denoising with kernel prediction networks,''
\newblock in {\em CVPR}, 2018, pp. 2502--2510.

\bibitem{romano2016raisr}
Yaniv Romano, John Isidoro, and Peyman Milanfar,
\newblock ``Raisr: rapid and accurate image super resolution,''
\newblock {\em IEEE TCI}, vol. 3, no. 1, pp. 110--125, 2016.

\bibitem{wang2004image}
Zhou Wang, Alan~C Bovik, Hamid~R Sheikh, Eero~P Simoncelli, et~al.,
\newblock ``Image quality assessment: from error visibility to structural
  similarity,''
\newblock {\em IEEE TIP}, vol. 13, no. 4, pp. 600--612, 2004.

\bibitem{zhao2016loss}
Hang Zhao, Orazio Gallo, Iuri Frosio, and Jan Kautz,
\newblock ``Loss functions for image restoration with neural networks,''
\newblock {\em IEEE TCI}, vol. 3, no. 1, pp. 47--57, 2016.

\bibitem{cho2017wavelet}
Ji~Hee Cho, Eunhee Kang, Jihye Byun, Gael Lee, and Hee Kho,
\newblock ``Wavelet residual network for low-dose ct via deep convolutional
  framelets,''
\newblock {\em IEEE TMI}, vol. 37, no. 6, 2017.

\bibitem{sheng2014denoised}
Ke~Sheng, Shuiping Gou, Jiaolong Wu, and Sharon~X Qi,
\newblock ``Denoised and texture enhanced mvct to improve soft tissue
  conspicuity,''
\newblock {\em Medical Physics}, vol. 41, no. 10, pp. 101916, 2014.

\bibitem{mccollough2016tu}
C~McCollough,
\newblock ``Tu-fg-207a-04: Overview of the low dose ct grand challenge,''
\newblock {\em Medical Physics}, vol. 43, no. 6Part35, pp. 3759--3760, 2016.

\bibitem{du2019methods}
Yihui Du, Yingru Zhao, Grigory Sidorenkov, Geertruida~H de~Bock, Xiaonan Cui,
  Yubei Huang, Monique~D Dorrius, Mieneke Rook, Harry~JM Groen, Marjolein~A
  Heuvelmans, et~al.,
\newblock ``Methods of computed tomography screening and management of lung
  cancer in tianjin: design of a population-based cohort study,''
\newblock {\em Cancer Biology and Medicine}, vol. 16, no. 1, pp. 181, 2019.

\bibitem{aharon2006k}
Michal Aharon, Michael Elad, and Alfred Bruckstein,
\newblock ``K-svd: An algorithm for designing overcomplete dictionaries for
  sparse representation,''
\newblock {\em IEEE TSP}, vol. 54, no. 11, pp. 4311--4322, 2006.

\bibitem{dabov2007image}
Kostadin Dabov, Alessandro Foi, Vladimir Katkovnik, and Karen Egiazarian,
\newblock ``Image denoising by sparse 3-d transform-domain collaborative
  filtering,''
\newblock {\em IEEE TIP}, vol. 16, no. 8, pp. 2080--2095, 2007.

\end{thebibliography}
\newpage
\begingroup
\let\clearpage\relax 
\onecolumn 
\appendix
\section{Discussion on KPN Kernel Size}\vspace{-3mm}
In this section we discuss the effect of denoising kernel size, the major hyperparameter of KPN, on the LDCT denoising performance. As can be seen in the table below, when the kernel size is small, the denoising filter is not expressive enough, and when kernel size is too large, the performance gain is negligible while the model has more parameters and can be harder to train.

\begin{table*}[!h]
	\begin{center}
		\caption{Denoising performance (PSNR/SSIM) on the NIH dataset by different KPN kernel sizes.}
		\label{table:ablation_k}
		\resizebox{\linewidth}{!}{
			\begin{tabular}{l|llllll}\hline\hline
				& $k$=15   & $k$=19 & $k$=21 & $k$=23 & $k$=27   \\ \hline
				NIH      & 32.50/0.890 & 32.55/0.893 & 32.56/0.893 & 32.58/0.894 & 32.53/0.891   \\ \hline\hline
		\end{tabular}}
		
	\end{center}
\end{table*}

\section{Additional Visual Results}\vspace{-3mm}
In this section we include additional visual comparisons of different methods on DOSE, PHANTOM and NIH datasets and provide qualitative analysis on clinically important regions.
\begin{figure*}[!ht]
	\begin{center}
		\begin{tabular}{cccccccc}
			\includegraphics[width=\swfiveh]{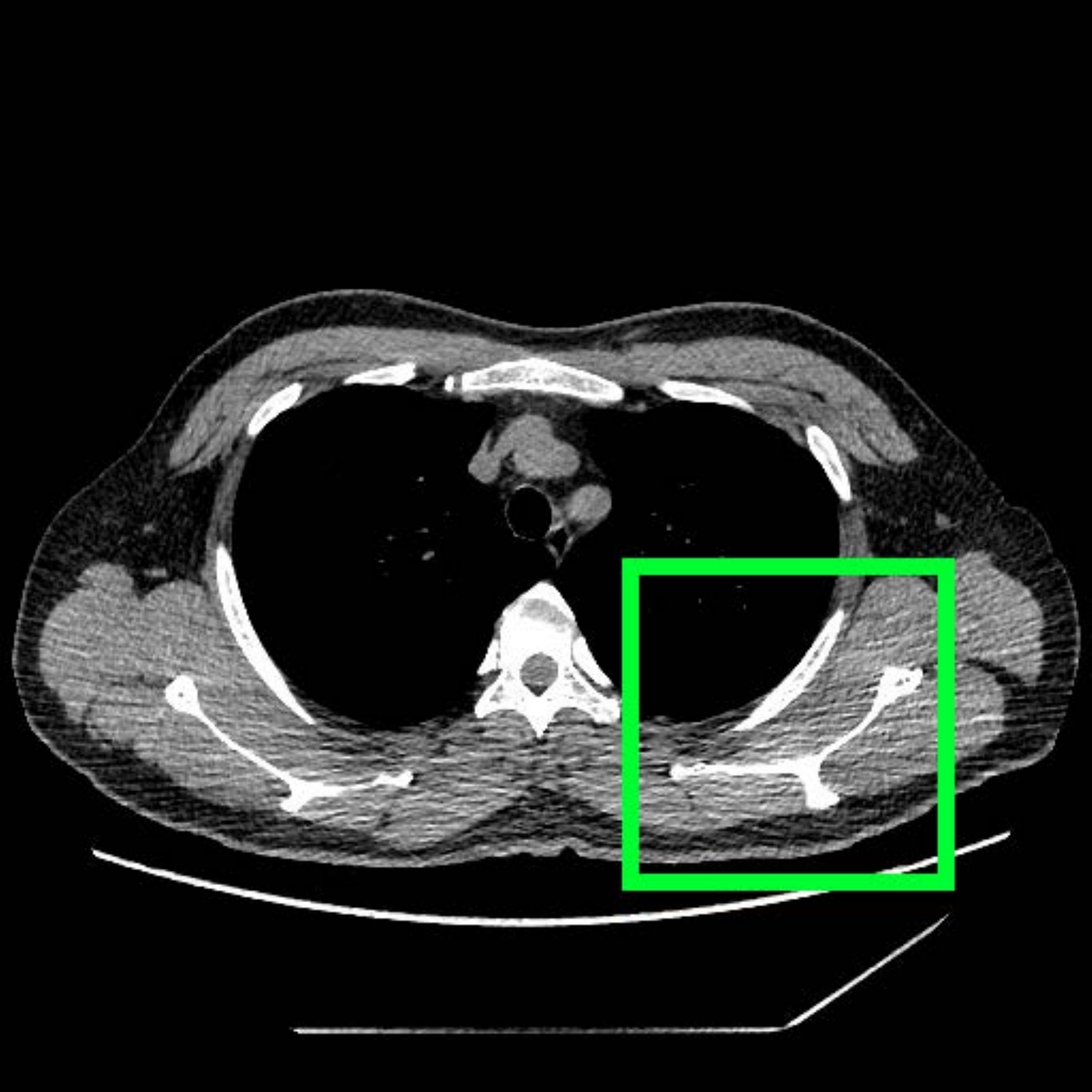} &
			\includegraphics[width=\swfiveh]{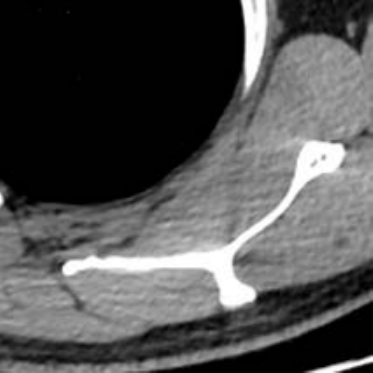} &
			\includegraphics[width=\swfiveh]{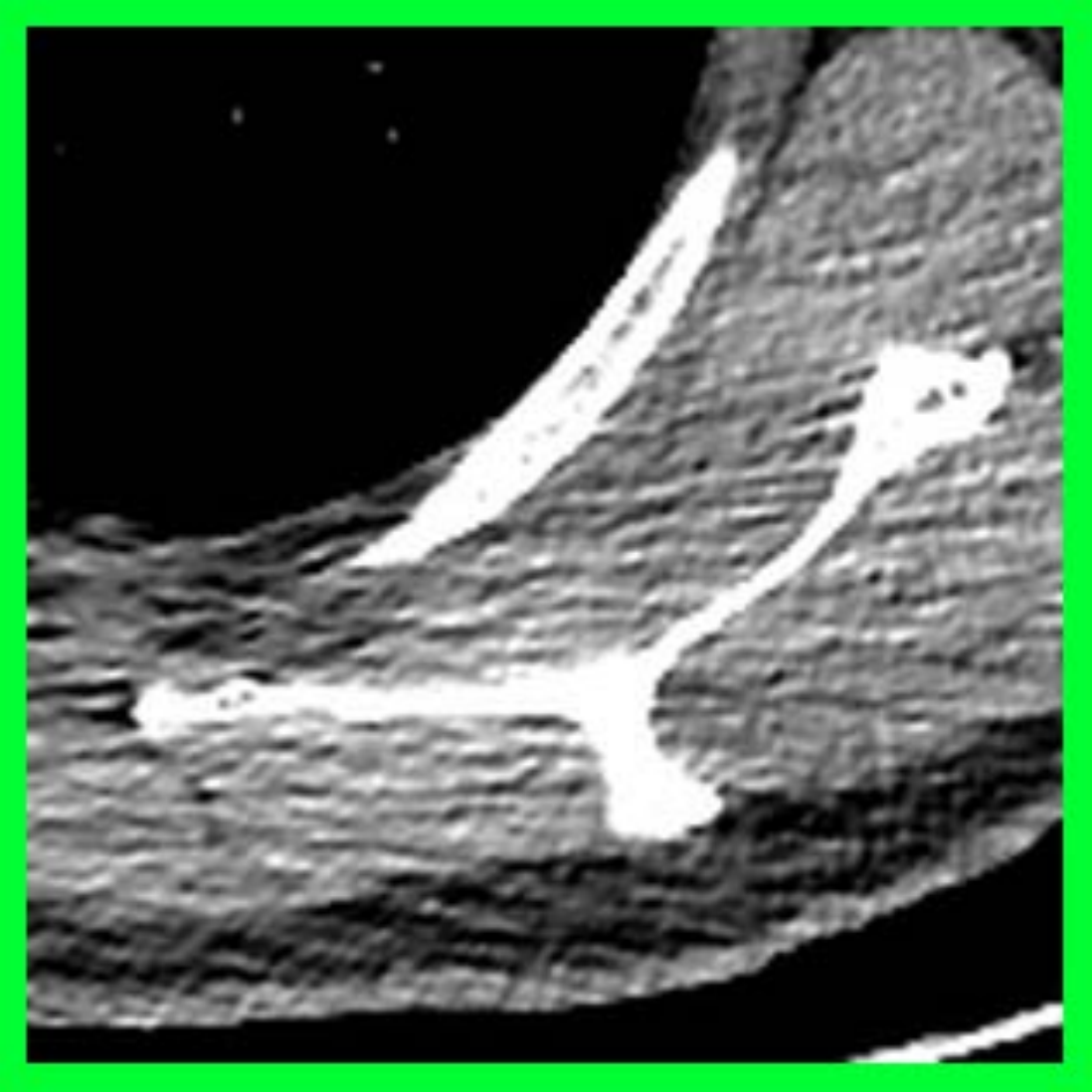} &
			\includegraphics[width=\swfiveh]{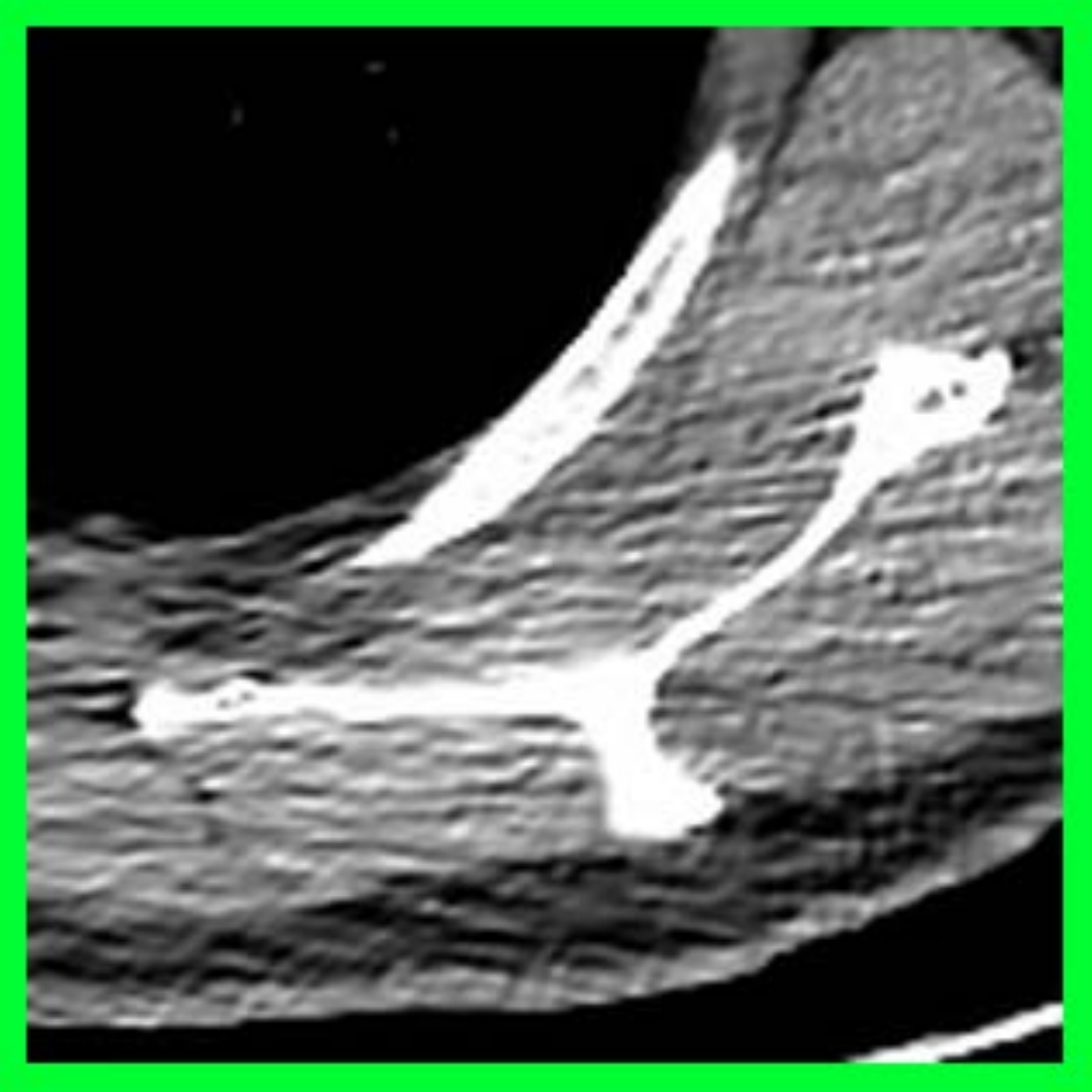} &\\
			full image & NDCT & LDCT & KSVD &\\
			\includegraphics[width=\swfiveh]{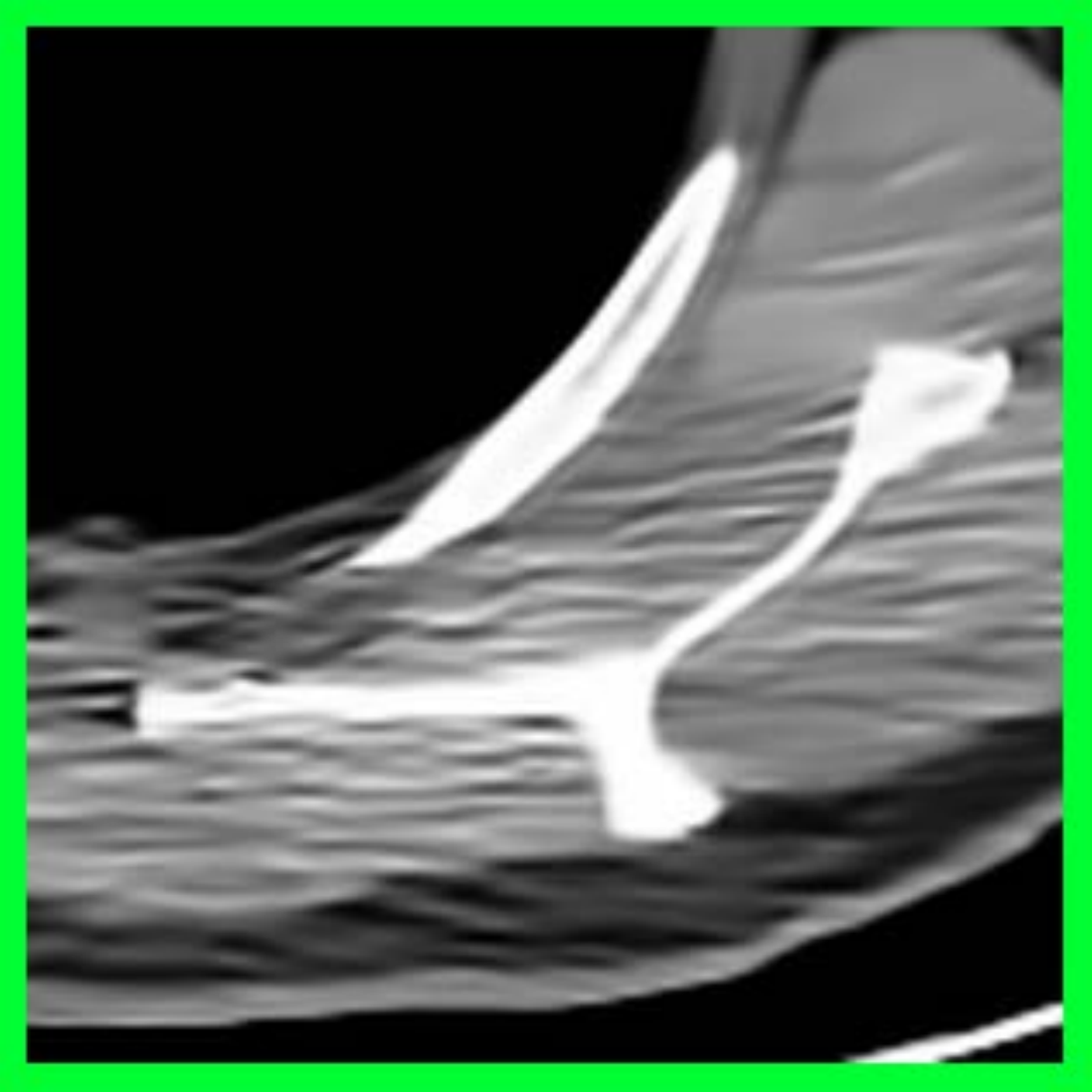} &
			\includegraphics[width=\swfiveh]{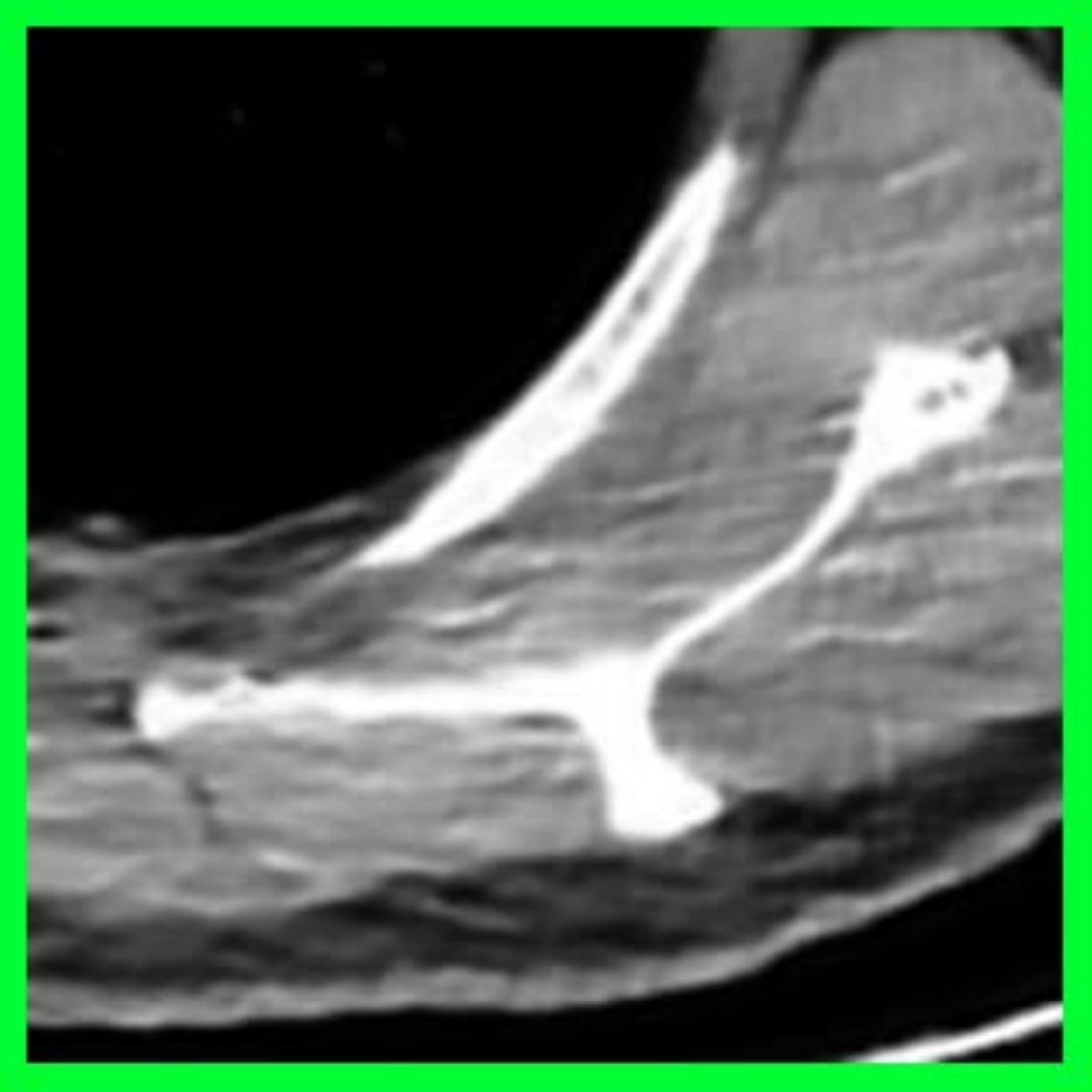} &
			\includegraphics[width=\swfiveh]{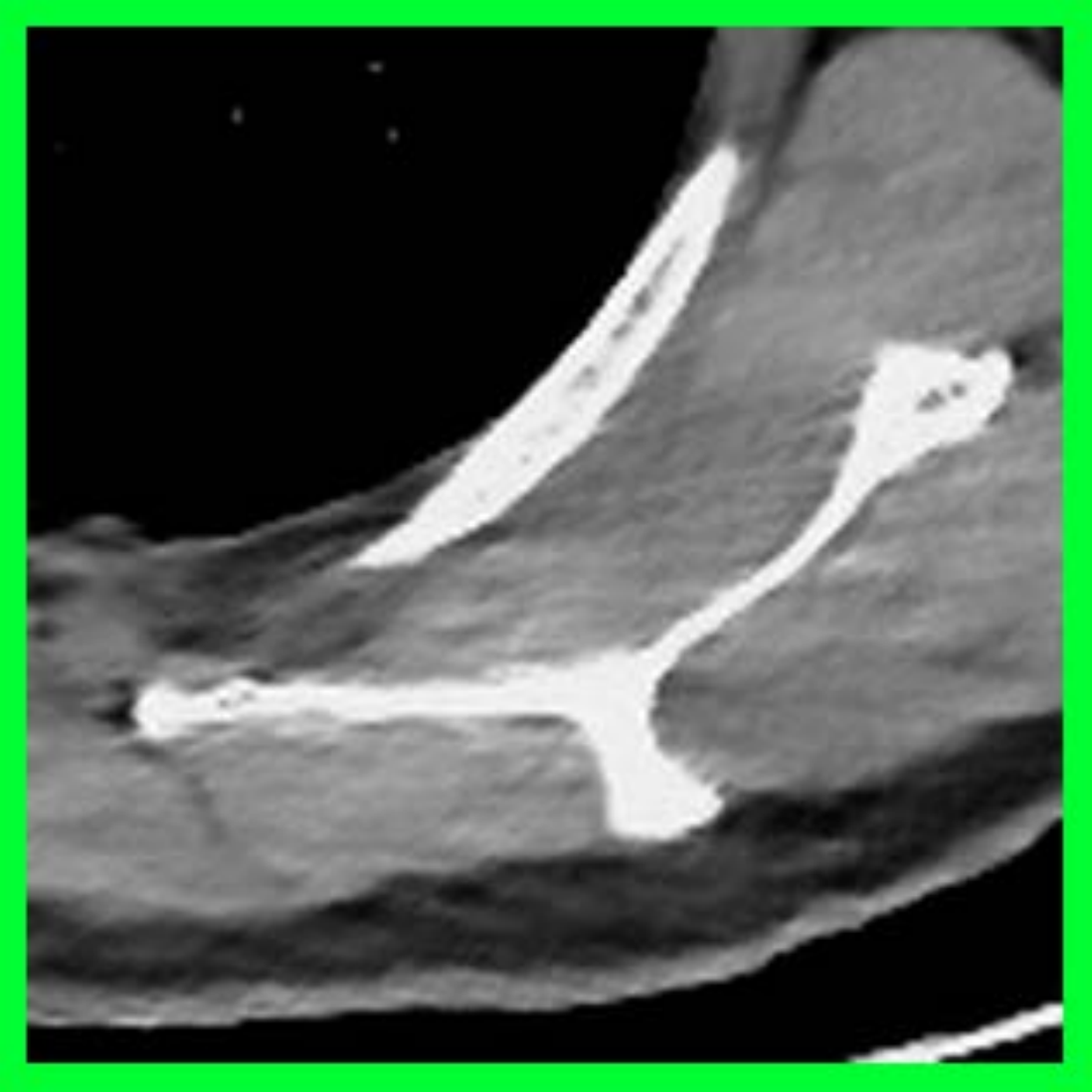} &
			\includegraphics[width=\swfiveh]{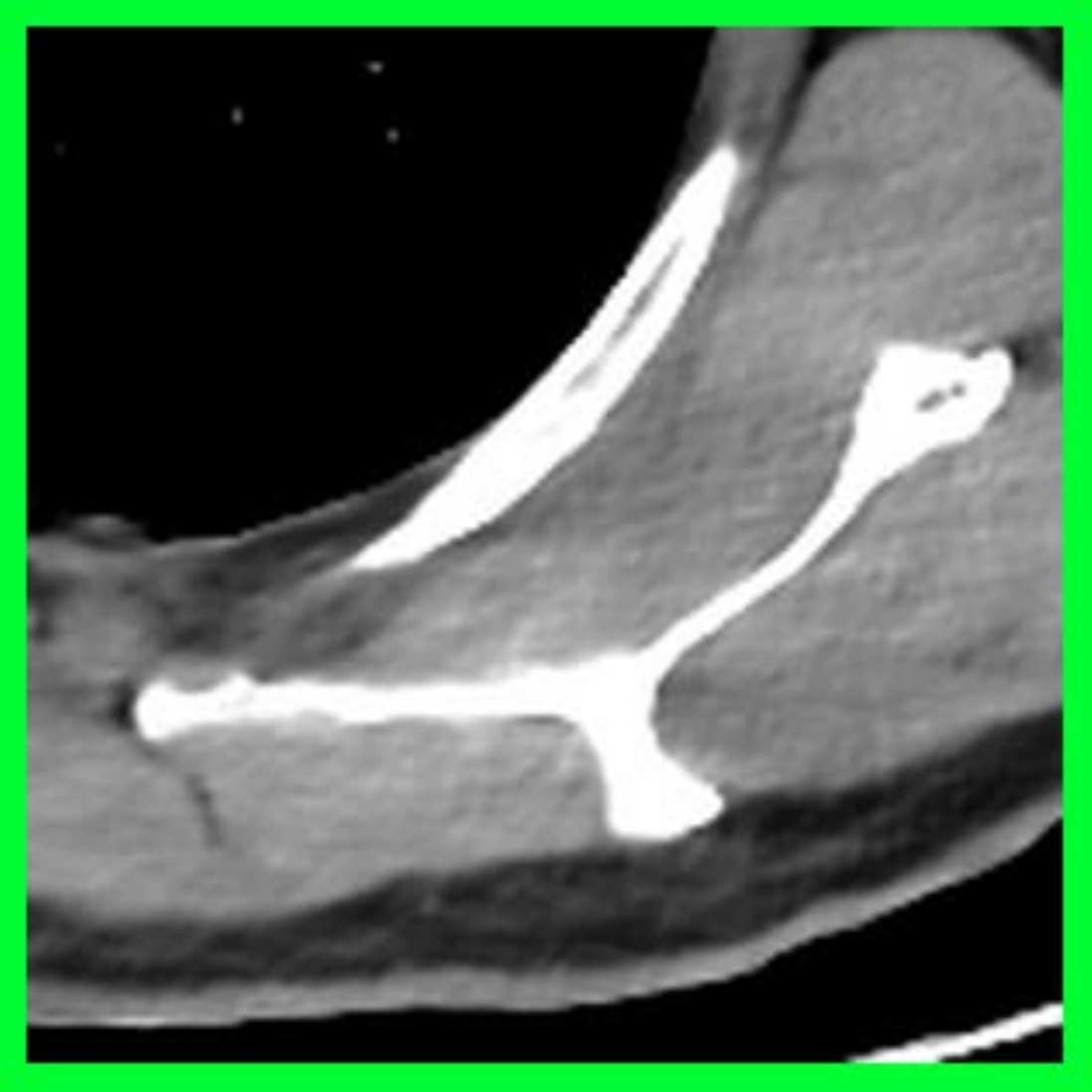} \\
			BM3D & WavResNet & RED-CNN & StructKPN\\
		\end{tabular}
	\end{center}
	\caption{
		Visual comparison on DOSE dataset. The NDCT image is picked by the radiologist and only serves as a reference as it is not perfectly aligned with LDCT. StructKPN results in better recovery of streaking artifacts caused by photon starvation and beam-hardening effect around bones.
	}
	\label{fig:doseres2}
\end{figure*}

\begin{figure*}[!ht]
	\begin{center}
		\begin{tabular}{cccccccc}
			\includegraphics[width=\swfiveh]{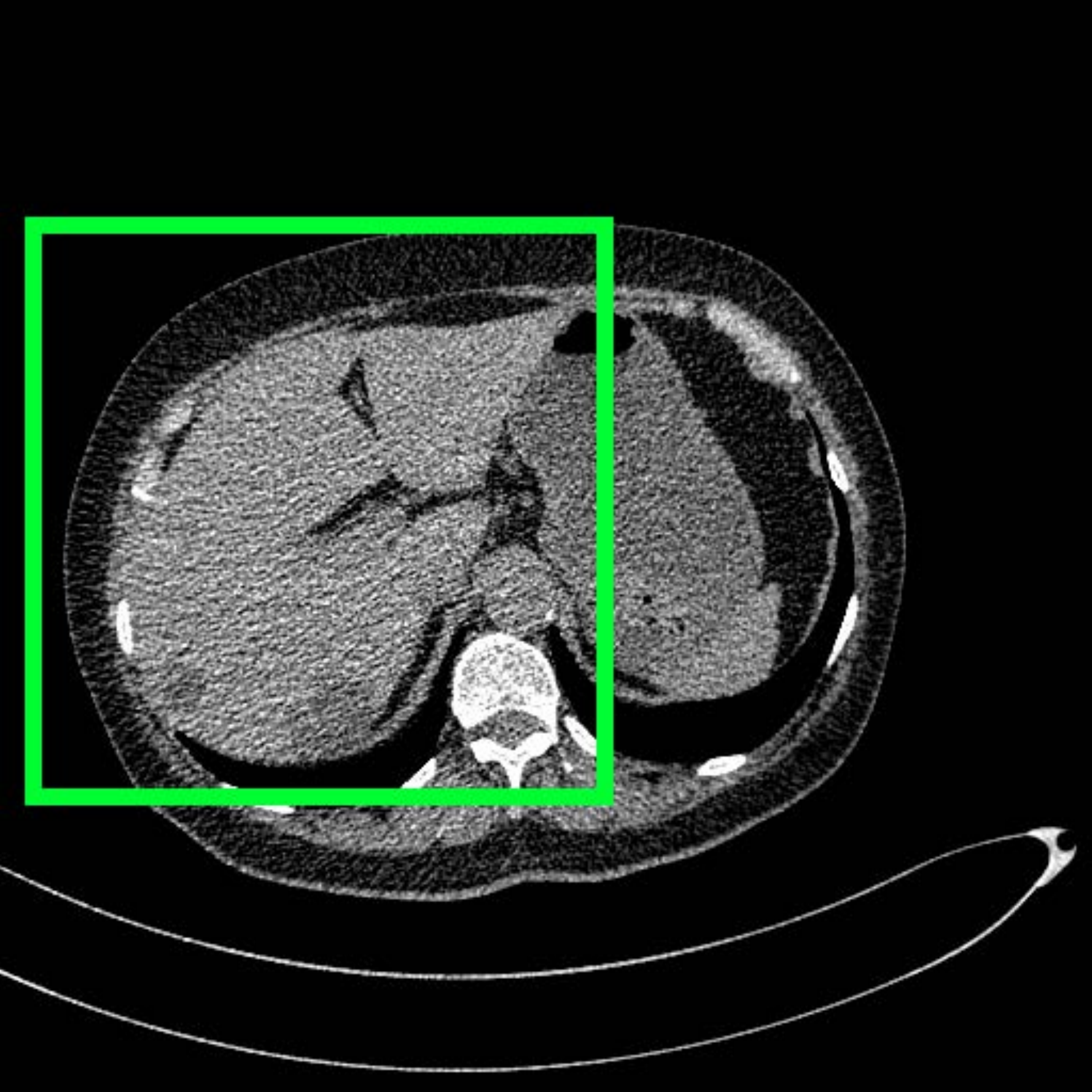} &
			\includegraphics[width=\swfiveh]{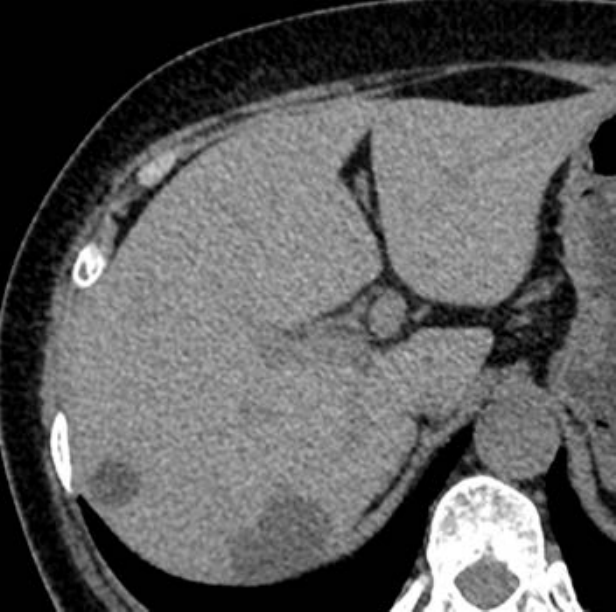} &
			\includegraphics[width=\swfiveh]{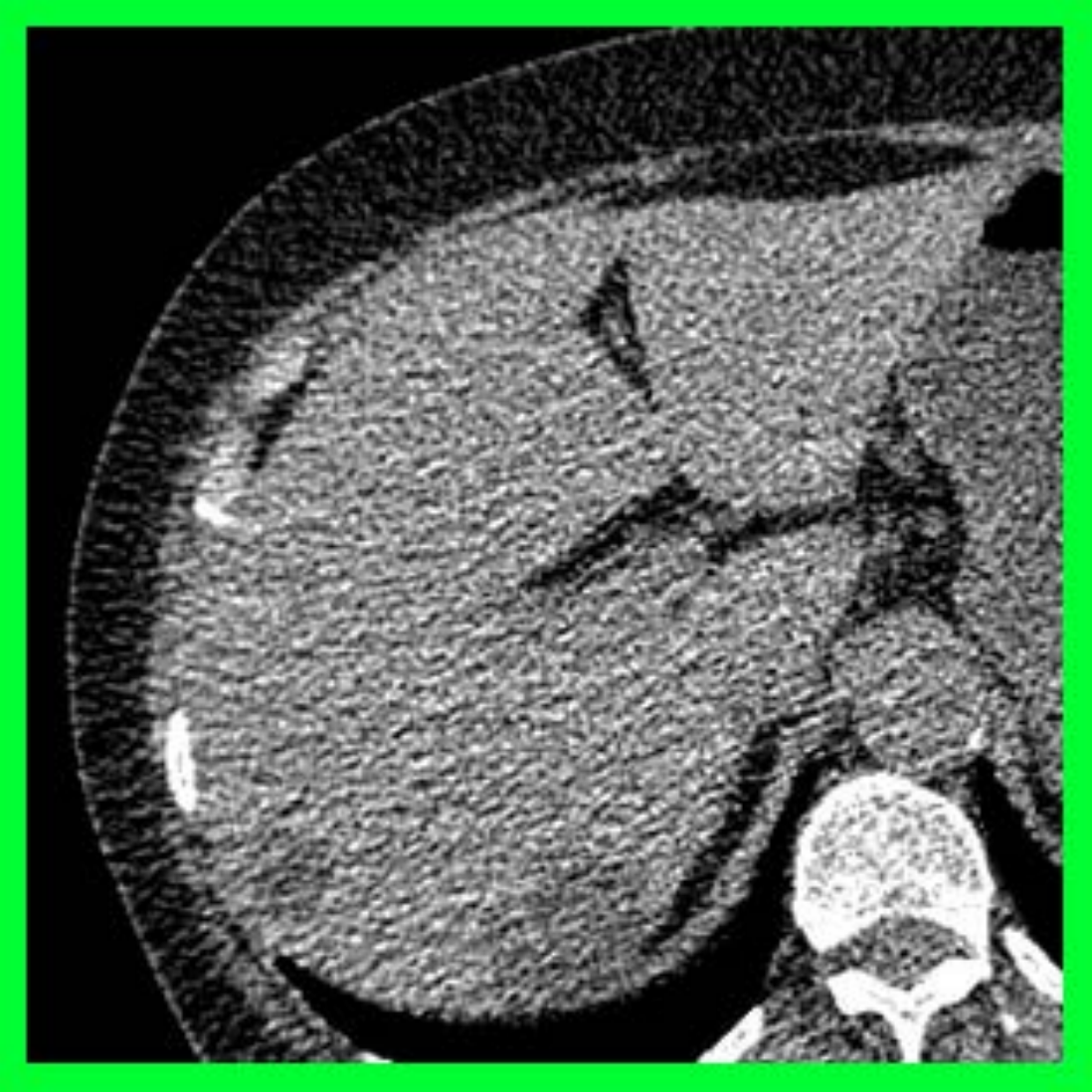} &
			\includegraphics[width=\swfiveh]{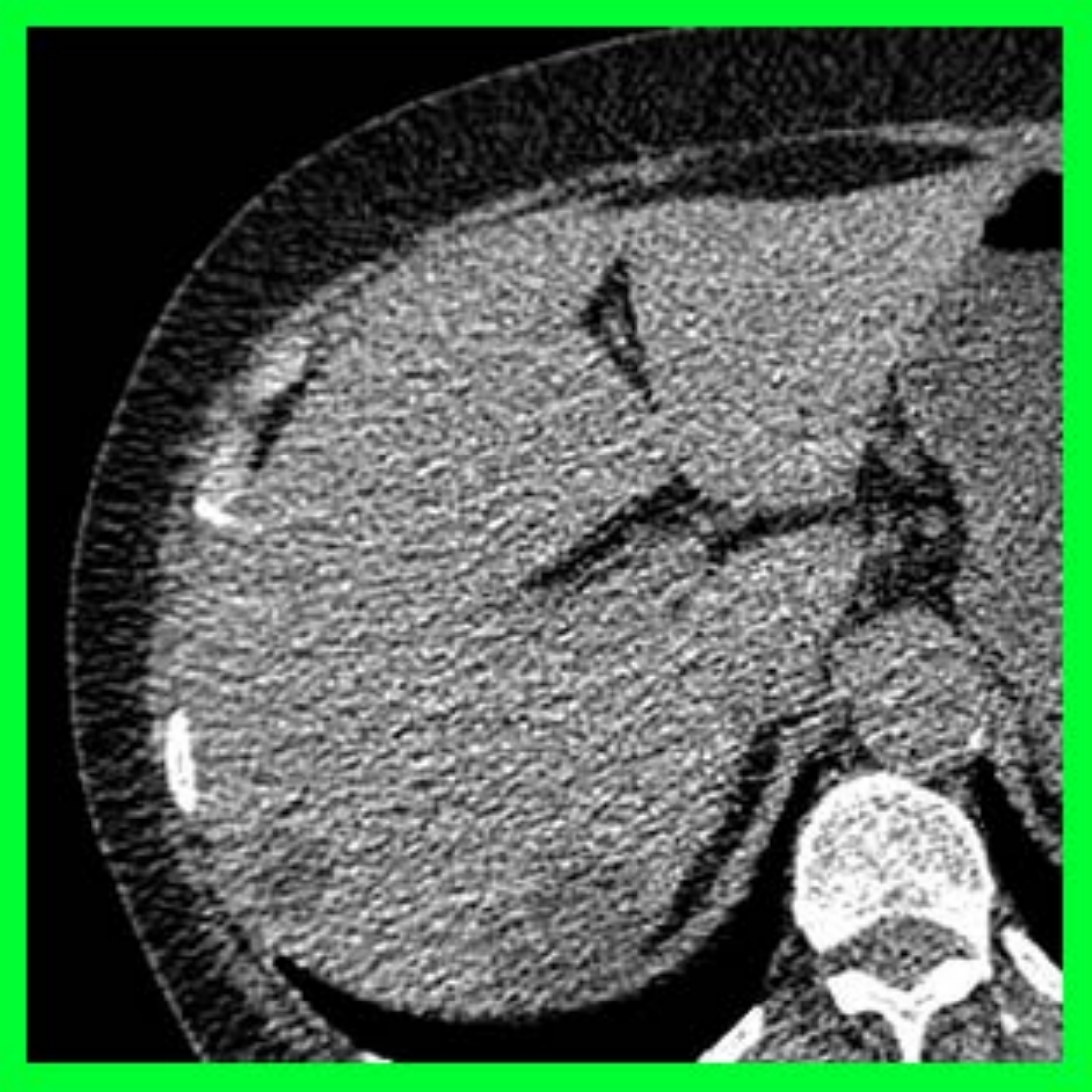} &\\
			full image & NDCT & LDCT & KSVD &\\
			\includegraphics[width=\swfiveh]{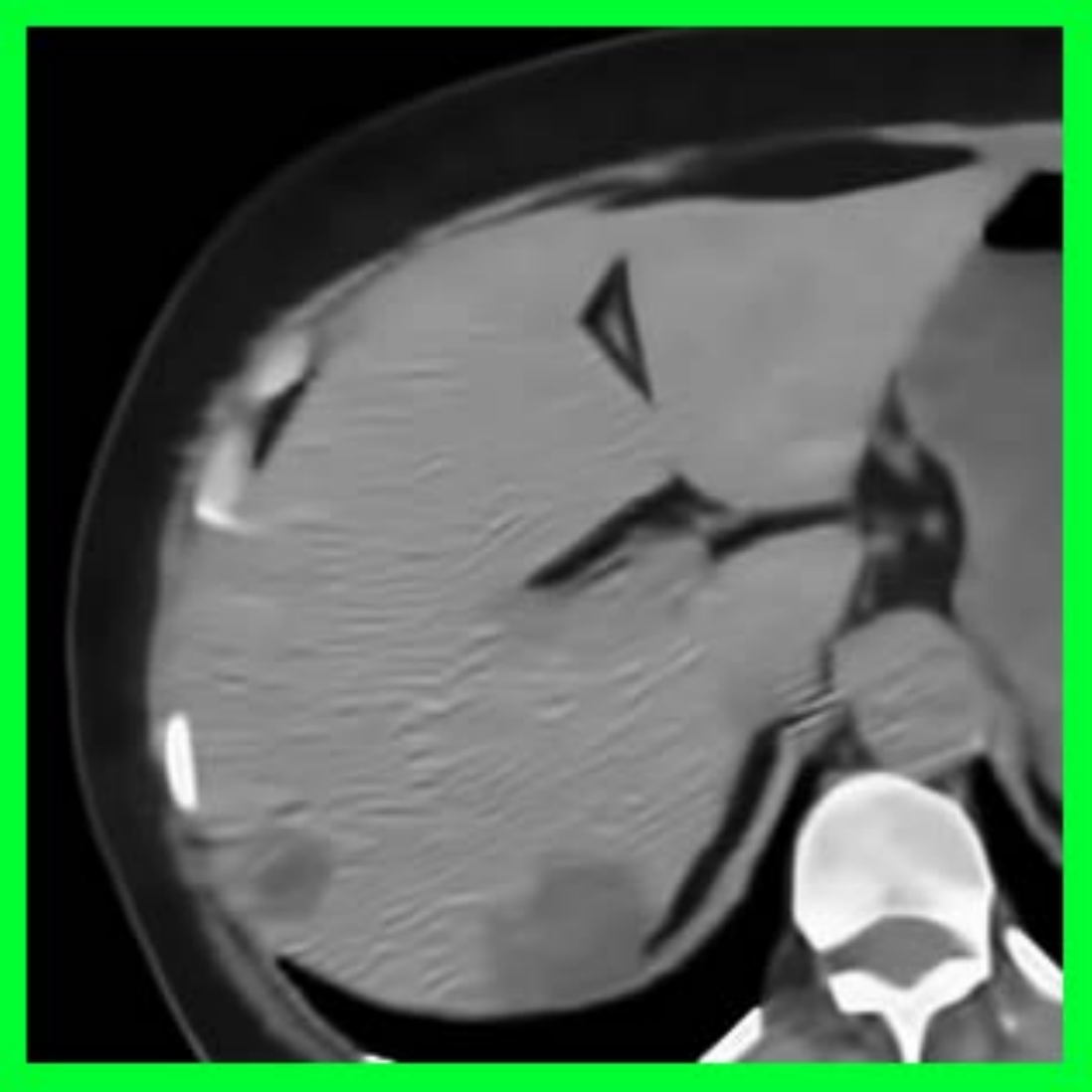} &
			\includegraphics[width=\swfiveh]{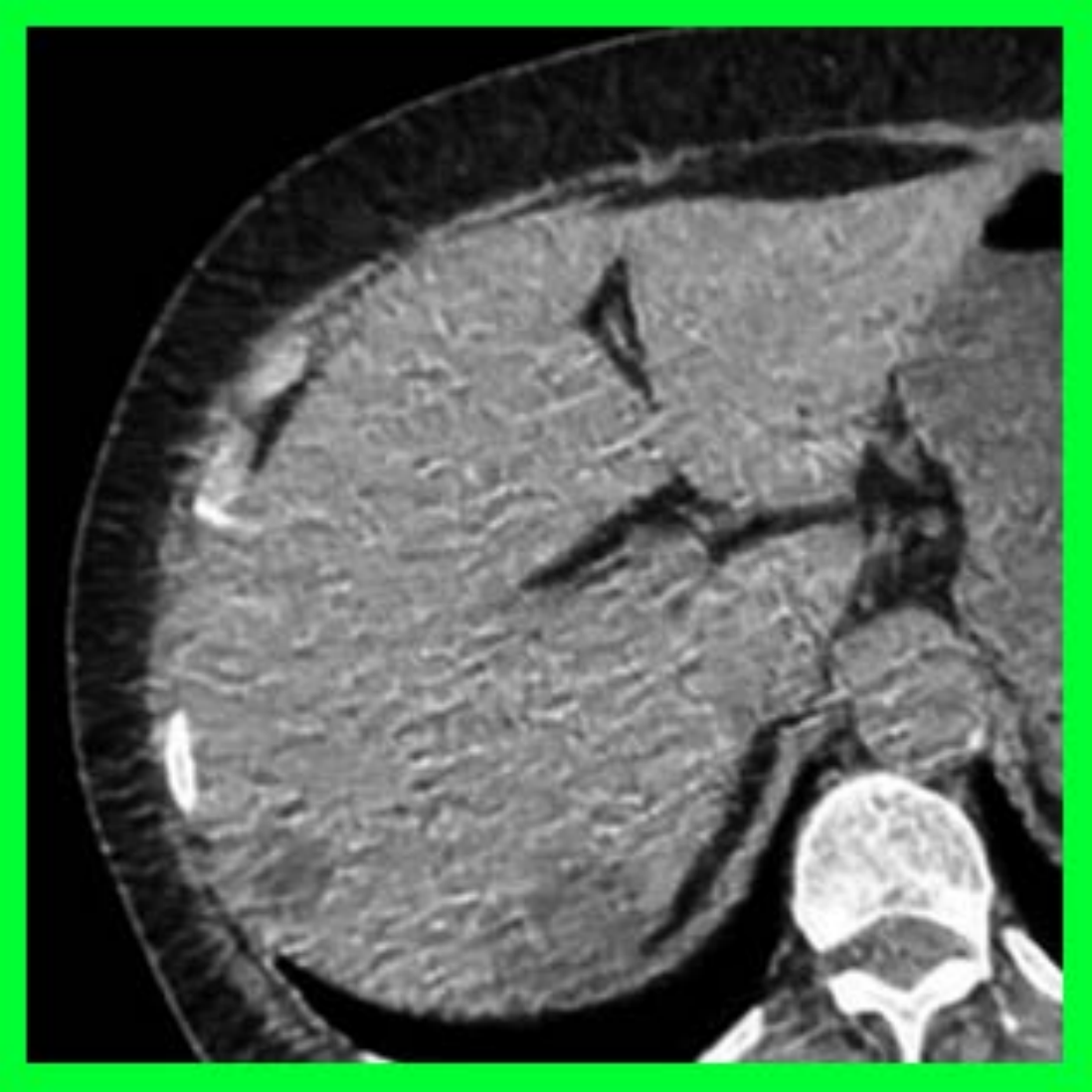} &
			\includegraphics[width=\swfiveh]{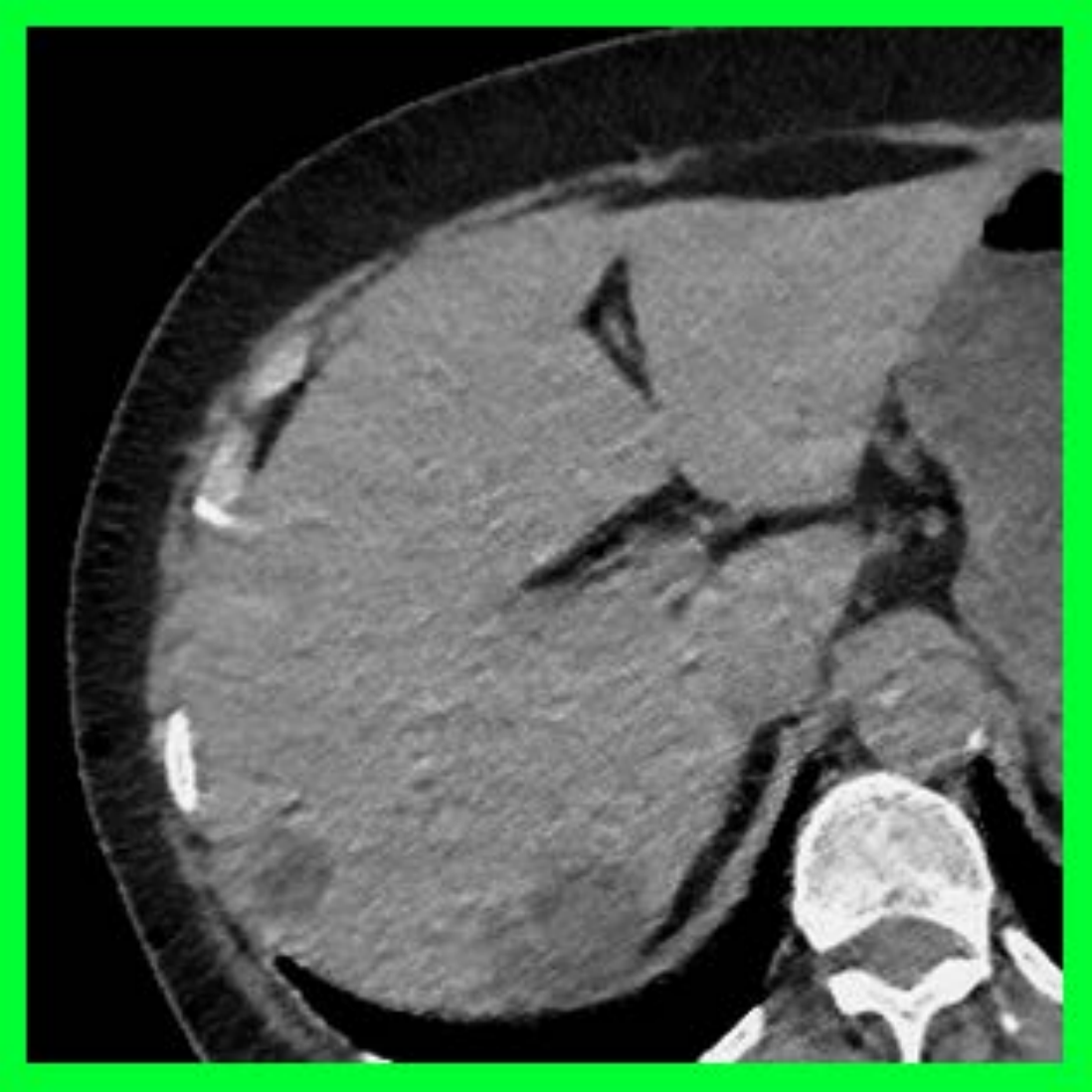} &
			\includegraphics[width=\swfiveh]{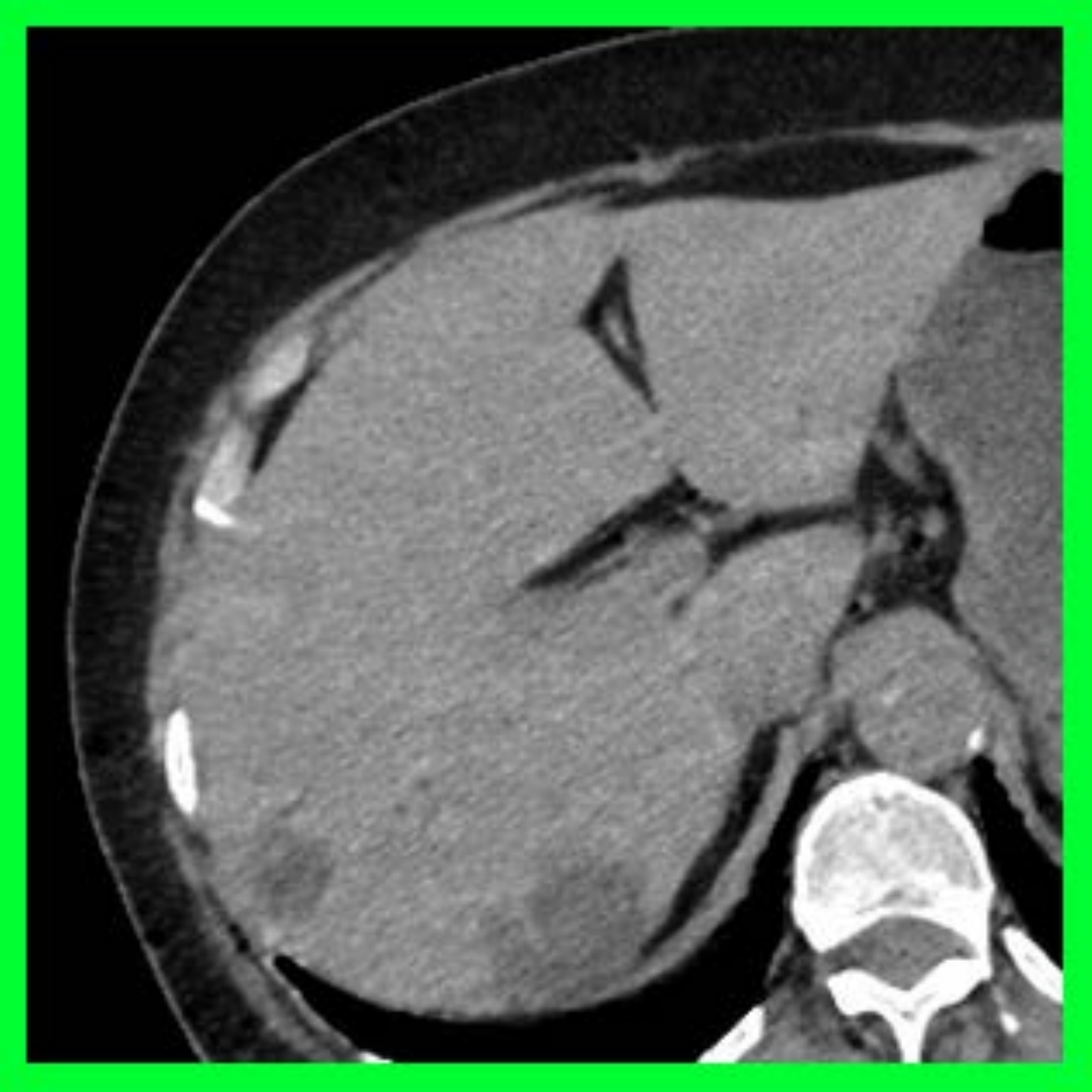} \\
			BM3D & WavResNet & RED-CNN & StructKPN \\
		\end{tabular}
	\end{center}
	\caption{
		Visual comparison on DOSE dataset. StructKPN results in more realistic and uniform liver parenchyma texture, more sharpened lesion edge. less painterly rendering sensation and quantum noise.
} 
\label{fig:doseres2}
\end{figure*}
\begin{figure*}[!ht]
	\begin{center}
		\begin{tabular}{cccccccc}
			\includegraphics[width=\swfiveh]{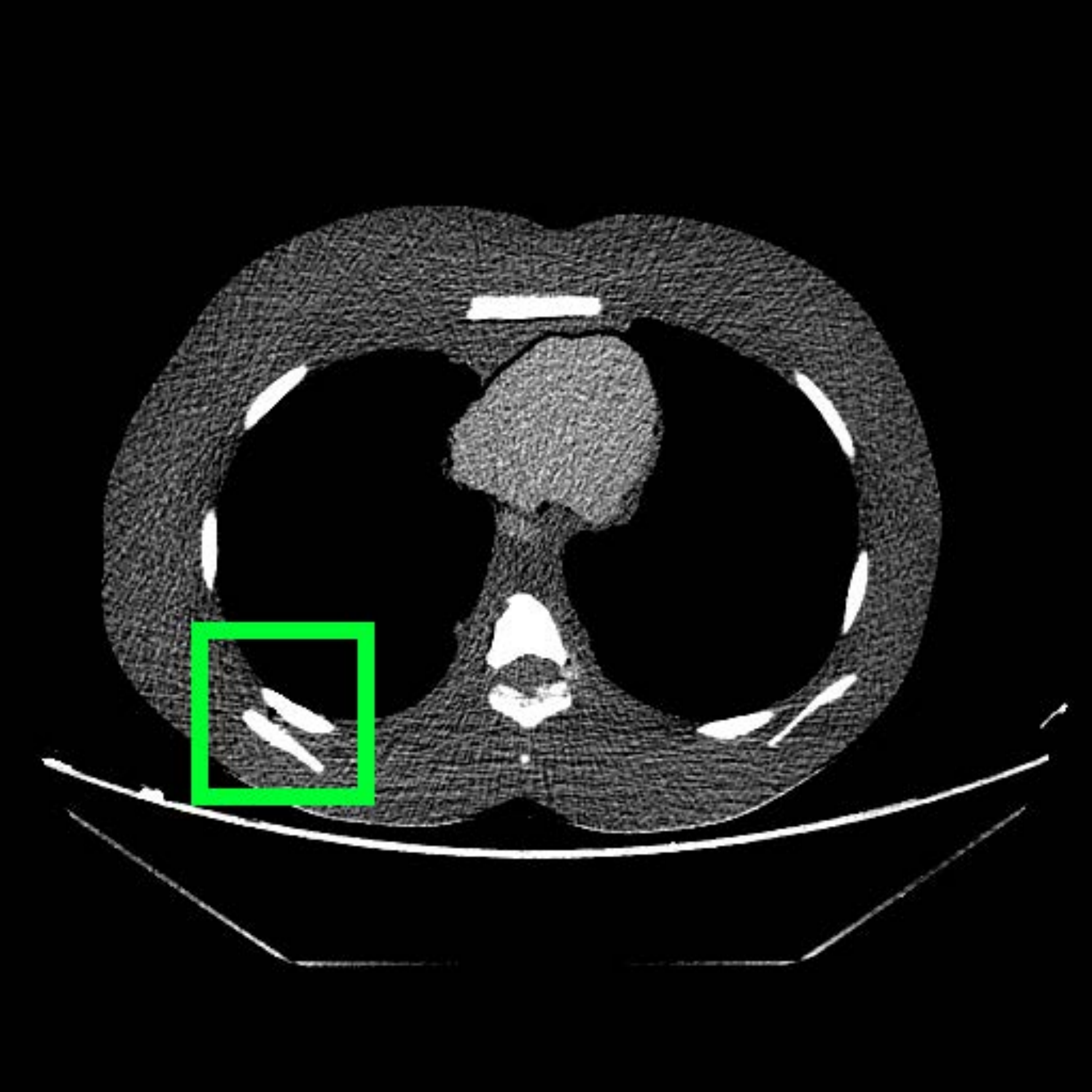} &
			\includegraphics[width=\swfiveh]{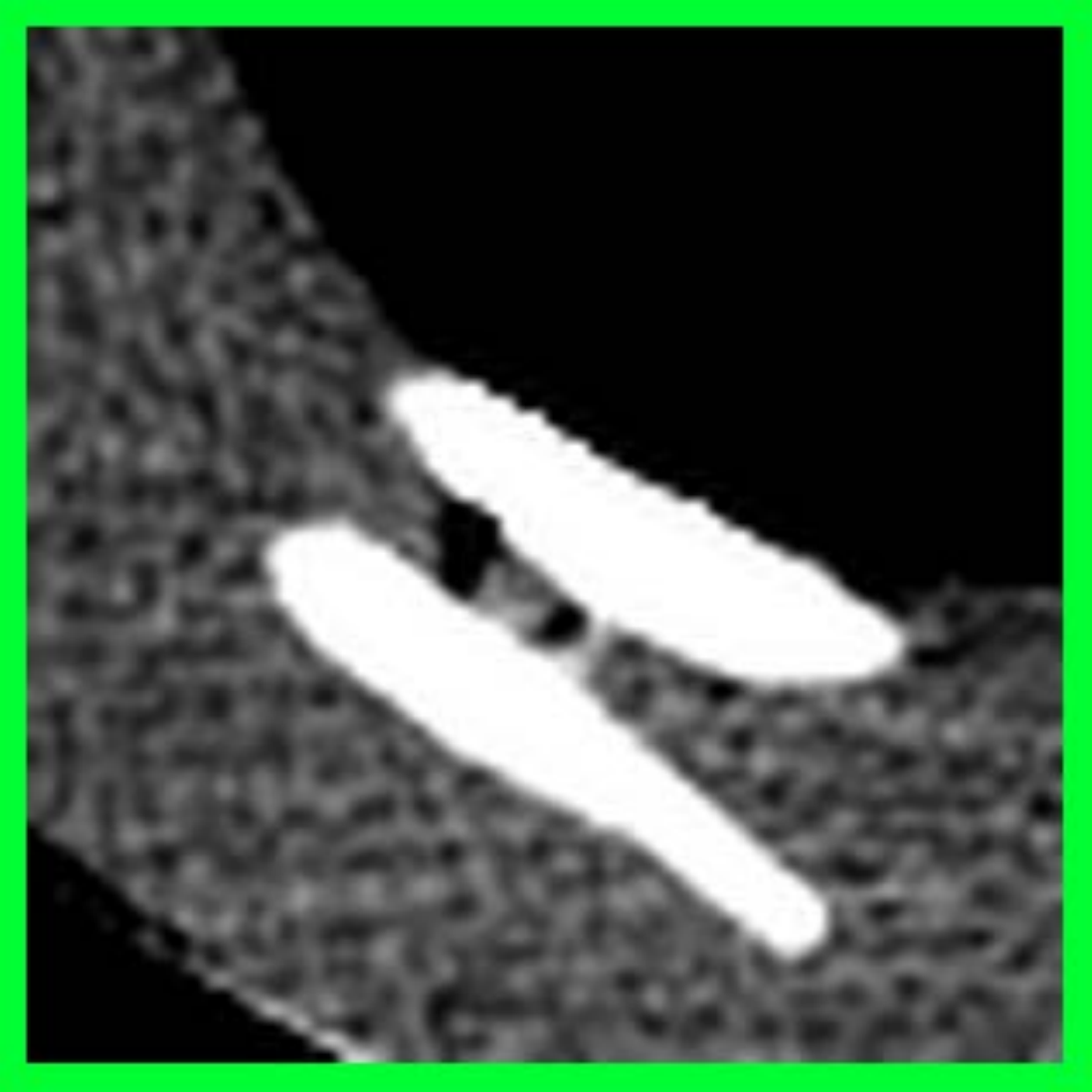} &
			\includegraphics[width=\swfiveh]{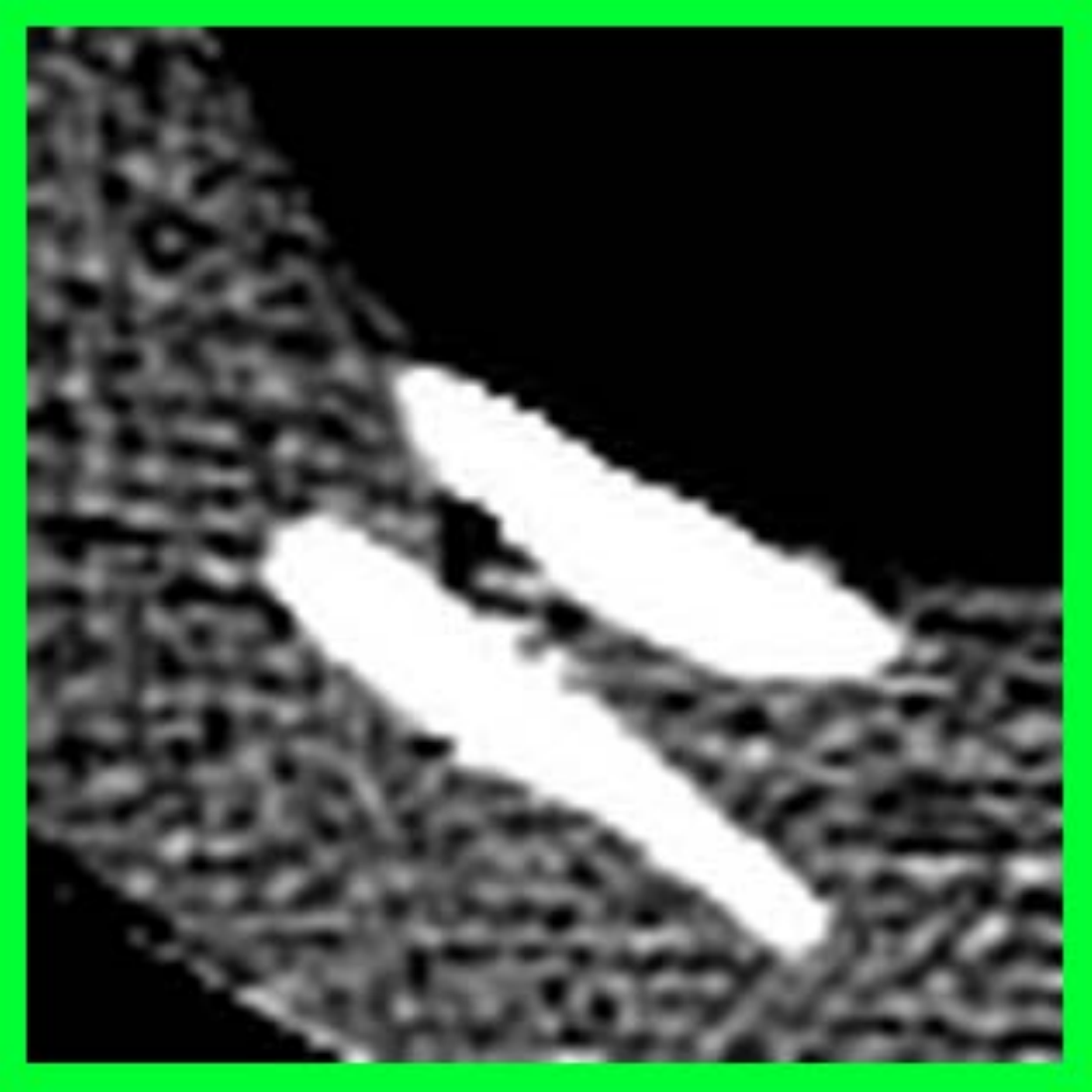} &
			\includegraphics[width=\swfiveh]{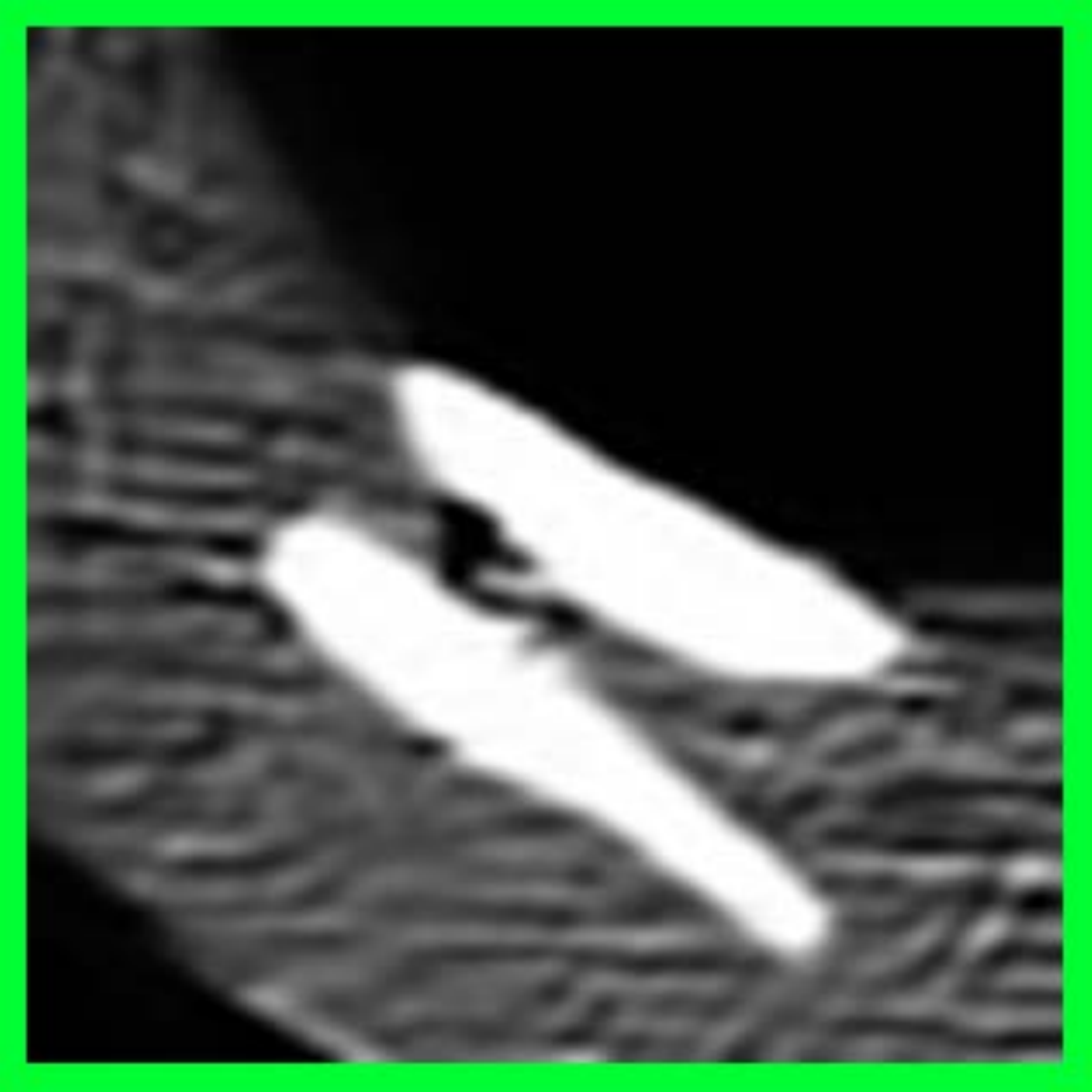} &\\
			full image & NDCT & LDCT & KSVD &\\
			& PSNR/SSIM & 30.36/0.677 & 33.32/0.817 &\\
			\includegraphics[width=\swfiveh]{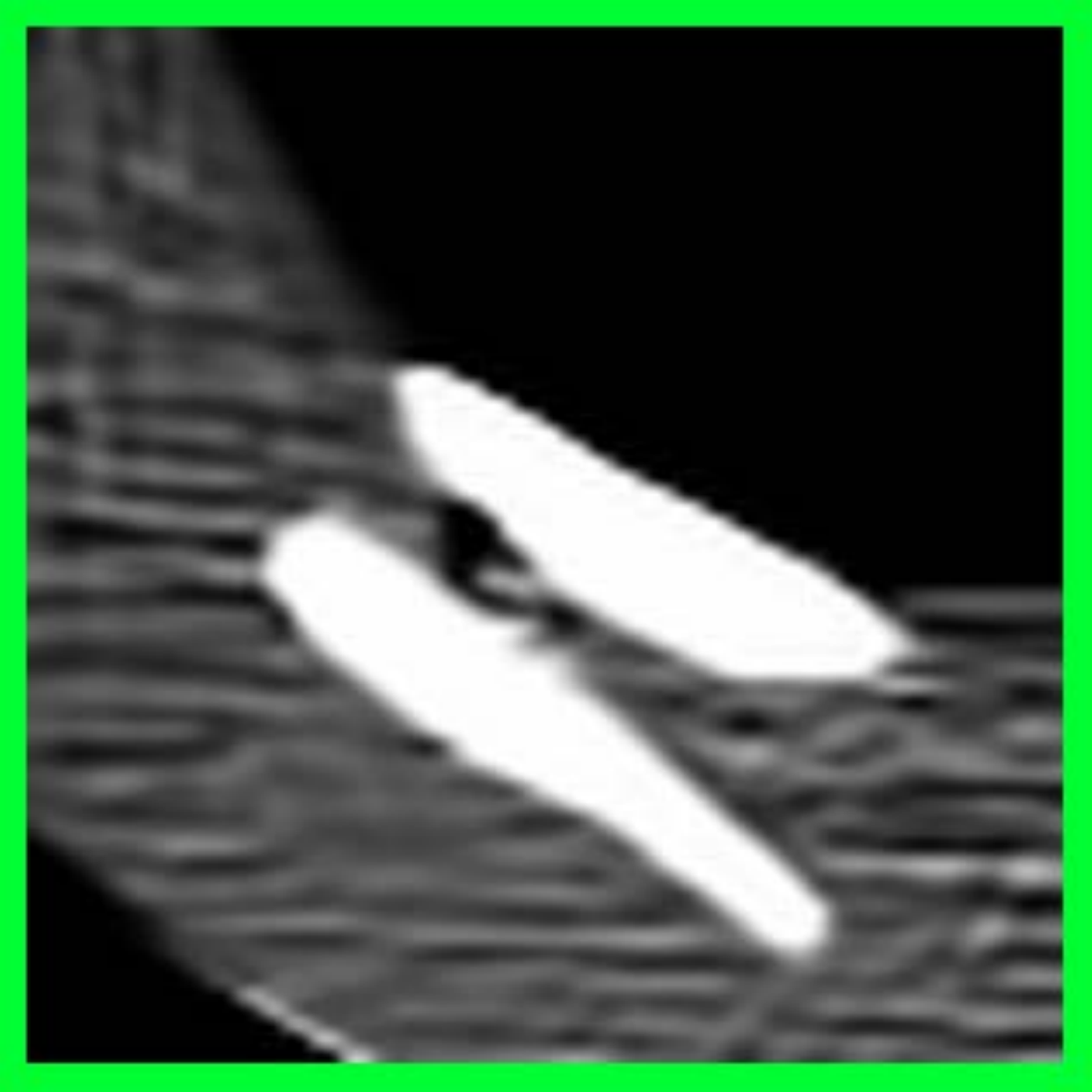} &
			\includegraphics[width=\swfiveh]{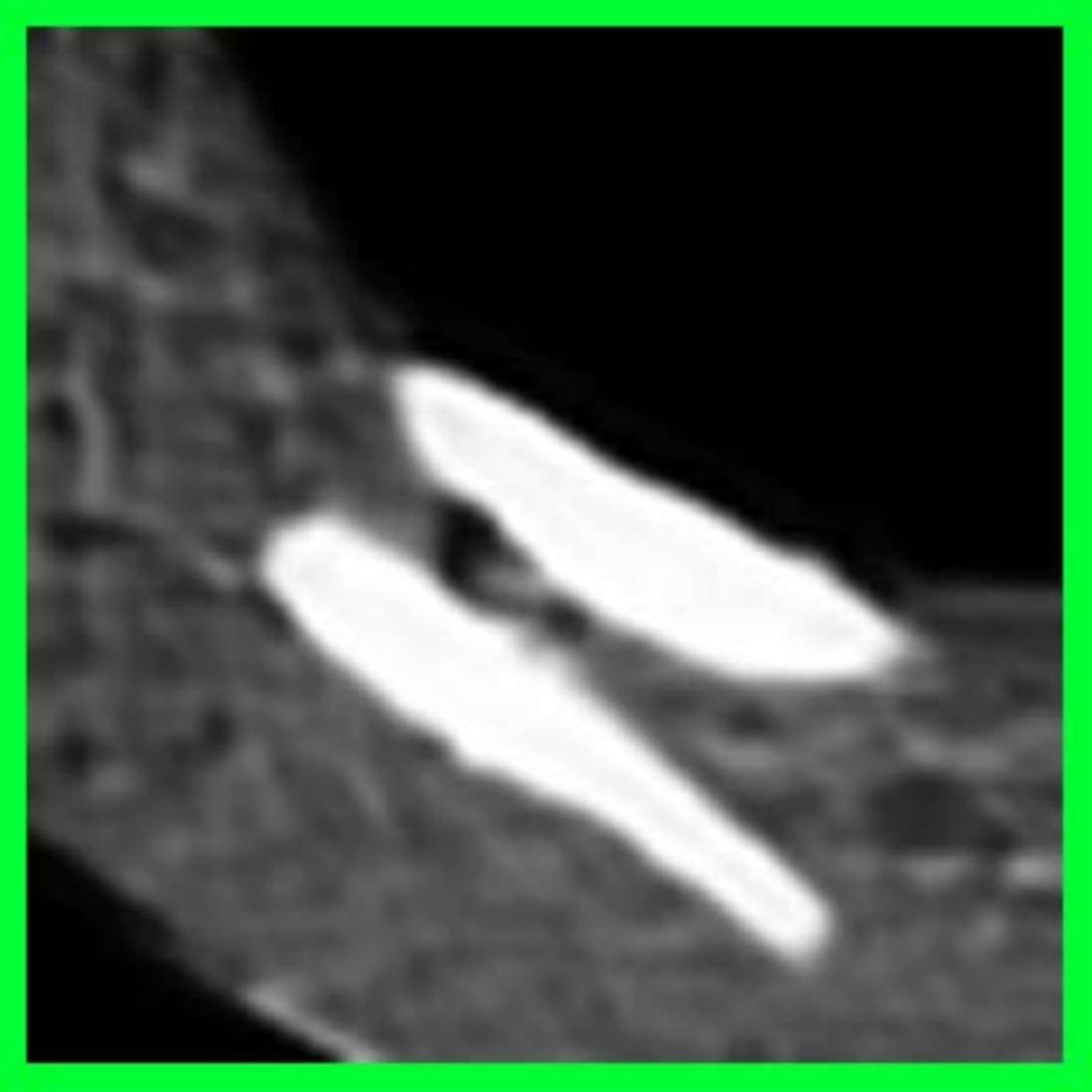} &
			\includegraphics[width=\swfiveh]{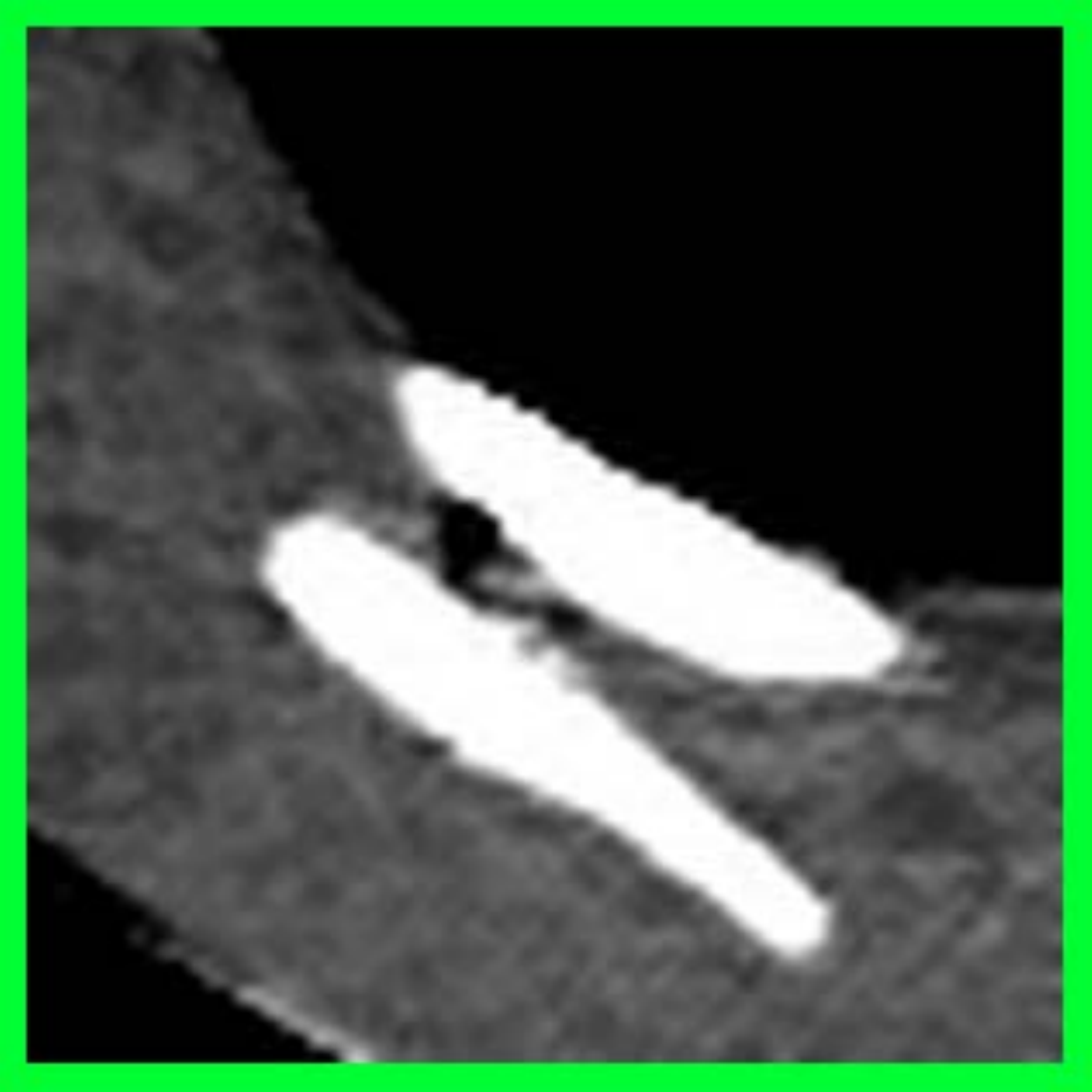} &
			\includegraphics[width=\swfiveh]{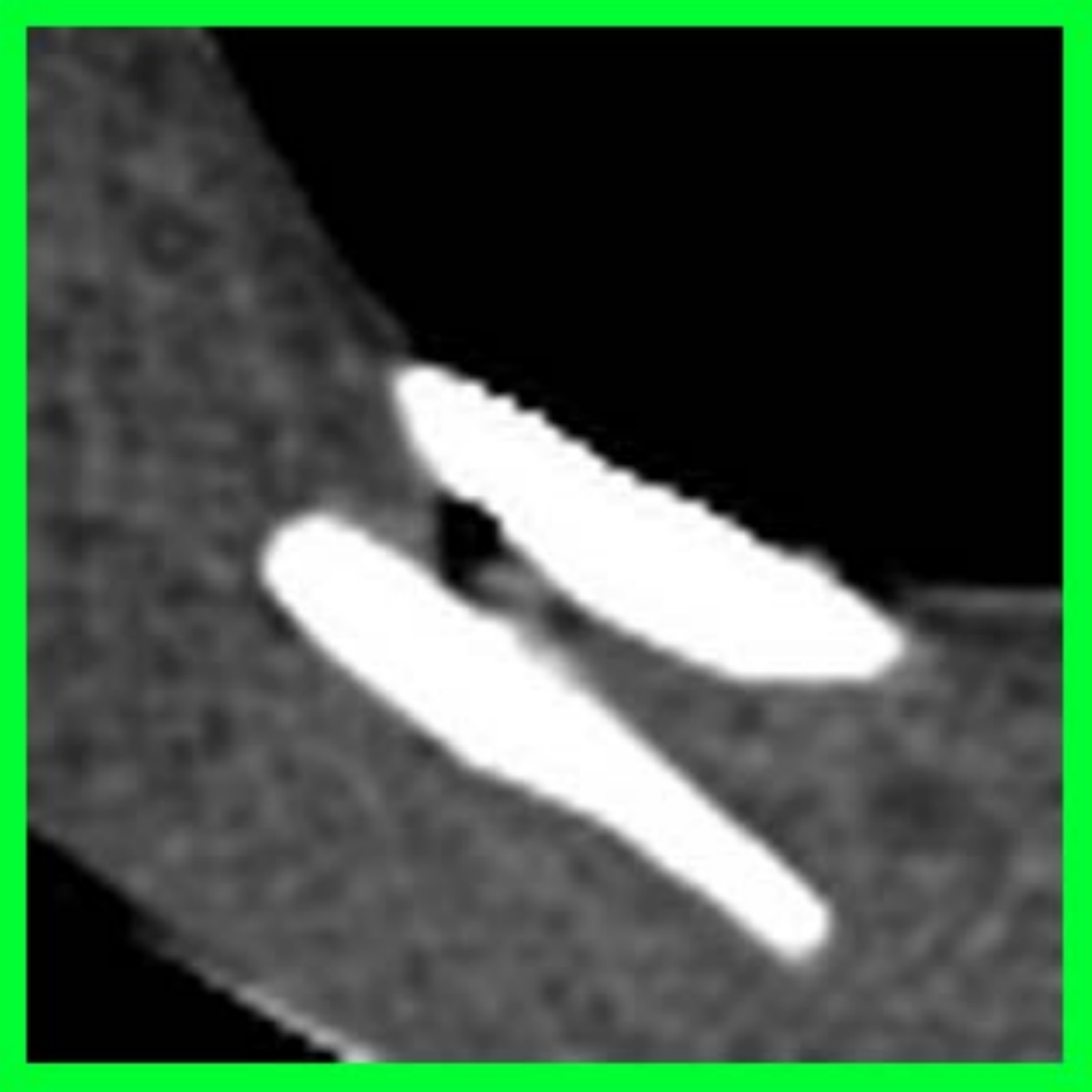} \\
			BM3D & WavResNet & RED-CNN & StructKPN\\
			33.50/0.832 & 33.15/0.828 & 33.64/0.828 & 34.74/0.838
		\end{tabular}
	\end{center}
	\caption{
		Visual comparison on PHANTOM dataset with LDCT voltage 80kV and tube current 90mA by different methods. Severe stripe artifacts can be seen in the BM3D \cite{sheng2014denoised} result. StructKPN results in better preservation of edge sharpness and more consistent texture.
	}
	\label{fig:phantomres0}
\end{figure*}

\begin{figure*}[!ht]
	\begin{center}
		\begin{tabular}{cccccccc}
			\includegraphics[width=\swfiveh]{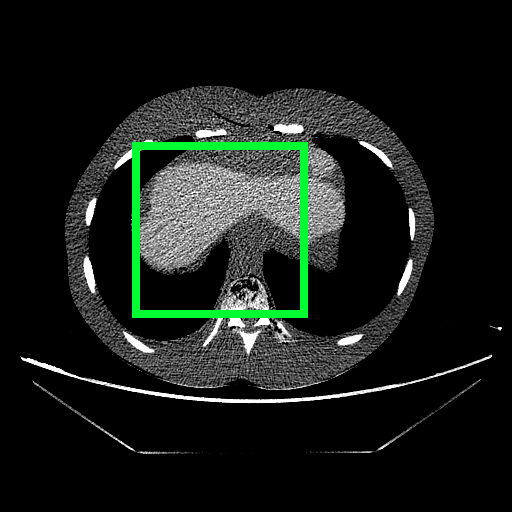} &
			\includegraphics[width=\swfiveh]{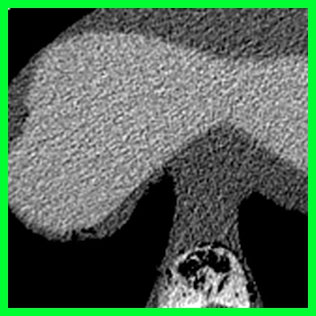} &
			\includegraphics[width=\swfiveh]{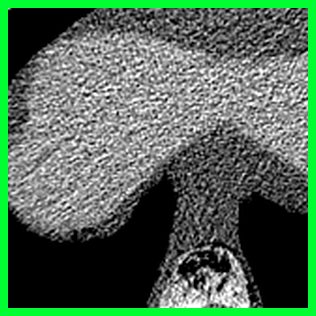} &
			\includegraphics[width=\swfiveh]{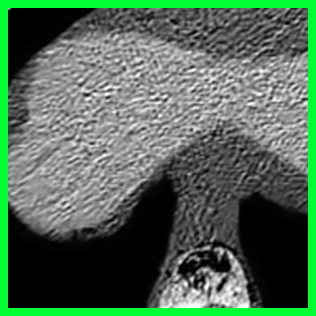} &\\
			full image & NDCT & LDCT & KSVD &\\
			& PSNR/SSIM & 21.66/0.802 & 24.34/0.803 &\\
			\includegraphics[width=\swfiveh]{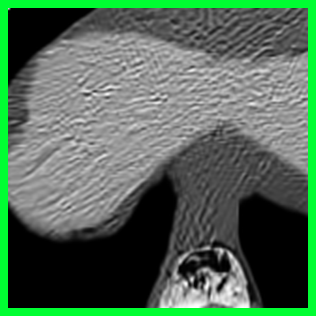} &
			\includegraphics[width=\swfiveh]{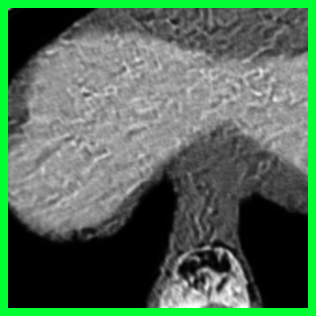} &
			\includegraphics[width=\swfiveh]{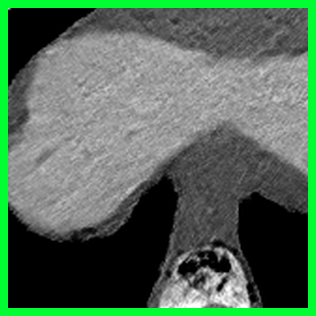} &
			\includegraphics[width=\swfiveh]{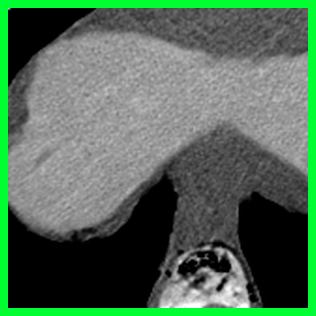} &\\
			BM3D & WavResNet & RED-CNN & StructKPN\\
			25.48/0.812 & 25.62/0.825 & 26.09/0.829 & 26.38/0.832
		\end{tabular}
	\end{center}
	\vspace{-2mm}
	\caption{
		Visual comparison on PHANTOM dataset with LDCT tube current 90mA by different methods. StructKPN results in better noise removal and more consistent texture.
	}
	\label{fig:phantomres1}
\end{figure*}
\begin{figure*}[!ht]
	\begin{center}
		\begin{tabular}{cccccccc}
			\includegraphics[width=\swfiveh]{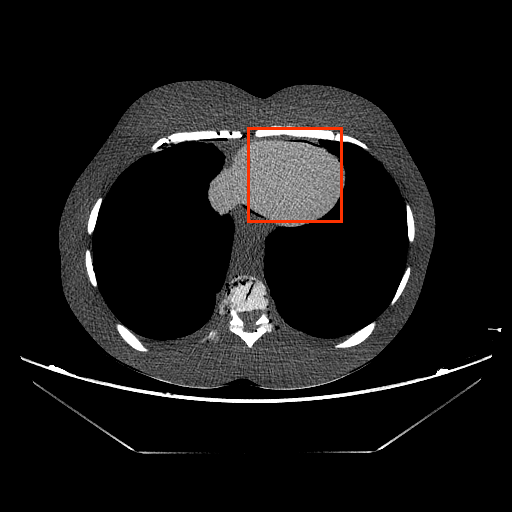} &
			\includegraphics[width=\swfiveh]{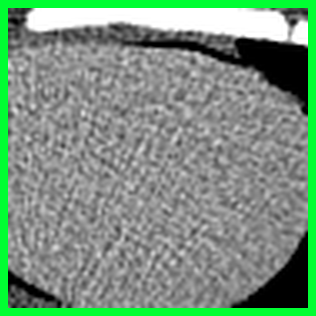} &
			\includegraphics[width=\swfiveh]{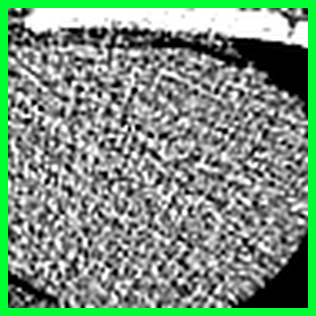} &
			\includegraphics[width=\swfiveh]{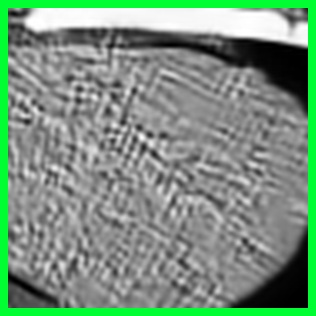} &\\
			full image & NDCT & LDCT & KSVD &\\
			& PSNR/SSIM & 19.56/0.807 & 24.80/0.795 &\\
			\includegraphics[width=\swfiveh]{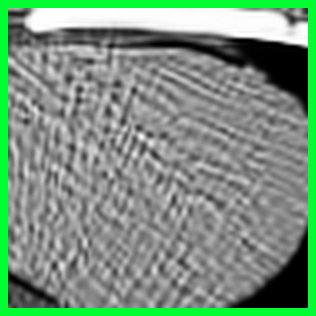} &
			\includegraphics[width=\swfiveh]{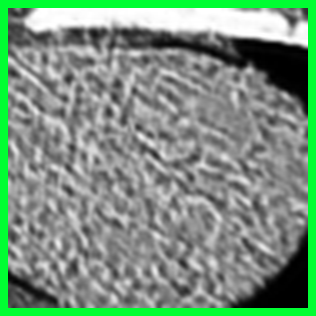} &
			\includegraphics[width=\swfiveh]{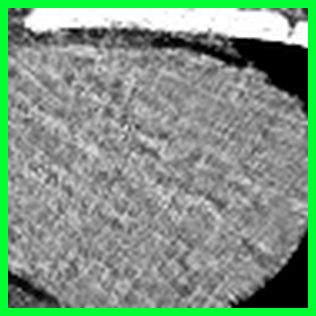} &
			\includegraphics[width=\swfiveh]{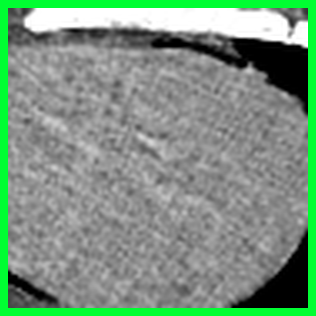} &\\
			BM3D & WavResNet & RED-CNN & StructKPN\\
			24.49/0.834 & 24.23/0.817 & 24.89/0.834 & 26.07/0.843
		\end{tabular}
	\end{center}
	\vspace{-2mm}
	\caption{
		Visual comparison on PHANTOM dataset with LDCT tube current 30mA by different methods. StructKPN results in better noise removal and fewer artifacts.
	}
	\label{fig:phantomres2}
\end{figure*}
\begin{figure*}[!ht]
	\begin{center}
		\begin{tabular}{cccccccc}
			\includegraphics[width=\swfiveh]{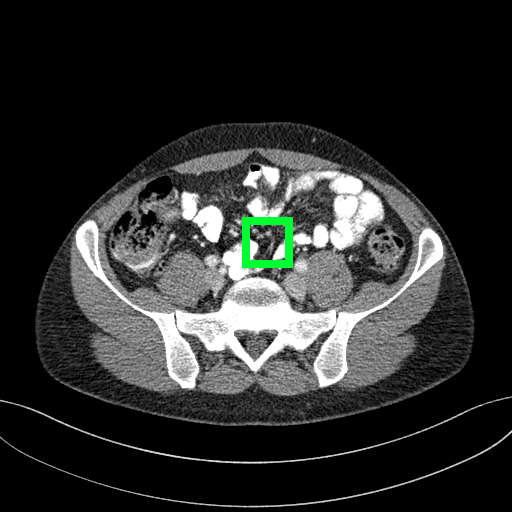} &
			\includegraphics[width=\swfiveh]{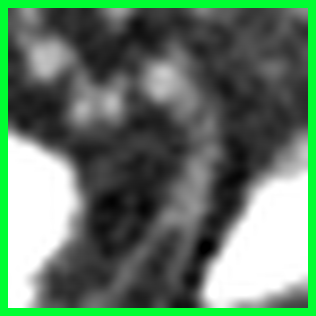} &
			\includegraphics[width=\swfiveh]{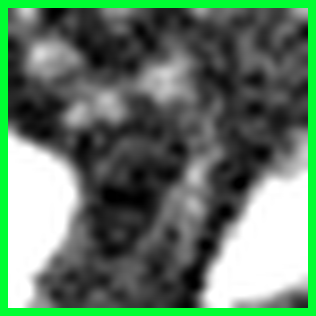} &
			\includegraphics[width=\swfiveh]{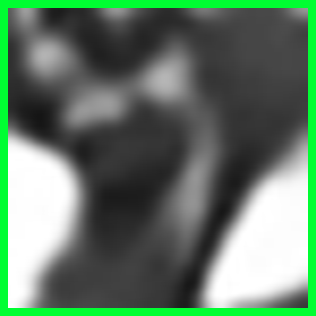} &\\
			full image & NDCT & LDCT & KSVD &\\
			& PSNR/SSIM & 28.99/0.886 & 31.19/0.887 &\\
			\includegraphics[width=\swfiveh]{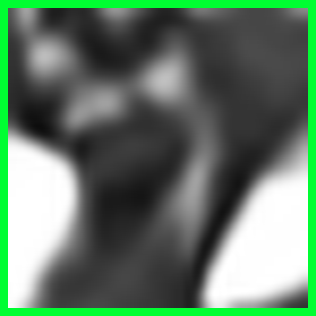} &
			\includegraphics[width=\swfiveh]{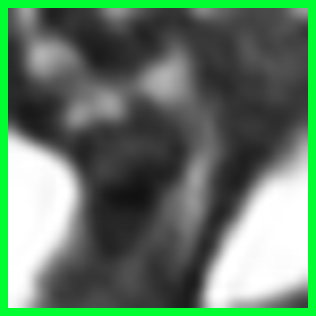} &
			\includegraphics[width=\swfiveh]{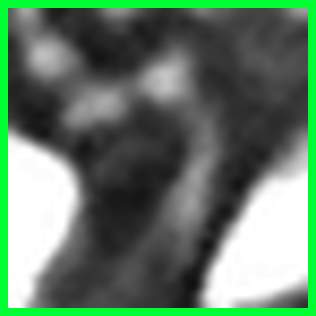} &
			\includegraphics[width=\swfiveh]{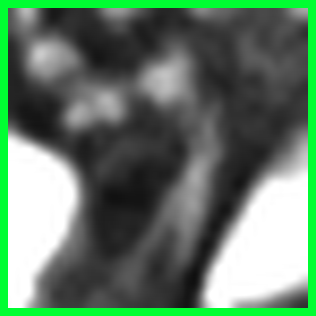} &\\
			BM3D & WavResNet & RED-CNN & StructKPN\\
			31.42/0.907 & 32.08/0.912 & 33.07/0.927 & 33.48/0.933
		\end{tabular}
	\end{center}
	\vspace{-3mm}
	\caption{
		Visual comparison on NIH dataset. StructKPN recovers clearer vein details, which is important for discriminating acute mesenteric ischemia from acute abdomen.
	}
	\label{fig:aapmres2}
\end{figure*}
\endgroup

\end{document}